\documentclass[fleqn,usenatbib]{mnras}

\usepackage[T1]{fontenc}

\DeclareRobustCommand{\VAN}[3]{#2}
\let\VANthebibliography\thebibliography
\def\thebibliography{\DeclareRobustCommand{\VAN}[3]{##3}\VANthebibliography}

\usepackage{graphicx}	
\usepackage{amsmath}	
\usepackage{amssymb}	
\usepackage{orcidlink}

\usepackage{newtxtext,newtxmath}

\newcommand{\OIII}{[\ion{O}{III}]}
\newcommand{\Ha}{H$\alpha$}
\newcommand{\Hb}{H$\beta$}
\newcommand{\Hg}{H$\gamma$}
\newcommand{\Hd}{H$\delta$}
\newcommand{\Lya}{Ly$\alpha$}
\newcommand{\He}{H$\epsilon$}

\newcommand{\OII}{[\ion{O}{II}]}
\newcommand{\SIII}{[\ion{S}{III}]}

\title[Extremely metal-poor galaxies at $2.5 < z < 6.5$]{Out of oxygen: Extremely metal-poor galaxy candidates identified at $2.5 < z < 6.5$ with deep JADES medium-band imaging}

\author[J.\@A.\@A.\@ Trussler et al.]{James A.\@ A.\@ Trussler,$^{1}$\thanks{E-mail: james.trussler@cfa.harvard.edu}
Daniel J.\@ Eisenstein,$^1$
Andrew J.\@ Bunker,$^2$
Alex J.\@ Cameron,$^{3,4}$
Stefano Carniani,$^5$
\newauthor
St\'ephane Charlot,$^6$
Jacopo Chevallard,$^2$
Christopher J.\@ Conselice,$^7$
Mirko Curti,$^8$
Emma Curtis-Lake,$^9$
\newauthor
Francesco D'Eugenio,$^{10, 11}$
Eiichi Egami,$^{12}$
Kevin Hainline,$^{12}$
Ryan Hausen,$^{13}$
Jakob M.\@ Helton,$^{14}$
\newauthor
Tiger Yu-Yang Hsiao,$^{15,16}$
Zhiyuan Ji,$^{12}$
Benjamin D.\@ Johnson,$^1$
Tobias J.\@ Looser,$^1$
Roberto Maiolino,$^{10,11,17}$
\newauthor
D\'avid Pusk\'as,$^{10, 11}$
Pierluigi Rinaldi,$^{18}$
Brant Robertson,$^{19}$
Fengwu Sun,$^1$
Sandro Tacchella,$^{10,11}$
\newauthor
Hannah \"Ubler,$^{20}$
Christina C.\@ Williams,$^{21}$
Christopher N.\@ A.\@ Willmer,$^{12}$
Joris Witstok$^{3,4}$
and Zihao Wu$^1$
\\
$^1$Center for Astrophysics $|$ Harvard \& Smithsonian, 60 Garden St., Cambridge MA 02138 USA\\
$^2$Department of Physics, University of Oxford, Denys Wilkinson Building, Keble Road, Oxford OX1 3RH, UK\\
$^3$Cosmic Dawn Center (DAWN), Copenhagen, Denmark\\
$^4$Niels Bohr Institute, University of Copenhagen, Jagtvej 128, DK-2200, Copenhagen, Denmark\\
$^5$Scuola Normale Superiore, Piazza dei Cavalieri 7, I-56126 Pisa, Italy\\
$^6$Sorbonne Universit\'e, CNRS, UMR 7095, Institut d'Astrophysique de Paris, 98 bis bd Arago, 75014 Paris, France\\
$^7$Jodrell Bank Centre for Astrophysics, University of Manchester, Oxford Road, Manchester M13 9PL, UK\\
$^8$European Southern Observatory, Karl-Schwarzschild-Strasse 2, 85748 Garching, Germany\\
$^9$Centre for Astrophysics Research, Department of Physics, Astronomy and Mathematics, University of Hertfordshire, Hatfield AL10 9AB, UK\\
$^{10}$Kavli Institute for Cosmology, University of Cambridge, Madingley Road, Cambridge, CB3 0HA, UK\\
$^{11}$Cavendish Laboratory, University of Cambridge, 19 JJ Thomson Avenue, Cambridge, CB3 0HE, UK\\
$^{12}$Steward Observatory, University of Arizona, 933 N. Cherry Avenue, Tucson, AZ 85721, USA\\
$^{13}$Department of Physics and Astronomy, The Johns Hopkins University, 3400 N. Charles St., Baltimore, MD 21218\\
$^{14}$Department of Astronomy \& Astrophysics, The Pennsylvania State University, University Park, PA 16802, USA\\
$^{15}$Department of Astronomy, The University of Texas at Austin, 2515 Speedway, Austin, Texas 78712, USA\\
$^{16}$Cosmic Frontier Center, The University of Texas at Austin, Austin, TX 78712, USA\\
$^{17}$Department of Physics and Astronomy, University College London, Gower Street, London WC1E 6BT, UK\\
$^{18}$Space Telescope Science Institute, 3700 San Martin Drive, Baltimore, Maryland 21218, USA\\
$^{19}$Department of Astronomy and Astrophysics University of California, Santa Cruz, 1156 High Street, Santa Cruz CA 96054, USA\\
$^{20}$Max-Planck-Institut f\"ur extraterrestrische Physik (MPE), Gie{\ss}enbachstra{\ss}e 1, 85748 Garching, Germany\\
$^{21}$NSF–DOE Vera C. Rubin Observatory/NSF NOIRLab, 950 N. Cherry Ave., Tucson, AZ 85719, USA\\
}

\date{Accepted XXX. Received YYY; in original form ZZZ}

\pubyear{2026}

\begin{document}
\label{firstpage}
\pagerange{\pageref{firstpage}--\pageref{lastpage}}
\maketitle

\begin{abstract}
\emph{JWST} is beginning to uncover a population of extremely metal-poor galaxies (EMPGs, $Z < 1\%~\mathrm{Z}_\odot$) at redshift $z > 3$, mostly through serendipitous NIRSpec discoveries and blind slitless spectroscopy. To accelerate our understanding of pristine star formation, we further develop a methodology to identify EMPG candidates from photometry, utilizing the extensive deep NIRCam medium-band imaging from JADES. Our EMPG candidates at $2.5 < z < 6.5$ exhibit strong photometric boosts by \Ha, yet correspondingly weak boosts by \OIII\ + \Hb, likely indicating extremely low metallicity to explain their lack of \OIII\ emission. We further demand our EMPG candidates to have strong Balmer jumps, as revealed by medium-band imaging, to ensure that they are young starbursts, as opposed to broad-line AGN/LRDs, though contamination by dusty/dense-gas starbursts and highly-obscured AGN remains a concern. SED-fitting with near-pristine models (${\sim}0.1$--$1\%~\mathrm{Z}_\odot$) indicates that our 14 EMPG candidates are low-mass (median $M_* \approx 10^{6.7}~\mathrm{M}_\odot$), faint dwarf galaxies ($M_\mathrm{UV} \approx -16.9$), with high ionizing photon production efficiencies ($\log\, (\xi_\mathrm{ion, obs}/\mathrm{(Hz\ erg^{-1})}) \approx 26.0$). Hence these are plausible sites of near-pristine star formation, constituting ${\sim}0.02$--$0.4\%$ of $2.5 < z < 6.5$ galaxies at $-19 < M_\mathrm{UV} < -16$. We discuss this extremely metal-poor extension to the mass--metallicity relation. We forecast that deep (${\sim}28$~h) NIRCam slitless spectroscopy can identify bright EMPGs through strong \Hb\ but lack of \OIII\ emission, or secure the redshifts of fainter systems through \Ha\ detections. Highly-multiplexed NIRSpec spectroscopy offers an alternate route to discovering the faintest pristine galaxies out to $z=10$, without requiring deep NIRCam medium-band and/or MIRI imaging to identify secure candidates.
\end{abstract}





\section{Introduction} \label{sec:intro}

\emph{JWST} is ever extending the observational frontier, progressively pushing to even earlier epochs, currently detecting galaxies beyond $z=14$ \citep[less than 300~Myr after the Big Bang,][]{Carniani2024, Carniani2025, Helton2025, Schouws2025b, Schouws2025, Naidu2026}. Yet the `first galaxies', composed purely of hydrogen and helium, remain elusive, with all $z > 10$ galaxies detected thus far exhibiting metal lines and thus representing non-pristine conditions \citep{Bunker2023, Castellano2024, D'Eugenio2024, Hainline2024, Zavala2024, Zavala2025, Helton2025, Schouws2025b, Schouws2025, Naidu2026, Witstok2026}. Owing to their metal-free nature, Population III galaxies have unique characteristics, resulting in distinct spectra and associated photometry, aiding their identification \citep{Schaerer2002, Schaerer2003, Nagao2008, Raiter2010, Inoue2011, Zackrisson2011, Nakajima2022, Nishigaki2023, Trussler2023, Maiolino2024, Maiolino2026, Fujimoto2025b, Fujimoto2025, Rusta2026, Ubler2026}. Due to their likely very hot stars, a consequence of pristine star formation \citep[though it is currently uncertain how top-heavy the Pop III initial mass function actually is, e.g.\@][]{Klessen2023}, Pop III galaxies have extremely high ionising photon production efficiencies powering exceptionally hot \ion{H}{II} region temperatures, resulting in strong hydrogen recombination emission, leading to e.g.\@ large \Ha\ equivalent widths \citep[$> 3000$~\AA\ for a few Myr old starburst,][]{Trussler2023} and nebular continuum emission that dominates over the starlight that is powering it \citep{Cameron2024, Katz2025, Trussler2026}. Moreover, Pop III galaxies lack metal emission lines that can otherwise be very prominent in metal-enriched galaxy spectra, such as \OIII\ $\lambda\lambda 4959, 5007$ and \SIII\ $\lambda\lambda 9069, 9531$, with the complete absence of metals thus being a key Pop III diagnostic feature \citep[e.g.\@][]{Trussler2023, Nishigaki2023, Fujimoto2025b, Fujimoto2025}. Finally, the extremely hot Pop III stars power a hard ionising spectrum that is very effective at doubly ionising helium, resulting in characteristically strong \ion{He}{II} recombination emission, the definitive Pop III indicator \citep[e.g.\@][]{Schaerer2002, Schaerer2003, Katz2023, Trussler2023}. Observations of Pop III galaxies and near-pristine extremely metal-poor galaxies (EMPGs, here defined as $Z_\mathrm{gas} < 1\%~\mathrm{Z}_\odot$) should provide unique insights on the nature of the first stars. Measurements of the \ion{He}{II} $\lambda 1640$ equivalent width can in principle help distinguish between different Pop III initial mass functions \citep[IMFs,][]{Trussler2023}. Additionally, the shape of the nebular continuum can constrain primordial \ion{H}{II} region temperatures \citep{Cameron2024, Katz2025, Trussler2026}, which together with the luminosity of the Balmer lines can give further indirect insights on the ionising spectrum. Moreover, metal--metal line ratios in EMPGs can deliver elemental abundance ratios, yielding the chemistry of the primordial nucleosynthesis pattern \citep[see e.g.\@][]{D'Eugenio2024, Scholtz2025b, Nakajima2026}.

\emph{JWST} has recently ushered in a new era, marked by a dramatic rise in the number of known $z > 3$ EMPGs with $Z \lesssim 1\%~\mathrm{Z}_\odot$, made possible by extensive spectroscopic programmes, via both blind NIRCam and NIRISS slitless spectroscopy, as well as targeted NIRSpec observations. Slitless spectroscopy requires no photometric pre-selection, allowing chance EMPG discoveries (albeit inefficiently in terms of line sensitivity) through deep integrations. In this way, \citet{Hsiao2025} discover 7 EMPG candidates with $Z_\mathrm{gas} < 0.02~\mathrm{Z}_\odot$ at $5 < z < 7$ from 9.4~h SAPPHIRES-EDR NIRCam F356W spectroscopy \citep{Sun2025}, identified by their distinctly weak \OIII\ $\lambda 5007$ emission relative to \Hb, resulting in low \OIII\ $\lambda 5007$/\Hb\ ratios $R_3 < 3$ \citep[compared to normal galaxies, which typically have $R_3 \approx 5$,][]{Matthee2023, Meyer2024} that can be used to infer gas-phase metallicities $12 + \log(\mathrm{O/H})$ via strong-line metallicity calibrations \citep[e.g.\@][]{Sanders2024, Sanders2026, Cataldi2025}. Combining deep F356W slitless spectroscopy from the ALT programme \citep{Naidu2024} with tremendous gravitational lensing ($\mu = 39, 78$), \citet{Morishita2025} identify the most pristine (i.e.\@ with the lowest inferred oxygen abundance) galaxy currently known, AMORE6 at $z=5.725$, with an $8\sigma$ \Hb\ detection combined with a \OIII\ $\lambda 5007$ non-detection resulting in a metallicity upper limit $12 + \log (\mathrm{O/H}) = 5.78$, i.e.\@ ${\sim}0.1\%~\mathrm{Z}_\odot$. The sensitivity of NIRSpec in principle enables the study of intrinsically less luminous galaxies via targeted observations, with the EMPGs found thus far selected via circumstantial evidence indirectly supporting a possibly metal-poor nature, or from serendipitous discoveries \citep[see the $z=3.19$ EMPG candidate reported by][]{Cai2025}. \citet{Chemerynska2024}, \citet{Willott2025} and \citet{Vanzella2026} find EMPGs amongst highly-magnified, intrinsically faint galaxies targeted for their great magnifications, thus having low stellar mass and hence may be expected to be relatively metal-poor. Moreover, \citet{Cullen2025} find evidence for very low metallicity in a $z=8.271$ galaxy originally targeted for its extremely blue UV slope, indicative of exotic stellar populations. Additionally, \citet{Vanzella2023} and subsequently \citet{Nakajima2026} establish weak \OIII\ emission and thus extremely low metallicity in the $z=6.639$ LAP1 galaxy targeted for its very high \Lya\ equivalent width emission in MUSE IFU spectroscopy \citep{Vanzella2020}, suggesting a very young starburst galaxy (with the comparatively bright \Lya\ greatly aiding in the detection of intrinsically very faint galaxies that are otherwise very challenging to detect with continuum photometry). Finally, the low $R_3$ ratio associated with the narrow component of \Hb\ for the lensed Little Red Dot \citep[LRD,][]{Furtak2023, Furtak2024, Furtak2025, Matthee2024} Abell2744-QSO1 at $7.04$ indicates an extremely metal-poor host galaxy \citep{Maiolino2026b} for this early growing AGN/LRD. Still, direct photometric evidence (weak oxygen emission relative to hydrogen suggesting a lack of oxygen) strongly favouring a metal-poor nature was absent for these EMPGs prior to their spectroscopic follow-up. Such direct evidence could guide future NIRSpec efforts, given the faintness and rarity of EMPGs.  

Indeed, targeted spectroscopy of promising EMPG candidates with the emission line sensitivity of the NIRSpec gratings would be highly beneficial, enabling statistical samples of spectroscopically-confirmed EMPGs to be built efficiently through optimised observations. \citet{Trussler2023} introduce various colour selections for identifying Pop III galaxy candidates at $z \sim 8, 10$ using NIRCam + MIRI imaging \citep[see also][]{Zackrisson2011, Nishigaki2023}. The characteristically strong \Ha\ line emission, combined with the complete lack of metal lines results in unique Pop III photometry, with strong photometric boosts by \Ha\ and correspondingly weak boosts by \OIII\ + \Hb\, leading to characteristically red $m_\mathrm{[O\ III]} - m_\mathrm{H\alpha}$ colours in the wide-band filters covering \OIII\ and \Ha. Moreover, the exceptionally strong \ion{He}{II} $\lambda 1640$ emission, the potential signature feature of Pop III galaxies, results in prominent photometric boosts in medium-band photometry tracing this line, yielding unique Pop III colours \citep[see also][]{Nagao2008}. However, the practical application of these colour selection principles out to $z \sim 8$ is limited, due to the much lower sensitivity (${\sim}2$~mag, i.e.\@ $40\times$ longer exposure time) and imaging field of view ($\approx 4\times$ smaller) of MIRI with respect to NIRCam, together with the currently much more limited area of MIRI F560W/F770W and NIRCam short wavelength medium-band imaging compared to the general NIRCam wide bands. 

Motivated by this, \citet{Fujimoto2025} push the Pop III selection to lower redshifts ($z < 6.6)$, so that the key selection principle of strong \Ha\ emission, yet weak \OIII\ emission, can be applied using the much greater sensitivity and area of NIRCam imaging alone. Moreover, they further develop the Pop III selection procedure, also demanding a Balmer jump, a free--bound hydrogen recombination feature associated with young starbursts \citep[see e.g.\@][]{Cameron2024, Katz2025, Trussler2026}, revealed via medium-band imaging through the deficit in rest-frame optical continuum level relative to the rest-frame UV. Applying these novel Pop III selections to a plethora of NIRCam imaging data, they identify one promising Pop III candidate GLIMPSE-16043, which is moderately lensed ($\mu = 2.9$) by the Abell S1063 galaxy cluster in the ultra-deep GLIMPSE \citep{Atek2025} imaging. Constituting the deepest effective NIRCam imaging yet, the GLIMPSE data likely provides the best current avenue with which to detect Pop III galaxies, due to their expected ultra-faint nature. However, deep 30~h follow-up NIRSpec G395M grating spectroscopy revealed moderately strong \OIII\ emission in GLIMPSE-16043 at $z=6.20$, with \OIII\ $\lambda 5007$/\Hb\ = 1.77, conclusively ruling out the metal-free Pop III scenario \citep{Fujimoto2025b}. Thus the photometric identification of Pop III galaxies and EMPGs can be challenging, requiring careful consideration of the data. The $m_\mathrm{[O\ III]} - m_\mathrm{H\alpha}$ colour for GLIMPSE-16043 is only modestly red (0.22~mag) compared to the highly red (${\sim}0.9$~mag) colour expected for Pop III galaxies \citep{Trussler2023}, its Pop III identification thus requiring a best-fit $z=6.50$ where \Ha\ is approaching the lower throughput edge of the F444W filter. Additionally, the strong inferred Balmer jump (exceeding those accessible by current models) suggests a low optical continuum level, implying very high emission line equivalent widths and photometric boosts, allowing for a relatively higher (but still low) metallicity to produce a given red $m_\mathrm{[O\ III]} - m_\mathrm{H\alpha}$ colour. Thus, \citet{Fujimoto2025b} advocate for high signal-to-noise ratio (SNR) medium-band continuum measurements to mitigate against non-Pop III contamination. Moreover, the $\chi^2$ for the best Pop III, metal-poor, and metal-enriched model fits are all less than or similar to $N_\mathrm{data}$, the expectation value of the $\chi^2$ in the case of a very good model fit, suggesting a possible non-pristine nature for GLIMPSE-16043 despite the $\Delta \chi^2$ preference for the Pop III model. 

Building on previous photometric Pop III selections \citep{Trussler2023, Fujimoto2025b, Fujimoto2025} targeting strong \Ha\ but weak \OIII\ emission, we extend the wide-band-only Pop III selection of \citet{Trussler2023} to a medium-band selection of extremely metal-poor galaxies, thus shifting from the extreme case of the youngest pristine (Pop III) starbursts to slightly older non-pristine (EMPG) starburst galaxies. We utilise the deep extensive NIRCam medium-band imaging from JADES to identify close-to-pristine (${\sim}0.001\textrm{--}0.01~\mathrm{Z}_\odot$) EMPG candidates at $2.5 < z < 6.5$, the exceptionally deep blank-field imaging helping to bring these very faint near-pristine candidates into view. We empirically select plausible EMPG candidates from medium-band constraints on the rest-frame optical/NIR continuum level, using photometric boost measurements to identify galaxies with strong \Ha\ but weak \OIII\ emission, through large \Ha\ equivalent widths relative to \OIII\ + \Hb. We employ multiple continuum estimates per medium-band filter to maximise their redshift range of applicability, increasing the EMPG search volume. Close-to-pristine SED model-fitting with \textsc{Bagpipes} \citep{Carnall2018} is used to establish the final 14 EMPG candidates that strongly favour the extremely metal-poor solution $(Z < 0.01~\mathrm{Z}_\odot)$. We report on this near-pristine extension to the mass--metallicity relation, the empirical properties of our EMPG candidates, and prospects for wide-band-only selections based off our EMPG colours. We put our EMPG candidates in a cosmological context through comparison of their observed number statistics with simulations, and provide forecasts for EMPG observability with future \emph{JWST} spectroscopy. 

This article is structured as follows. In Section~\ref{sec:data} we describe the data used in our analysis. In Section~\ref{sec:selection} we outline our medium-band selection procedure for identifying preliminary EMPG candidates, our close-to-pristine SED-fitting procedure, and a brief overview of our final EMPG candidates. In Section~\ref{sec:properties} we discuss the inferred properties of our EMPG candidates. Finally we conclude in Section~\ref{sec:conclusions}. We adopt the AB magnitude system \citep{Oke1983}, and assume a flat $\Lambda$CDM cosmology with $\Omega _\mathrm{m}$ = 0.3 and $H_0 = 70~\mathrm{km}~\mathrm{s}^{-1}~\mathrm{Mpc}^{-1}$.

\section{Data} \label{sec:data}

We utilise imaging data in the GOODS-S and GOODS-N fields \citep{Giavalisco2004}, incorporating the JADES DR5 footprint. For details on the imaging area and depth, as well as the data reduction procedure (including wisp removal), cataloging process, and photometric redshift determination, we refer the reader to the following DR5 papers, respectively: \citet{Johnson2026}, \citet{Wu2026}, \citet{Robertson2026}, \citet{Hainline2026}. We restrict our analysis to regions of the DR5 footprint where medium-band imaging is available. As we shall discuss, medium bands greatly aid in the secure identification of EMPG candidates, by increasing the confidence in their selection, also providing tighter photometric constraints on their gas-phase metallicities. We make use of the F162M, F182M, F210M, F250M, F300M, F335M, F410M, F430M, F460M, F480M imaging that is available. The extensive medium-band filter set allows for the identification of EMPGs across a broad redshift range ($2.5 < z < 6.5$). Most of the medium-band depth and area stems from JADES-related programmes, namely the JADES GTO \citep{Bunker2020, Bunker2024, Rieke2020, Rieke2023, Hainline2024, D'Eugenio2025b, Eisenstein2026}, the JADES Origins Field \citep{Eisenstein2025}, JEMS \citep{Williams2023} and OASIS (Looser et al., in prep.). This is supplemented by data from FRESCO \citep{Oesch2023}, PANORAMIC \citep{Williams2025} and BEACON \citep{Morishita2025b}. We further use  NIRCam wide-band imaging (F070W, F090W, F115W, F150W, F200W, F277W, F356W, F444W) from the previously-mentioned programmes, as well as HST/ACS imaging in F435W, F606W, F775W, F814W \citep{Giavalisco2004, Beckwith2006, Grogin2011, Koekemoer2011, Ellis2013, Whitaker2019}.

We use 0.2~arcsec diameter circular aperture photometry at each filter's native PSF, adopting this narrow aperture to push to the faintest possible systems, among which we might expect to find near-pristine galaxies. Photometry is corrected for Galactic dust extinction using the \citet{Schlafly2011} $\mathrm{E(B-V)}$ dust map, adopting the \citet{Fitzpatrick1999} dust attenuation law with total-to-selective extinction ratio $R_\mathrm{V}=3.1$. We scale absolute rest-frame UV magnitudes $M_\mathrm{UV}$, stellar masses $M_*$, star formation rates $\mathrm{SFR}$ and \Ha\ luminosities $L_\mathrm{H\alpha}$ determined from the 0.2~arcsec diameter photometry to total, through the ratio of flux densities measured by the Kron elliptical aperture and the 0.2~arcsec diameter circular aperture in the appropriate wide-band filter, adopting a minimum ratio of 1. The appropriate wide-band filters are determined based off the source's best-fit EAZY photometric redshift, derived following the procedures described in \citet{Hainline2024} and \citet{Hainline2026}. For $M_\mathrm{UV}$, we use the first wide-band filter fully redward of \Lya\ to scale the circular aperture measurement to total. For the circular aperture-to-total scaling of $M_*$, $\mathrm{SFR}$ and $L_\mathrm{H\alpha}$, we use the wide-band filter covering \Ha. We note that, owing to the deep imaging, resulting in more precise measurements of the Ly($\alpha$) break, together with the extensive medium bands, enabling further redshift constraints from line boosts/continuum deficits \citep[i.e.\@ the Balmer jump, see][]{Trussler2026}, the JADES photometric redshifts are very precise, with fractional redshift uncertainties $\Delta z /(1+z) \approx 0.01\textrm{--}0.02$ \citep{Hainline2024, Hainline2026, Helton2024}. 

To increase the EMPG search volume, we employ three different continuum estimates per medium-band filter, covering three different rest-frame wavelengths, thus maximising the redshift range probed by each filter.
\begin{enumerate}
    \item To estimate the rest-frame optical continuum, we utilise the medium band at redshifts where it resides between \OIII\ $\lambda 5007$ and \Ha\ \citep[for more details, see][]{Trussler2026}. 
    \item To estimate the rest-frame NIR continuum, we require the medium band to reside between \Ha\ and \SIII\ $\lambda 9069$.
    \item Finally, to estimate the continuum around \Ha, we require \Ha\ to reside in the wavelength range spanned by the medium-band filter. 
\end{enumerate}

For each medium-band filter and continuum estimate, we select all galaxies in the appropriate redshift range for that filter+continuum combination. We further require $>5\sigma$ detections in the wide-band filters fully redward of the \Lya\ break, though allowing up to 4 filters with $2.5 < \mathrm{SNR} < 5$, to account for variations in depth and noise in the photometry. Additionally, we require the source to be undetected ($<3 \sigma$) in all filters fully blueward of the \Lya\ break (at higher redshift), or the Lyman break (at lower redshift, where the \Lya\ break is much weaker). The appropriate redshifts for this dropout requirement are determined by establishing when the bandpass-averaged flux density in a given filter drops out by $> 5$~mag, assuming a flat $f_\nu$ spectrum with the \citet{Inoue2014} IGM prescription. Combining the galaxies from all the medium-band filter + continuum combinations yields the general high-redshift galaxy sample. In later comparisons against our EMPG candidates, we further require our non-EMPG high-redshift galaxy comparison (and Balmer-jump) sample to have reduced $\chi^2_\mathrm{red}$ < 6.25 from the JADES EAZY fitting, ensuring adequate fits so that the photometric redshifts and thus empirically-determined parameters are reasonable.

We further utilise the 7~h NIRSpec/PRISM spectrum of a $z=5.39$ EMPG candidate (ID 123650) selected as a MSA filler target as part of program 8060 (PI: Egami), which is conducting both multi-epoch imaging and deep NIRSpec follow-up of transients in the JADES Deep field in GOODS-S. We discuss this spectrum in Section~\ref{subsec:123650}.

\section{Extremely metal-poor galaxy selection} \label{sec:selection}

\begin{figure*}
\centering
\includegraphics[width=.7\linewidth]{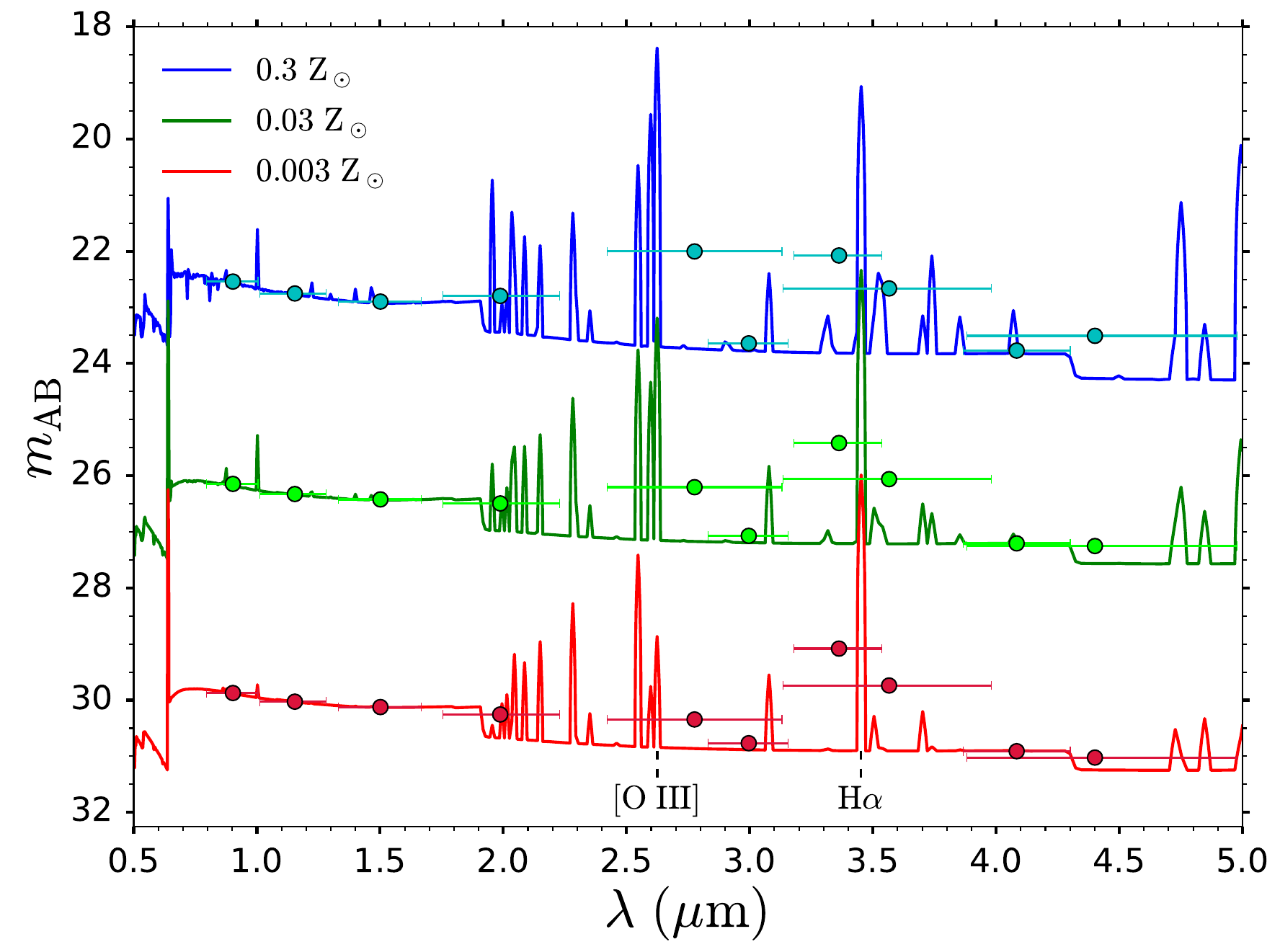}
\caption{SEDs and associated NIRCam photometry for 1~Myr old instantaneous starbursts at $z=4.25$, generated using \textsc{Bagpipes}. The \OIII\ + \Hb\ and \Ha\ emission lines (denoted) boost the bandpass-averaged flux densities in the F277W and F356W filters above the rest-frame optical continuum level. For typical galaxies, with moderate metallicities (e.g.\@ $Z = 0.3~\mathrm{Z}_\odot$, blue), this results in blue $m_\mathrm{[O\ III]}$ - $m_\mathrm{H\alpha}$ colours in the wide-band filters tracing these emission lines. With decreasing metallicity, the \OIII\ boost wanes, resulting in level photometry at low metallicity ($0.03~\mathrm{Z}_\odot$, green), and characteristically red $m_\mathrm{[O\ III]}$ - $m_\mathrm{H\alpha}$ colours for extremely metal-poor galaxies (e.g.\@ $0.003~\mathrm{Z}_\odot$, red), reflecting strong boosts by \Ha, but correspondingly weak boosts by \OIII. Medium bands disentangle continuum and line emission, enabling EMPG candidates to be identified with greater confidence and with tighter metallicity constraints. Medium bands can directly trace the rest-frame optical (F300M) and NIR (F410M) continuum levels, also providing independent measurements of line boosts (F335M) which can be combined with wide-band measurements to obtain indirect continuum constraints. These continuum measurements enable empirical estimates of emission line equivalent widths from the line boosts, as well as a determination of the Balmer jump, a free--bound hydrogen recombination feature (at 3646~\AA\ rest-frame, roughly at 1.9~\textmu m observed-frame) associated with young starbursts, through the deficit in rest-frame optical/NIR continuum level compared to the UV. Note that the $Z = [0.3,\, 0.03,\,  0.003]~\mathrm{Z}_\odot$ models correspond to stellar masses $M_* = [10^9,\, 10^{7.5}, 10^6]~\mathrm{M}_\odot$, respectively, and hence are vertically offset for visual clarity.}
\label{fig:Ha_vs_OIII}
\end{figure*}

In this section we identify extremely metal-poor galaxy candidates using NIRCam photometry, the key principle being that EMPGs have very little oxygen, resulting in young starbursts exhibiting strong photometric boosts by \Ha, but correspondingly weak boosts by \OIII, yielding characteristically red colours between the wide-band filters tracing \OIII\ and \Ha, respectively. Medium-band photometry disentangles line and continuum emission, enabling the identification of EMPGs with greater confidence, also placing tighter constraints on the photometrically-inferred gas-phase metallicity. We discuss our empirical pre-selection of plausible EMPG candidates in Section~\ref{subsec:empirical_sel}. We fit close-to-pristine SED models to identify the most secure EMPG candidates in Section~\ref{subsec:sed}. We introduce our final EMPG candidates in Section~\ref{subsec:candidates}. Finally, motivated by the NIRSpec PRISM spectrum of EMPG candidate ID 123650, we discuss how dusty starbursts and starbursts radiating into dense gas can masquarade as EMPG candidates in photometry in Section~\ref{subsec:123650}.

\subsection{Empirical selection: Strong \Ha, weak \OIII} \label{subsec:empirical_sel}

\subsubsection{Photometric features of EMPGs}

We begin by outlining the principle behind identifying EMPG candidates from photometric data. We show SEDs for 1~Myr old instantaneous starbursts at $z=4.25$ of various gas-phase metallicities (various colours) generated by \textsc{Bagpipes} \citep{Carnall2018} in Fig.~\ref{fig:Ha_vs_OIII}, as well as their associated bandpass-averaged flux densities in NIRCam wide- and medium-band filters. The \Ha\ and \OIII\ $\lambda\lambda 4959, 5007$ + \Hb\ lines (wavelengths denoted by the vertical dashed lines) boost the bandpass-averaged flux densities in the F356W and F277W filters above the rest-frame optical continuum level, respectively. For the vast majority of galaxies, which have moderate-to-high metallicities, the \OIII\ $\lambda 5007$ emission is strong relative to \Ha, resulting in an \OIII\ + \Hb\ photometric boost that exceeds that of \Ha, leading to blue $m_\mathrm{[O\ III]} - m_\mathrm{H\alpha}$ colours in typical galaxies, such as the $0.3~\mathrm{Z}_\odot$ model (blue). However, as the metallicity declines to very low levels, roughly the metallicity floor currently encountered by general high-redshift galaxy surveys \citep[see e.g.\@][]{Nakajima2023, Curti2024, Sarkar2025}, the \OIII\ emission wanes, with the once prominent blue gap between the \OIII\ and \Ha\ filters now becoming level, as shown for the $0.03~\mathrm{Z}_\odot$ model (green). Note that in this case the combined \OIII\ + \Hb\ equivalent width is already below that of \Ha, despite the similar flux densities in the two wide-band filters, due to the narrower filter used to cover \OIII. This behaviour can be seen from the approximate relation between the magnitude boost $\Delta m$ by an emission line and its rest-frame equivalent width $\mathrm{EW_{rest}}$: $\Delta m = -2.5\log_{10}(1 + \mathrm{EW_{rest}}(1+z)/\Delta\lambda))$ where the narrower filter width $\Delta\lambda$ stems from the roughly constant spectral resolution $R$ of the NIRCam filters, with $\Delta\lambda = \lambda / R$. It is only in the case of extremely low metallicity, currently marking the metallicity frontier probed by \emph{JWST} \citep{Morishita2025}, that the \OIII\ emission is so weak that \Hb\ dominates, with the photometric boost by \OIII\ + \Hb\ being comparatively weak compared to \Ha, resulting in a characteristically red $m_\mathrm{[O\ III]} - m_\mathrm{H\alpha}$ colour, such as the $0.003~\mathrm{Z}_\odot$ model (red). With decreasing \OIII\ emission (i.e.\@ lower metallicity), the $m_\mathrm{[O\ III]} - m_\mathrm{H\alpha}$ colour becomes progressively redder, resulting in a prominent red gap in wide-band filters tracing \OIII\ and \Ha.

Now the scenario shown in Fig.~\ref{fig:Ha_vs_OIII} is the extreme case, with maximally young 1~Myr old instantaneous starbursts. Here the line equivalent widths are at their strongest, maximising the line boosting effect, resulting in the most prominent colour gaps between the \OIII\ and \Ha\ filters. In the case of older starbursts, or with emission from an underlying older stellar population present, the line equivalent widths will be lower, resulting in a smaller colour gap between the \OIII\ and \Ha\ filters for a given metallicity, tending to zero in the limit of zero equivalent width.\footnote{We note that dust attenuation similarly affects emission lines and their neighbouring continuum, preserving line equivalent widths in the case of a single stellar population, though will increase $m_\mathrm{[O\ III]} - m_\mathrm{H\alpha}$. In the extreme case of a very dusty starburst, together with an underlying older stellar population with much less dust attenuation, the \OIII\ equivalent widths will decrease by a greater factor than \Ha. We discuss this scenario in detail in Section~\ref{subsec:123650}.} Hence a given intermediate red $m_\mathrm{[O\ III]} - m_\mathrm{H\alpha}$ colour could be attributable to an older extremely metal-poor starburst, or a younger starburst of slightly higher metallicity. Considering the $m_\mathrm{[O\ III]} - m_\mathrm{H\alpha}$ colour alone there is therefore an ambiguity in the metallicity, with a given colour requiring lower metallicity as the equivalent width decreases.

The benefit of medium bands is that they can disentangle line and continuum emission \citep[see e.g.\@][]{Trussler2025b, Trussler2026}, therefore removing this ambiguity, resulting in tighter constraints on the inferred gas-phase metallicities. As shown in Fig.~\ref{fig:Ha_vs_OIII}, medium bands (at $z=4.25$, F300M) are narrow enough to reside in the wavelength interval between the prominent \OIII\ $\lambda5007$ and \Ha\ lines, providing a clean measurement of the rest-frame optical continuum level. Hence the line photometric boosts above the continuum level are now clearly evident, and the \OIII\ + \Hb\ and \Ha\ equivalent widths can be determined, both empirically (through the observed photometric boosts) and through full SED-fitting. Through the medium bands we can now identify strong \Ha\ but weak \OIII\ emission from the ratio of the \OIII\ + \Hb\ to \Ha\ equivalent widths $\mathrm{EW_{[O\ III] + H\beta}} / \mathrm{EW_{H\alpha}}$, rather than just the more ambiguous $m_\mathrm{[O\ III]} - m_\mathrm{H\alpha}$ colour. This thus extends the Pop III wide-band selection of \citet{Trussler2023} that was limited to the youngest pristine ($Z_\mathrm{gas} = 0$) starbursts, to older (i.e.\@ potentially with an underlying older stellar population also present) non-pristine ($Z_\mathrm{gas} > 0$) EMPG starbursts through the added medium-band constraints. 

Now, medium bands (at $z=4.25$, F410M) can also provide continuum constraints through the wavelength interval between \Ha\ and \SIII $\lambda 9069$ in the NIR, which is valuable if the rest-frame optical continuum measurement is not available. Furthermore, the photometric boost by \Ha\ in a medium band (at $z=4.25$, F335M) also encodes indirect information about the continuum level. Due to their narrower spectral range, medium bands have their bandpass-averaged flux density more strongly boosted by emission lines than their respective wide-band filters, as can be seen for F335M and F356W by \Ha\ at this redshift. This differential photometric boost increases with increasing equivalent width, so the gap in line boost by medium--wide filter pairs can be used to assess the line equivalent width and thus indirectly determine the continuum level. These different medium-band continuum measurements \citep[for more details, see][]{Trussler2026} enable a more comprehensive empirical search for EMPG candidates, maximising the redshift range over which a given filter provides information on the continuum and line equivalent widths, enhancing the EMPG search volume. Additionally, the medium-band continuum measurements can establish the presence of a Balmer jump \citep[see e.g.\@][]{Cameron2024, Katz2025, Trussler2026}, a free--bound hydrogen recombination feature associated with young starbursts, through the deficit in rest-frame optical/NIR continuum level compared with the rest-frame UV, as seen for the models in Fig.~\ref{fig:Ha_vs_OIII}. The blue UV--optical/NIR continuum colour associated with the Balmer jump is in contrast to the red colour associated with the Balmer break, helping to remove Little Red Dot \citep[LRD,][]{Matthee2024, Kokorev2024, Kocevski2025} contaminants which can otherwise also exhibit red $m_\mathrm{[O\ III]} - m_\mathrm{H\alpha}$ colours due to their red optical continua \citep[see e.g.\@][]{deGraaff2025, Barro2026}.

\subsubsection{EMPG pre-selection procedure}

We empirically select plausible EMPG candidates by requiring strong \Ha, yet weak \OIII\ emission. This is mostly achieved by requiring the \OIII\ + \Hb\ equivalent width to be small with respect to \Ha. However, when the empirical equivalent width determination is deemed unreliable, we instead base the selection off red $m_\mathrm{[O\ III]} - m_\mathrm{H\alpha}$ colours. For each galaxy in our initial high-redshift sample, we determine the rest-frame optical continuum $m_\mathrm{Opt}$ and/or rest-frame NIR continuum level $m_\mathrm{NIR}$ using the appropriate medium band (if available), given the source's EAZY redshift. Following the procedure outlined in \citet{Trussler2026}, we then determine the \Ha\ and \OIII\ + \Hb\ equivalent widths using the photometric boosts in the wide-band filters tracing these lines, accounting for the filter throughput at the emission line observed-frame wavelength. The \Ha\ flux $F_\mathrm{H\alpha}$ is related to the bandpass-averaged flux density measured in the wide-band filter tracing \Ha\ $\langle f_{\lambda, \mathrm{H}\alpha}\rangle$ and the medium-band continuum measurement $\langle f_{\nu, \mathrm{Opt/NIR}}\rangle$ via
\begin{equation} \label{eq:boost}
\langle f_{\lambda,\mathrm{H\alpha}} \rangle = \frac{\int f_\lambda (\lambda)\lambda T(\lambda)\mathrm{d}\lambda}{\int \lambda T(\lambda)\mathrm{d}\lambda} = 
\langle f_{\lambda,\mathrm{cont}}\rangle + \frac{F_\mathrm{H\alpha}\lambda_\mathrm{H\alpha, obs}T(\lambda_\mathrm{H\alpha, obs})}{\int \lambda T(\lambda)\mathrm{d}\lambda},
\end{equation}
\noindent where $f_\lambda = f_{\lambda, \mathrm{cont}} + f_{\lambda, \mathrm{H\alpha}}$ is the combined continuum+\Ha\ flux density per unit wavelength interval, $T(\lambda)$ is the throughput of the wide-band filter, $\lambda_\mathrm{H\alpha, obs}$ is the observed wavelength of \Ha, and \citep[assuming a flat $f_\nu$ spectrum, as would be expected for the combined stellar+nebular rest-frame optical continuum of a young starburst, see Fig.~\ref{fig:Ha_vs_OIII} and][]{Trussler2026} $\langle f_{\lambda,\mathrm{cont}}\rangle = \langle f_{\nu,\mathrm{Opt/NIR}}\rangle \frac{c}{\lambda^2_\mathrm{pivot,H\alpha}}$, where $\lambda_\mathrm{pivot,H\alpha}$ is the pivot wavelength of the wide-band filter covering \Ha. Dividing $F_\mathrm{H\alpha}$ by the flat $f_\nu$ continuum evaluated at $\lambda_\mathrm{H\alpha, obs}$ then yields the observed-frame equivalent width, with $\mathrm{EW_{H\alpha, rest}} = F_{\mathrm{H}\alpha} /[f_\lambda(\lambda_\mathrm{H\alpha, obs})(1+z_\mathrm{EAZY})]$, where $z_\mathrm{EAZY}$ is the best-fit EAZY redshift. 

In the case of the \OIII\ + \Hb\ line complex, we must assume a \OIII\ $\lambda 5007$ / \Hb\ ratio to obtain a $\mathrm{EW_{[O\ III] + H\beta}}$ estimate, accounting for and weighting correctly the filter throughputs at the \OIII\ and \Hb\ lines. Since we are considering the possibility of extremely metal-poor galaxies but the line ratio of a given galaxy is not known in advance, we estimate $\mathrm{EW_{[O\ III] + H\beta}}$ assuming either \OIII\ $\lambda 5007$ / \Hb\ = [1, 5], the former and latter ratio values corresponding to the metal-poor and normal metallicity \citep{Matthee2023, Meyer2024} scenarios, respectively. We accept either $\mathrm{EW_{[O\ III] + H\beta}}$ value in our EMPG selection criteria (though we note that the ratio assumed has a very negligible effect on the EW determined, unless the lines are near a filter edge). 

We determine the Balmer jump $\Delta m _\mathrm{jump}$ \citep[for more details, see][]{Trussler2026} by the magnitude difference between the continuum level measured by the first wide-band filter fully blueward of the Balmer jump $m_\mathrm{UV,\ jump}$ at 3646~\AA\, and the optical/NIR continuum level traced by the medium band (either $m_\mathrm{opt}$ or $m_\mathrm{NIR}$): $\Delta m_\mathrm{jump} = m_\mathrm{UV,\ jump} - m_\mathrm{Opt/NIR}$.

Our base EMPG selection criteria are as follows.

\begin{figure*}
\centering
\includegraphics[width=.65\linewidth]{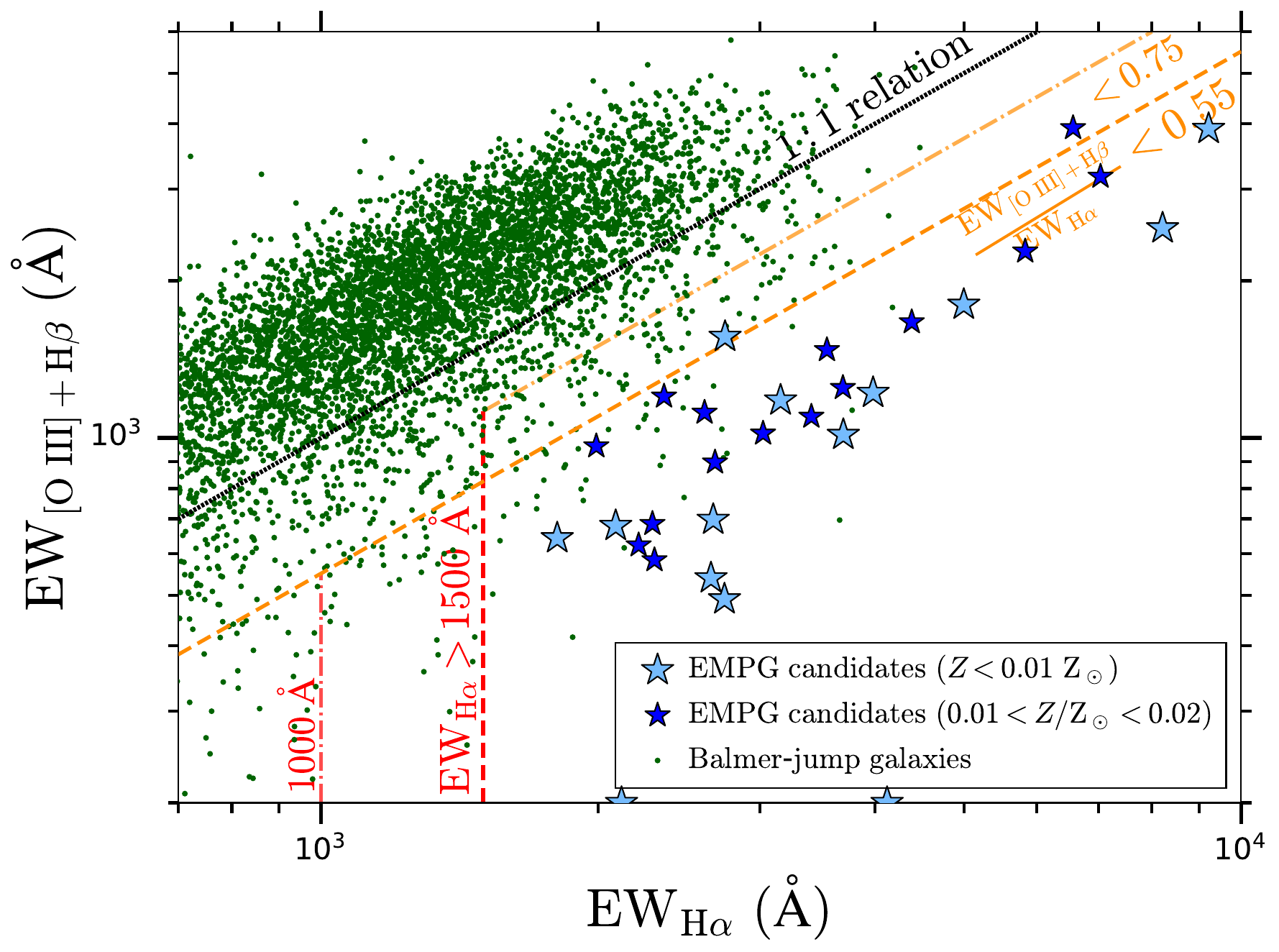}
\caption{Empirical selection of EMPG candidates from their strong \Ha, but weak \OIII\ emission, via equivalent width constraints from medium-band imaging. Most Balmer-jump galaxies (with $\Delta m_\mathrm{jump} < -0.15$ and $\mathrm{SNR_{Opt/NIR}} > 4$, green dots, of which there are 7956 in total from the JADES DR5 footprint) have high \OIII\ + \Hb\ rest-frame equivalent widths compared to \Ha, generally being well above the 1:1 relation (black dotted line). Galaxies with low \OIII\ + \Hb\ equivalent widths relative to \Ha\ are rare. Our strict (dashed) and looser (dash-dotted) EMPG selection criteria, demanding strong \Ha\ emission ($\mathrm{EW_{H\alpha}} > 1500, 1000$~\AA, red lines), but correspondingly weak \OIII\ emission ($\mathrm{EW_{[O\ III] + H\beta}}/\mathrm{EW_{H\alpha}} < 0.55, 0.75$, orange) are shown. Our final 14 EMPG candidates with $Z < 0.01~\mathrm{Z}_\odot$ that strongly favour close-to-pristine SED model fits are displayed (light blue stars). For reference, the 15 EMPG candidates with $0.01 < Z/\mathrm{Z}_\odot < 0.02$ are also displayed (we do not consider these further, dark blue stars). Balmer-jump galaxies located in the EMPG selection region that are not considered EMPGs (green dots) typically have insufficient evidence to strongly support the very metal-poor scenario (i.e.\@ $\Delta \chi^2$ between low and intermediate metallicity \textsc{Bagpipes} model fits is small), sometimes are inferred to have relatively weak \Ha\ equivalent widths (below 1500~\AA), or in some cases are inferred to have higher photometric redshifts during \textsc{Bagpipes} fitting than as estimated via EAZY, with what was once considered strong \Ha\ emission now attributed to \OIII\ + \Hb. Furthermore the choice of $R_3$ -- $12 + \log(\mathrm{O/H})$ metallicity calibration, either empirical or theoretical (here via \textsc{Bagpipes}), also affects the metallicity inference and hence photometric selection of EMPG candidates.}
\label{fig:ew_ew_selection}
\end{figure*}

\begin{itemize}

\item{$\mathrm{EW_{[O\ III] + H\beta}} < 0.55 \times \mathrm{EW_{H\alpha}}$: The \OIII\ emission is weak relative to \Ha, likely implying extremely low metallicity. This EW ratio threshold roughly corresponds to \OIII\ $\lambda 5007$/\Hb\ = 1.5, resulting in $12 + \log(\mathrm{O/H}) \approx 6.7$ using the \citet{Sanders2024} strong-line calibration.}

\item{$\mathrm{EW_{H\alpha}} > 1500$~\AA: The \Ha\ equivalent width is very high, so the \Ha\ photometric boost, and lack thereof for \OIII, is evident in the data.}

\item{$\Delta m_\mathrm{jump} < -0.15$: The galaxy exhibits a prominent Balmer jump \citep[the $-0.15$ threshold being the smallest (in absolute value) jump seen in current Pop III models, for more details see][]{Trussler2026}, so is likely a young starburst galaxy, rather than a LRD which typically would exhibit a Balmer break with $\Delta m_\mathrm{jump} > 0$ \citep[see e.g.\@][]{deGraaff2025}.}

\item{$\mathrm{SNR_{Opt/NIR}} > 4$: The SNR on the medium-band continuum measurement is high, so the resulting empirical line equivalent width determinations are deemed reliable.}

\item{$m_{1500\ \text{\AA}} - m_\mathrm{F444W} < 1.5$: The colour between the flux density traced by the first wide-band filter fully redward of \Lya\ $m_{1500\ \text{\AA}}$, and the F444W filter $m_\mathrm{F444W}$ is not too red. This cut helps remove very dusty galaxies, AGN, LRDs and quiescent galaxies from the sample, which can exhibit extremely red colours, given their red continuum slopes. In principle these contaminants should generally also be removed by the above Balmer-jump cut.}

\end{itemize}

To widen our EMPG search and accommodate for uncertainties in the data, we loosen at most one of the above stricter base criteria with the corresponding cut below. 

\begin{itemize}

\item{$\mathrm{EW_{[O\ III] + H\beta}} < 0.75 \times \mathrm{EW_{H\alpha}}$: This EW ratio threshold roughly corresponds to \OIII\ $\lambda 5007$/\Hb\ = 2.5, resulting in $12 + \log(\mathrm{O/H}) \approx 7.0$ using the \citet{Sanders2024} strong-line calibration.}

\item{$\mathrm{EW_{H\alpha}} > 1000$~\AA.}

\item{$\Delta m_\mathrm{jump} < 0$: This corresponds to the threshold between the Balmer jump and Balmer break.}

\item{$\mathrm{SNR_{Opt/NIR}} > 3$: We consider this the minimum data quality threshold for which empirically determined equivalent widths are reliable.}

\item{$\mathrm{SNR_{Opt/NIR}} < 3$ \& $\Delta m_\mathrm{jump,\ 3\sigma} < -0.15$ \& $m_\mathrm{[O\ III]} - m_\mathrm{H\alpha} > 0.2$ \& $m_\mathrm{H\alpha} - m_\mathrm{Opt/NIR,\ 3\sigma} < -0.6$: Due to the difficulty in detecting the rest-frame optical/NIR continuum with medium bands, we relax the SNR requirement, pushing to the faintest systems. For a reliable Balmer-jump detection, we require the $3\sigma$ upper limit on the optical/NIR continuum $m_\mathrm{Opt/NIR,\ 3\sigma}$ to still yield a sufficiently large Balmer jump. As we deem the empirical equivalent width determination unreliable (due to the imprecise continuum constraint) in this regime, due to the uncertain continuum level, we replace our EW ratio criteria and $\mathrm{EW_\mathrm{H\alpha}}$ criteria with colour selections.}

\end{itemize}

We supplement this optical/NIR continuum selection with a selection based off the \Ha\ photometric excess in a medium band, to widen the EMPG search volume and to possibly push the selection to intrinsically fainter systems. For a given galaxy, we determine the medium--wide filter pair that is boosted by \Ha, given the EAZY redshift. If the corresponding medium band is available, we estimate the \Ha\ equivalent width by assuming a flat $f_\nu$ continuum and iteratively varying the \Ha\ equivalent width of this mock spectrum, determining the value that best matches the magnitude difference between the medium--wide filter pair boosted by \Ha\ \citep[for an early application of this procedure, see][]{Bunker1995}. From the resulting continuum estimate $m_\mathrm{cont,\ H\alpha}$ we determine the Balmer jump and photometric boost by \Ha. Given this less reliable approach for determining the continuum \citep[see][]{Trussler2026}, we adopt the following selection criteria for identifying further EMPG candidates from our general high-redshift galaxy sample.

\begin{itemize}

\item{$m_\mathrm{[O\ III]} - m_\mathrm{H\alpha} > 0.2$.}

\item{$m_\mathrm{H\alpha} - m_\mathrm{cont,\ H\alpha} < -0.6$.}

\item{$\Delta m_\mathrm{jump} < 0$.}

\item{$\mathrm{SNR_{H\alpha, med/wide}} > 7$: The SNR for the medium- and wide-bands tracing \Ha\ are high, so that the indirect determination of the rest-frame optical continuum is as reliable as possible.}

\item{$m_{1500\ \text{\AA}} - m_\mathrm{F444W} < 1.5$.}

\end{itemize}

This empirical selection procedure (employing the combination of the base/looser optical/NIR continuum and \Ha\ selections) results in 377 preliminary EMPG candidates, drawn out of the 52888 galaxies comprising the general high-redshift galaxy parent sample. We subsequently fit their photometry with close-to-pristine SED models (see next section) to establish the most secure EMPG candidates. To put our EMPG equivalent width selection procedure in context and highlight the extraordinarily low $\mathrm{EW_{[O\ III] + H\beta}}$ of our final EMPG candidates (described in more detail in Section~\ref{subsec:candidates}), we show the $\mathrm{EW_{[O\ III] + H\beta}}$ -- $\mathrm{EW_{H\alpha}}$ plane in Fig.~\ref{fig:ew_ew_selection}. Our base criteria of strong \Ha\ ($\mathrm{EW_{H\alpha}} < 1500$~\AA) but correspondingly weak \OIII\ ($\mathrm{EW_{[O\ III] + H\beta}} < 0.55 \times \mathrm{EW_{H\alpha}}$) are shown as the red and orange dashed lines, respectively. The looser selection criteria (1000~\AA\ threshold and EW ratio < 0.75) are shown as dot-dashed lines. Regular Balmer-jump galaxies (7956 in total, green dots) with $\Delta m_\mathrm{jump} < -0.15$ and $\mathrm{SNR_{Opt/NIR}} > 4$ generally have much higher $\mathrm{EW_{[O\ III] + H\beta}}$ than $\mathrm{EW_{H\alpha}}$, typically well above the 1:1 relation (black dotted line). Thus, galaxies with comparatively weak \OIII\ emission (below the dashed / dot-dashed orange lines) are exceedingly rare. 

Our final 14 EMPG candidates with $Z < 0.01~\mathrm{Z}_\odot$ (inferred photometrically using \textsc{Bagpipes} fitting, see next section) are shown as light blue stars, and are found primarily within the lowest right region of the diagram, exhibiting very strong \Ha\ emission but weak \OIII, generally satisfying our strictest empirical selection criteria. Some of these are empirically inferred to have unphysically large $\mathrm{EW_{H\alpha}}$ ($\gg 4000$~\AA), highlighting the limitation of the empirical EW estimate in the case of very low optical/NIR continuum SNR. For reference, we also show 15 EMPG candidates with $0.01 < Z/\mathrm{Z}_\odot< 0.02$ as dark blue stars, though do not consider these further in our subsequent analysis. Galaxies located in the strictest region of the selection diagram tend to be inferred to be very metal-poor ($Z < 0.02~\mathrm{Z}_\odot$). For the few exceptions (green dots), typically the very metal-poor interpretation is not strongly demanded by the data (i.e.\@ $\Delta \chi^2$ with intermediate metallicity models is small), sometimes the inferred \Ha\ equivalent width is smaller than our final selection criterion (below 1500~\AA), or in a few cases the galaxies are inferred to have higher photometric redshifts in the \textsc{Bagpipes} fitting (relative to EAZY), now attributing what was originally considered to be a strong photometric boost by \Ha\ and weak \OIII\ (so low metallicity), to a strong boost by \OIII\ and a weak boost by the weaker rest-frame optical lines (e.g.\@ \Hg, \Hd, \He, \OII, etc.\@) and so regular metallicity. Furthermore, we note that the \textsc{Bagpipes} models used to make the photometric metallicity inference follow a $R_3 = $ \OIII\ / \Hb\ -- $12 + \log(\mathrm{O/H})$ relation that is more similar to local strong-line calibrations \citep[e.g.\@][]{Curti2020, Nakajima2022b}, rather than high-redshift calibrations \citep[e.g.\@][]{Sanders2024, Sanders2026, Cataldi2025, Isobe2026}, with a lower $R_3$ for a given $12 + \log(\mathrm{O/H})$. Hence the metallicity inferred for a given equivalent width ratio (and thus $R_3$) is higher than would be expected given the high-redshift \citet{Sanders2024} $R_3$ calibration used to motivate our equivalent width ratio selection cuts, explaining some of the non-EMPG candidates (green dots) in the strictest region of the selection diagram, with our metallicity estimates therefore being rather conservative. Reformulating the $R_3$ ratios inferred from \textsc{Bagpipes} into empirically-motivated gas-phase metallicities using high-redshift calibrations would result in lower metallicities and thus more EMPG candidates than presented here.

Despite the available filter set allowing for the identification of EMPG candidates at $z < 2.5$, we find no secure EMPG candidates at $z < 2.5$, suggesting that extremely metal-poor star formation is very rare in dwarf galaxies at these later cosmic epochs. Hence we focus on $2.5 < z < 6.5$ in the remainder of the analysis. 

\subsection{Fitting close-to-pristine SED models} \label{subsec:sed}

Having identified plausible preliminary EMPG candidates from our empirical pre-selection, we now fit their photometry with \textsc{Bagpipes} \citep{Carnall2018} to establish the most secure final EMPG candidates (discussed in Section~\ref{subsec:candidates}), where the extremely metal-poor solution is heavily favoured over regular metallicity models. Here we describe this fitting procedure.

Our \textsc{Bagpipes} fitting configuration is as follows. We fit the 0.2~arcsec diameter circular aperture NIRCam + HST/ACS photometry, adopting a minimum error of 5 per cent in each filter, to account for possible flux calibration issues and/or template mismatch. We use the v2.2 BPASS \citep{Stanway2018} stellar templates with a 300~$\mathrm{M}_\odot$ upper mass threshold. We adopt a non-parametric star-formation history to describe the likely extremely bursty nature of our EMPGs, adopting the continuity model from \citet{Leja2019}. We set the following age bins, where the star formation rate can take on a distinct constant value in each bin: [0, 1.7, 3, 10, 30, 100, 300, $a_\mathrm{max}$]~Myr, where $a_\mathrm{max} = \min(t_\mathrm{Univ}(z_\mathrm{EAZY}) - t_\mathrm{Univ}(z=20),\ 1000)$, with $t_\mathrm{Univ}(z)$ being the age of the Universe at redshift $z$. The subdivision into very young [0, 1.7, 3]~Myr age bins is motivated by the extremely high \Ha\ EWs of our EMPG candidates, demanding exceptionally young stellar populations to produce. We assume a bursty-continuity prior \citep{Tacchella2022}, following a Student's t-distribution with $\sigma = 1.0$ and $\nu = 2$.

We adopt uniform priors on all fitted parameters, with the logarithm of the stellar mass $\log_{10}(M_*/\mathrm{M}_\odot) \in [1, 15]$ and dust attenuation $A_\mathrm{V} \in [0, 5]$. We adopt a \citet{Calzetti2000} dust attenuation law, with greater extinction in < 10~Myr old \ion{H}{II} regions than outside, with $\mathrm{E(B-V)}_\mathrm{out} = 0.44 \times \mathrm{E(B-V)}_\mathrm{in}$. We set $z \in [0.55, 15]$, where the non-zero lower limit is chosen to avoid fitting the photometry with low-$z$ galaxies exhibiting extreme 3.3~\textmu m PAH emission, which can sometimes resemble the \Ha\ photometric excess in EMPGs, unphysically requiring substantial dust emission and absorption in low-mass, ultra-faint low-$z$ dwarf galaxies ($M_* < 10^{6}~\mathrm{M}_\odot$). We find that the low metallicity fit (see next paragraph) is heavily favoured ($\Delta \chi^2 > 9$) over the PAH fit for all-but-one of our final EMPG candidates (ID 123650, which we subsequently spectroscopically confirmed to be at $z=5.39$), where $\Delta \chi^2 \approx 4$, so we do not consider PAH contamination further. Finally, we set the logarithm of the ionisation parameter $\log_{10}U \in [-2, -1]$. We restrict our fitting to high ionisation parameters ($\log U > -2$) so that a lack of \OIII\ emission in the photometry is attributed to a lack of oxygen (so low metallicity), rather than very low ionisation parameter (e.g.\@ $\log U = -4$). This is motivated by the high ionisation parameters typically seen in high-redshift galaxies \citep[see e.g.][]{Cameron2023}. It is conceivable that there can be high-$z$ galaxies with exceptionally low $U$, though given the very bursty nature of our EMPGs, with prominent Balmer jumps, large $\mathrm{EW}_\mathrm{H\alpha}$ and extremely high ionising photon production efficiencies $\xi_\mathrm{ion, obs}$, this seems very unlikely to be the case. 

In order to identify the most secure EMPG candidates, we perform three distinct SED fits per galaxy. Firstly, we fit using low metallicity models, with $Z_\mathrm{star} \in [0.0005, 0.03\times 10^{0.22}]~\mathrm{Z}_\odot$, where $Z_\odot = 0.02$ for the BPASS models, The stellar and gas-phase metallicities are tied together, with the factor of $10^{0.22}$ accounting for depletion of gas-phase oxygen onto dust grains \citep{Dopita2000}. When discussing the metallicity of our EMPG candidates we are referring to that traced by gas-phase oxygen, i.e.\@ relative to $12 + \log(\mathrm{O/H})_\odot = 8.69$. \footnote{We note that \textsc{Bagpipes} by default adopts the solar abundances of \citet{Anders1989}, with solar metallicity in the models therefore corresponding to $12 + \log(\mathrm{O/H})_\odot = 8.93$. Hence we shift all of our photometrically-inferred metallicities by $+0.24$~dex with respect to the \textsc{Bagpipes} output, to relate this to the solar abundance scale of \citet{Asplund2009}, which instead has $12 + \log(\mathrm{O/H})_\odot = 8.69$.} We note that adopting a uniform prior on the logarithm of the metallicity would bias the inferred metallicities lower than from our uniform prior on the linear metallicity. Although we are mostly interested in EMPGs with close-to-pristine metallicities (${\sim}0.001\textrm{--}0.01~\mathrm{Z}_\odot$), the relatively high $0.03~\mathrm{Z}_\odot$ upper limit is chosen to serve as an adequate buffer between the low and intermediate metallicity models, so that genuine EMPGs (with metallicity close to $1\%~\mathrm{Z}_\odot$) are not discarded because of a comparable quality fit in the low/intermediate metallicity models (which might occur if the upper limit was set to e.g.\@ $0.015~\mathrm{Z}_\odot$). We do acknowledge that the coarse sampling ([$0.05,\ 0.5,\ 5$]\%~$\mathrm{Z}_\odot$) of the BPASS models at very low metallicity could potentially cause the photometrically-inferred metallicities to be inaccurate.

\begin{figure*}
\centering
\includegraphics[width=.81\linewidth] {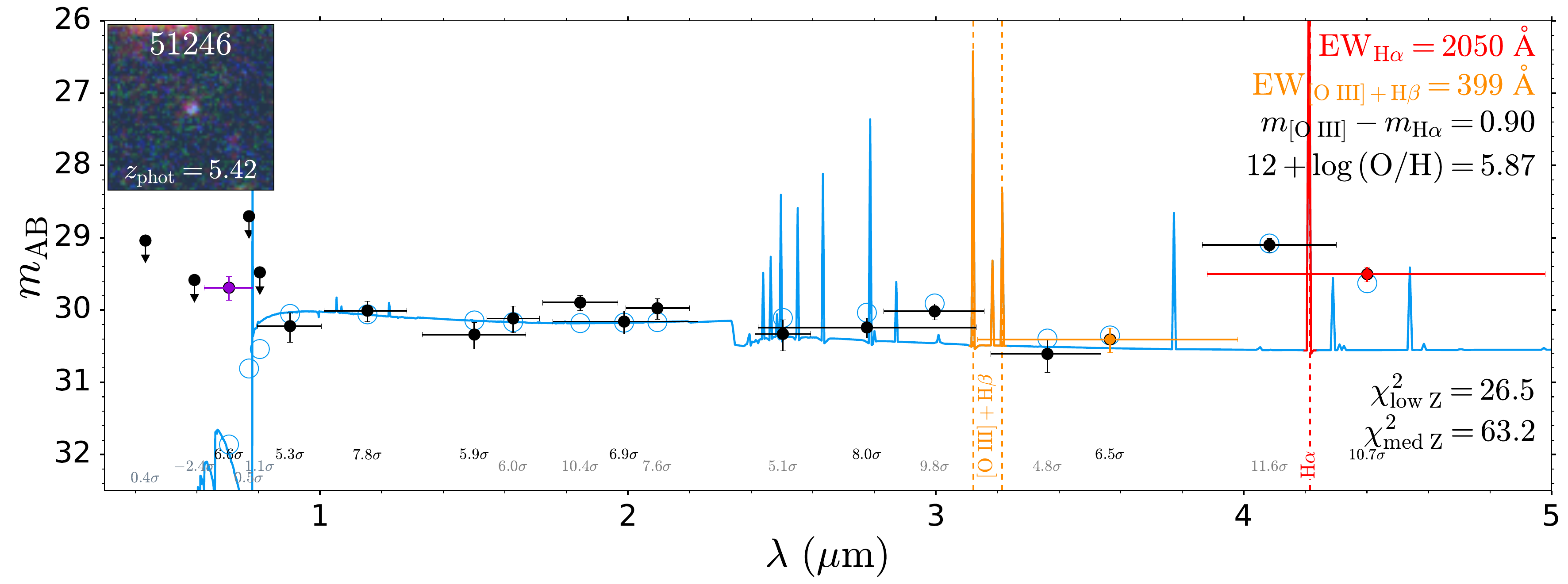}\\
\includegraphics[width=.81\linewidth]{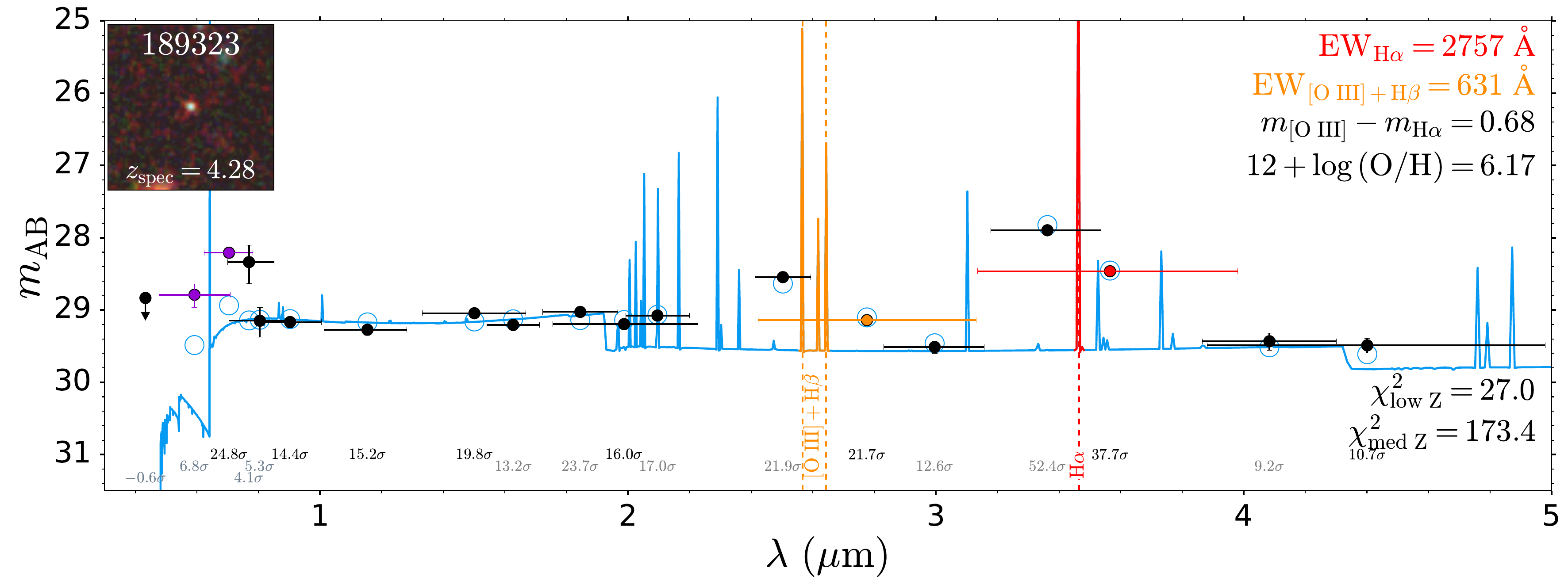} \\
\includegraphics[width=.81\linewidth] {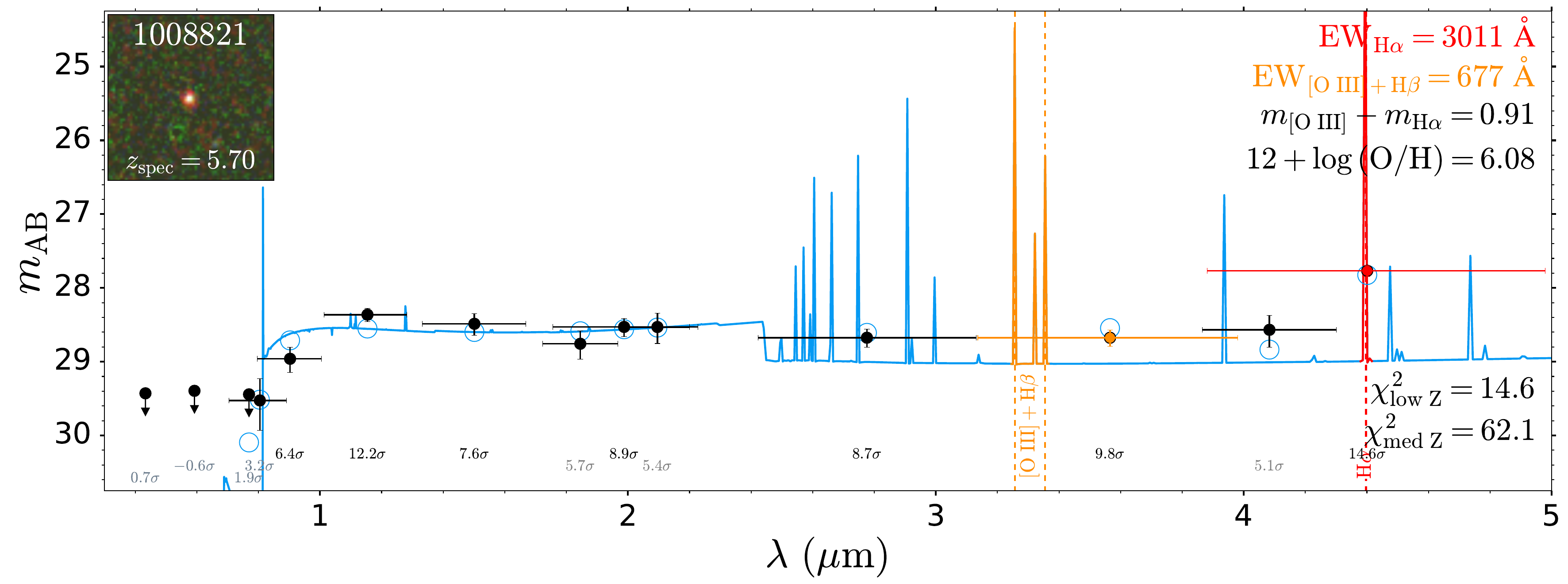} 
\includegraphics[width=.81\linewidth]{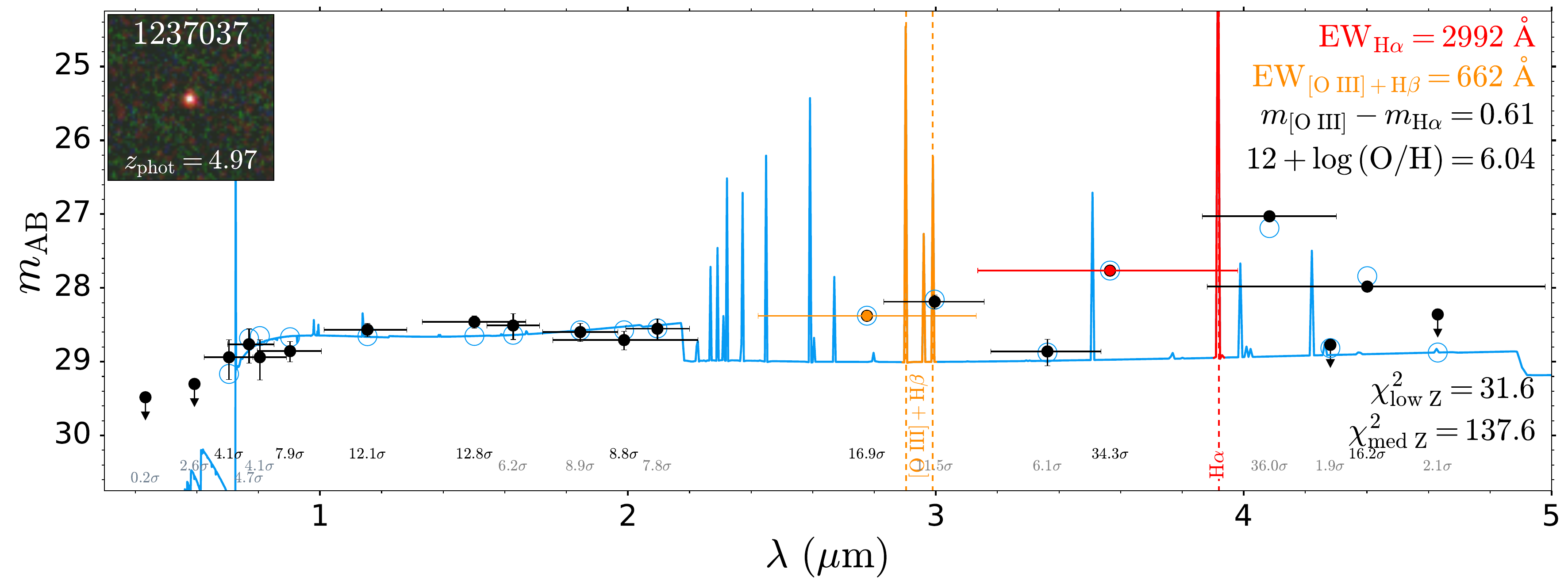}  
\caption{Example SEDs for 4 of our most pristine EMPG candidates. NIRCam + HST/ACS photometry are displayed (black datapoints), with wide-band filters covering \OIII\ and \Ha\ highlighted in orange and red, respectively. The SNR in the 0.2~arcsec diameter apertures are also shown. Low metallicity \textsc{Bagpipes} models (blue) provide the best possible description of the data, with much lower $\chi^2$ values ($\Delta \chi^2 > 9$) compared to intermediate metallicity models. The median inferred \Ha\ and \OIII\ + \Hb\ rest-frame equivalent widths from \textsc{Bagpipes} are displayed, together with the very red observed $m_\mathrm{[O\ III]} - m_\mathrm{H\alpha}$ colour, as well as the median inferred oxygen abundance $12 + \log (\mathrm{O/H})$. $2 \times 2$ arcsec RGB (F444W, F200W, F115W) cutouts are shown, also displaying the galaxy ID and photometric or spectroscopic (for 189323, 1008821) redshift. The EMPG candidates exhibit strong photometric boosts by \Ha, yet correspondingly weak boosts by \OIII, also displaying prominent Balmer jumps, as required by our selection procedure. Potential photometric boosts by strong \Lya\ emission are visible in the dropout filters (purple) of IDs 51246 and 189323.} 
\label{fig:empg_candidates}
\end{figure*}

Secondly, we fit with intermediate metallicity models, with $Z_\mathrm{star} \in [0.03, 0.20]\times 10^{0.22}~\mathrm{Z}_\odot$. The $0.20~\mathrm{Z}_\odot$ upper limit is adopted so that a lack of \OIII\ emission is not attributed to extremely high metallicities ($Z\sim \mathrm{Z}_\odot$), which again we would not expect in such faint, low-mass dwarf galaxies at high redshift \citep[see e.g.\@][]{Nakajima2023, Curti2024, Sarkar2025}.

Thirdly, we fit an edge case scenario, if applicable. Indeed, edge cases can be a concern in the identification of EMPG candidates. For example, the galaxy GLIMPSE-16043 was identified as a Pop III candidate by \citet{Fujimoto2025}, though this required a best-fit photometric redshift at $z=6.5$, where \Ha\ is approaching the decreasing throughput of F444W at its long wavelength limit, to explain the modest observed $m_\mathrm{[O\ III]} - m_\mathrm{H\alpha} = 0.22$ colour, despite the intrinsically very red ($\approx0.9$~mag) colour for Pop III galaxies \citep{Trussler2023}. This motivates our restriction to EAZY redshifts where the \Ha\ and \OIII\ $\lambda 5007$ lines are at $>75\%$ of their respective wide-band filter's maximum throughput. Indeed, an additional concern is that a galaxy with intrinsically strong \OIII\ emission is perceived to have weak \OIII\ emission if this resides at a wide-band filter edge with low throughput or within the NIRCam dichroic gap between F200W and F277W. We account for the latter scenario by performing a third SED fit for a subset of galaxies, with intermediate metallicity $Z_\mathrm{star} \in [0.03, 0.20]\times 10^{0.22}~\mathrm{Z}_\odot$. For galaxies with \Ha\ in F277W (F356W) we set $z \in [3.45, 3.77]$ ($z \in [3.77, 3.87]$), corresponding to the redshift range where \OIII\ resides in the dichroic gap. In principle, the photometry can effectively rule out the edge case scenario based off the Ly($\alpha$) break strength, as this is redshift dependent (though can be affected by Ly$\alpha$ emission and DLA absorption so can bias the redshift inference, unless accounted for), as well as from medium bands tracing the Balmer jump / photometric boosts from emission lines. Without medium bands, it would be difficult to convincingly photometrically identify the spectroscopically-observed $z=3.19$ EMPG candidate CR3 \citep{Cai2025}, needing F210M+F300M to unambiguously establish that \OIII\ is not in the dichroic gap. We also account for an additional edge case where \Ha\ simultaneously boosts F356W and F444W, due to the narrow overlap of the two filters. In the EMPG scenario, the SED displays weak F277W, but strong F356W and F444W. This could be confused for a normal metallicity galaxy with \OIII\ boosting F356W and \Ha\ boosting F444W. Thus for galaxies in the filter overlap region with $4.90 < z < 5.07$, we perform an additional SED fit with $z \in [5.20, 5.70]$ to check against this scenario. 

Finally, some EMPG candidates exhibit a notable excess in their dropout filter, being brighter than they should be given their redshift. This photometric excess is likely attributable to line boosting by \Lya, which has been observed to be strong in several spectroscopically-confirmed EMPGs \citep{Vanzella2023, Cai2025, Morishita2025, Willott2025, Messa2026}. Upon visual inspection of the resulting low metallicity fits, the models can in some cases yield poor matches to some of the photometric data points, being skewed in an attempt to match the excess in the dropout filter, which is not possible without incorporating strong \Lya\ emission in the \textsc{Bagpipes} templates. We mitigate this by rerunning the \textsc{Bagpipes} fitting procedure for these sources, but excluding the dropout filter photometry.

\subsection{Extremely metal-poor galaxy candidates} \label{subsec:candidates}

We require our final EMPG candidates to be best described by low metallicity models, also exhibiting strong \Ha\ emission. Our selection criteria are as follows:

\begin{itemize}

\item{$[12 + \log(\mathrm{O/H})]_{84} < 6.7$: The 84$^\mathrm{th}$ percentile of the metallicity posterior for the low metallicity model is below $1\%~\mathrm{Z}_\odot$, implying close-to-pristine metallicity.}

\item{$\Delta \chi_{\mathrm{med/edge}}^2 > 9$: The $\chi^2$ difference between the low metallicity and intermediate metallicity fits $\chi_\mathrm{med}^2 - \chi_\mathrm{low}^2 > 9$ and the low metallicity and edge case fits (if applicable) $\chi_\mathrm{edge}^2 - \chi_\mathrm{low}^2 > 9$ are greater than 9, implying the low metallicity solution is strongly favoured.}

\item{$\mathrm{EW_{H\alpha}} > 1500$~\AA: The median inferred \Ha\ rest-frame equivalent width is greater than 1500~\AA, implying the \Ha\ line boost (and lack thereof for \OIII) is evident.}
\end{itemize}

Applying this final selection results in 14 extremely metal-poor galaxy candidates (out of the 377 preliminary EMPG candidates from the empirical pre-selection). We show example SEDs for 4 of our most pristine candidates in Fig.~\ref{fig:empg_candidates}. The NIRCam + HST/ACS photometry is shown in black, with the wide-band filters covering \Ha\ and \OIII\ highlighted in red and orange, respectively. We show the close-to-pristine \textsc{Bagpipes} model fits (blue curve) and associated bandpass-averaged flux densities (blue circles). Upper limits are displayed as downward arrows at the 3$\sigma$ limit. The median inferred \Ha\ and \OIII\ rest-frame equivalent widths from \textsc{Bagpipes} are denoted, as well as the observed $m_\mathrm{[O\ III]} - m_\mathrm{H\alpha}$ colour, together with the median inferred oxygen abundance $12 + \log(\mathrm{O/H})$. We also display the $\chi^2$ values for the low and intermediate metallicity models, the SNR in the 0.2~arcsec diameter circular aperture photometry, and a (F444W, F200W, F115W at native resolution) RGB cutout \citep[generated using Trilogy,][]{Coe2012} spanning $2\times2$ arcsec denoting the source ID and redshift.

Galaxy IDs 189323 (top-right panel) and 1008821 (bottom-left) have \Ha-based NIRCam slitless spectroscopic redshifts from JWST GTO 4540 (Sun et al., in prep.) and FRESCO, respectively. The grism-based \Ha\ flux for ID 189323 ($1.85\times10^{-18}$~erg~s$^{-1}$~cm$^{-2}$) is in excellent agreement with our SED-based flux ($[1.96 \pm 0.08] \times 10^{-18}$~erg~s$^{-1}$~cm$^{-2}$). ID 123650 has a \Ha-based spectroscopic redshift from JWST program 8060, as well as a MUSE \Lya\ detection through the MUSE eXtremely Deep Field survey \citep{Bacon2023}. 
The remaining EMPG candidates, such as ID 51246 (top-left) and 1237037 (bottom right), have precise photometric redshifts from the deep, extensive medium-band imaging.

The EMPG candidates shown in Fig.~\ref{fig:empg_candidates} all display prominent photometric boosts by \Ha, but correspondingly weak boosts by \OIII, with a prominent red colour gap between the two wide-band filters probing these emission lines. Furthermore, the galaxies exhibit strong Balmer jumps, all as demanded by our empirical selection procedure. We further note the photometric excess in the F070W and F606W+F070W dropout filters (shown in purple) for IDs 51246 and 189323, respectively. As discussed earlier, this excess is likely attributable to line boosting by strong \Lya\ emission. Indeed, \Lya\ is detected for ID 189323 with SNR = 11.4 in the MUSE spectrum taken by the MUSE-Wide survey \citep{Herenz2017, Urrutia2019}.

We do note that for galaxy ID 1008821 (bottom-left panel), and perhaps other EMPG candidates with \Ha\ in F444W, we cannot currently conclusively establish that the prominent photometric excess in F444W is attributable to a line boost by strong \Ha\ emission, rather than a suddenly steeply rising rest-frame optical continuum. For ID 1008821 in particular, the grism-based \Ha\ flux ($[1.17\pm0.27]\times10^{-18}$~erg~s$^{-1}$~cm$^{-2}$) is substantially smaller ($0.42\times$) than the SED-based flux from our extremely metal-poor model fit ($[2.79\pm0.18]\times10^{-18}$~erg~s$^{-1}$~cm$^{-2}$), with the seemingly prominent photometric boost in the F444W filter perhaps being due to a combination of relatively weak \Ha\ emission and a strongly rising continuum. Indeed, this is a potential concern, as the $z=6.64$ galaxy Virgil \citep{Iani2025, Rinaldi2025b} appears as an otherwise normal star-forming galaxy with NIRCam and NIRSpec, but displays a strongly rising continuum beyond \Ha, suggesting either a heavily dust-obscured AGN \citep{Iani2025, Rinaldi2025b} or an extreme LRD \citep{Barro2026} that begins to dominate the SED in the MIRI bands. While this hidden AGN/LRD represents an extreme scenario, we conclude that ID 1008821 may not necessarily have strong \Ha\ emission relative to \OIII\, and thus could possibly not be extremely metal-poor. Unlike our lower redshift EMPG candidates, where \Ha\ is in the F277W/F356W filters, we cannot currently probe the continuum beyond \Ha\ when it resides in F444W, so cannot rule out the hidden AGN/LRD scenario through a lack of rising continuum (though complementary grism \Ha\ flux measurements provide valuable constraints) . This would require additional NIRCam medium-band (F430M, F460M, F480M) and/or MIRI imaging. The strong F410M photometric boost relative to F444W for ID 51246 (top-left panel) indicates high equivalent width \Ha\ emission, supporting the EMPG scenario.

\begin{table*}
\begin{center}
\begin{tabular}{|c|c|c|c|c|c|c|c|c|}
\hline
ID & R.\@A.\@ & Dec.\@ & $z$ & $\log M_*$ & $\mathrm{SFR}_{10}$ & $M_\mathrm{UV}$ & $\log\, \xi_\mathrm{ion,obs}$ & $\log\, L_\mathrm{H\alpha}$\\
 & & & & ($\mathrm{M}_\odot$) & ($\mathrm{M}_\odot~\mathrm{yr}^{-1}$) & (mag) & (Hz~erg$^{-1}$) & (erg~s$^{-1}$)\\
\hline
116605 & $53.14108$ & $-27.79267$ & $2.84^{+0.01}_{-0.01}$ & $6.65^{+0.03}_{-0.02}$ & $0.44^{+0.02}_{-0.02}$ & $-16.57^{+0.04}_{-0.03}$ & $26.13^{+0.02}_{-0.02}$ & $41.53^{+0.01}_{-0.01}$\\
189323 & $53.07826$ & $-27.84267$ & $4.28$ & $6.60^{+0.07}_{-0.05}$ & $0.39^{+0.04}_{-0.04}$ & $-17.06^{+0.04}_{-0.04}$ & $25.96^{+0.02}_{-0.03}$ & $41.55^{+0.02}_{-0.02}$\\
467299 & $52.99664$ & $-27.81685$ & $4.43^{+0.01}_{-0.01}$ & $7.47^{+0.19}_{-0.21}$ & $1.48^{+0.38}_{-0.27}$ & $-18.61^{+0.03}_{-0.03}$ & $25.69^{+0.04}_{-0.04}$ & $41.88^{+0.03}_{-0.04}$\\
487533 & $53.00348$ & $-27.72757$ & $4.66^{+0.08}_{-0.07}$ & $7.42^{+0.15}_{-0.14}$ & $1.06^{+0.21}_{-0.30}$ & $-17.55^{+0.06}_{-0.06}$ & $25.85^{+0.04}_{-0.05}$ & $41.60^{+0.04}_{-0.05}$\\
1237037 & $189.34859$ & $62.26441$ & $4.97^{+0.02}_{-0.01}$ & $7.05^{+0.04}_{-0.03}$ & $1.10^{+0.07}_{-0.06}$ & $-17.90^{+0.06}_{-0.04}$ & $26.06^{+0.03}_{-0.03}$ & $41.96^{+0.01}_{-0.02}$\\
117051 & $53.13204$ & $-27.79198$ & $5.10^{+0.02}_{-0.02}$ & $6.76^{+0.20}_{-0.19}$ & $0.31^{+0.09}_{-0.10}$ & $-16.77^{+0.06}_{-0.06}$ & $25.59^{+0.06}_{-0.06}$ & $41.21^{+0.05}_{-0.05}$\\
546658 & $52.96106$ & $-27.81957$ & $5.15^{+0.06}_{-0.03}$ & $6.21^{+0.06}_{-0.04}$ & $0.16^{+0.02}_{-0.01}$ & $-16.71^{+0.09}_{-0.07}$ & $25.77^{+0.03}_{-0.02}$ & $41.24^{+0.03}_{-0.03}$\\
100687 & $53.11694$ & $-27.81649$ & $5.23^{+0.05}_{-0.06}$ & $6.39^{+0.08}_{-0.05}$ & $0.23^{+0.02}_{-0.02}$ & $-16.41^{+0.10}_{-0.08}$ & $25.97^{+0.05}_{-0.04}$ & $41.32^{+0.02}_{-0.03}$\\
123650 & $53.16853$ & $-27.78411$ & $5.39$ & $6.46^{+0.07}_{-0.05}$ & $0.28^{+0.02}_{-0.03}$ & $-16.34^{+0.11}_{-0.09}$ & $26.05^{+0.04}_{-0.05}$ & $41.36^{+0.02}_{-0.02}$\\
51246 & $53.10175$ & $-27.86269$ & $5.42^{+0.04}_{-0.18}$ & $6.97^{+0.09}_{-0.10}$ & $0.21^{+0.06}_{-0.03}$ & $-16.48^{+0.07}_{-0.09}$ & $25.79^{+0.04}_{-0.04}$ & $41.20^{+0.04}_{-0.03}$\\
487127 & $53.00016$ & $-27.82951$ & $5.63^{+0.01}_{-0.01}$ & $7.33^{+0.02}_{-0.02}$ & $2.11^{+0.10}_{-0.09}$ & $-18.54^{+0.02}_{-0.03}$ & $26.07^{+0.02}_{-0.02}$ & $42.25^{+0.01}_{-0.01}$\\
1008821 & $189.08432$ & $62.25145$ & $5.70$ & $7.06^{+0.04}_{-0.05}$ & $1.11^{+0.11}_{-0.11}$ & $-18.22^{+0.06}_{-0.06}$ & $25.99^{+0.04}_{-0.04}$ & $41.99^{+0.03}_{-0.03}$\\
121273 & $53.12764$ & $-27.78665$ & $5.77^{+0.11}_{-0.06}$ & $6.56^{+0.06}_{-0.05}$ & $0.36^{+0.04}_{-0.03}$ & $-16.18^{+0.10}_{-0.10}$ & $26.17^{+0.05}_{-0.06}$ & $41.41^{+0.03}_{-0.03}$\\
102686 & $53.15069$ & $-27.81392$ & $5.97^{+0.06}_{-0.05}$ & $6.35^{+0.48}_{-0.12}$ & $0.18^{+0.05}_{-0.02}$ & $-17.35^{+0.06}_{-0.06}$ & $25.74^{+0.04}_{-0.11}$ & $41.23^{+0.05}_{-0.10}$\\

\hline
\end{tabular}
\caption{General properties of the 14 EMPG candidates, ordered by increasing redshift. Listed are the JADES DR5 ID, right ascension, declination, \textsc{Bagpipes} photometric redshift (or spectroscopic redshift, if listed without error bars), median stellar mass $M_*$ from \textsc{Bagpipes} SED-fitting, average star formation rate over the past 10~Myr $\mathrm{SFR}_{10}$, absolute magnitude in the rest-frame UV $M_\mathrm{UV}$, observed ionising photon production efficiency $\xi_\mathrm{ion, obs}$, and \Ha\ luminosity $L_\mathrm{H\alpha}$. Error bars correspond to the 16--84 percentiles.}
\label{tab:empg_candidates}
\end{center}
\end{table*}

\begin{table*}
\begin{center}
\begin{tabular}{|c|c|c|c|c|c|c|c|} 
\hline
ID & $m_\mathrm{[O\ III]} - m_\mathrm{H\alpha}$ & $\mathrm{EW_{H\alpha}}$ & $\mathrm{EW_{[O\ III]+H\beta}}$ & $R_3$ & $12 + \log(\mathrm{O/H})$ & $\chi^2_\mathrm{low\ Z}$ & $\Delta \chi^2_\mathrm{med\ Z}$\\
 & (mag) & (\AA) & (\AA) & & & \\
\hline
116605 & $0.55^{+0.04}_{-0.03}$ & $2948^{+48}_{-98}$ & $851^{+63}_{-72}$ & $0.45^{+0.09}_{-0.09}$ & $6.51^{+0.11}_{-0.13}$ & $50.8$ & $98.8$\\
189323 & $0.68^{+0.06}_{-0.06}$ & $2757^{+193}_{-204}$ & $631^{+127}_{-102}$ & $0.21^{+0.14}_{-0.09}$ & $6.17^{+0.25}_{-0.34}$ & $27.0$ & $146.4$\\
467299 & $0.40^{+0.04}_{-0.04}$ & $1575^{+135}_{-133}$ & $345^{+51}_{-44}$ & $0.26^{+0.13}_{-0.12}$ & $6.30^{+0.21}_{-0.28}$ & $34.6$ & $24.7$\\
487533 & $0.50^{+0.08}_{-0.08}$ & $1563^{+181}_{-169}$ & $318^{+86}_{-44}$ & $0.17^{+0.32}_{-0.08}$ & $6.08^{+0.56}_{-0.34}$ & $11.5$ & $35.8$\\
1237037 & $0.61^{+0.07}_{-0.07}$ & $2992^{+81}_{-195}$ & $662^{+71}_{-59}$ & $0.18^{+0.10}_{-0.07}$ & $6.04^{+0.22}_{-0.27}$ & $31.6$ & $106.0$\\
117051 & $0.21^{+0.16}_{-0.17}$ & $1753^{+315}_{-247}$ & $328^{+81}_{-43}$ & $0.12^{+0.07}_{-0.03}$ & $5.82^{+0.32}_{-0.23}$ & $24.8$ & $19.5$\\
546658 & $0.58^{+0.34}_{-0.31}$ & $2933^{+133}_{-246}$ & $700^{+152}_{-96}$ & $0.25^{+0.23}_{-0.13}$ & $6.24^{+0.37}_{-0.41}$ & $24.9$ & $31.7$\\
100687 & $0.49^{+0.12}_{-0.12}$ & $2888^{+119}_{-293}$ & $734^{+157}_{-117}$ & $0.32^{+0.22}_{-0.15}$ & $6.36^{+0.28}_{-0.33}$ & $13.5$ & $63.2$\\
123650 & $0.93^{+0.22}_{-0.20}$ & $2884^{+126}_{-309}$ & $754^{+177}_{-155}$ & $0.37^{+0.20}_{-0.19}$ & $6.43^{+0.24}_{-0.35}$ & $15.4$ & $54.7$\\
51246 & $0.90^{+0.20}_{-0.19}$ & $2050^{+215}_{-243}$ & $399^{+72}_{-55}$ & $0.14^{+0.04}_{-0.04}$ & $5.87^{+0.15}_{-0.21}$ & $26.5$ & $36.7$\\
487127 & $1.14^{+0.03}_{-0.03}$ & $3131^{+16}_{-69}$ & $619^{+7}_{-12}$ & $0.07^{+0.01}_{-0.01}$ & $5.52^{+0.11}_{-0.07}$ & $142.9$ & $72.9$\\
1008821 & $0.91^{+0.14}_{-0.13}$ & $3011^{+85}_{-186}$ & $677^{+108}_{-65}$ & $0.18^{+0.18}_{-0.08}$ & $6.08^{+0.35}_{-0.36}$ & $14.6$ & $47.4$\\
121273 & $0.74^{+0.12}_{-0.12}$ & $2951^{+72}_{-168}$ & $787^{+118}_{-111}$ & $0.37^{+0.17}_{-0.15}$ & $6.43^{+0.23}_{-0.29}$ & $20.2$ & $51.7$\\
102686 & $0.42^{+0.18}_{-0.18}$ & $2512^{+441}_{-742}$ & $559^{+168}_{-193}$ & $0.21^{+0.16}_{-0.09}$ & $6.19^{+0.30}_{-0.33}$ & $18.1$ & $22.1$\\
\hline
\end{tabular}
\caption{Metallicity-related properties of the 14 EMPG candidates. Listed are the JADES DR5 ID, the observed $m_\mathrm{[O\ III]} - m_\mathrm{H\alpha}$ colour in the wide-band filters tracing \OIII\ and \Ha\ (characteristically red for EMPGs), the \Ha\ rest-frame equivalent width $\mathrm{EW_{H\alpha}}$, the \OIII\ + \Hb\ rest-frame equivalent width $\mathrm{EW_{[O\ III] + H\beta}}$, $R_3 = \mathrm{[O\ III]\ \lambda 5007 /H\beta}$ line ratio, oxygen abundance $12 + \log(\mathrm{O/H})$, chi-squared statistic for the low metallicity model $\chi^2_\mathrm{low\ Z}$, difference in chi-squared statistic for the intermediate and low metallicity models $\Delta \chi^2_\mathrm{med\ Z} = \chi^2_\mathrm{med\ Z} - \chi^2_\mathrm{low\ Z}$.}
\label{tab:empg_candidates2}
\end{center}
\end{table*}

\begin{figure*}
\centering
\includegraphics[width=.9\linewidth]{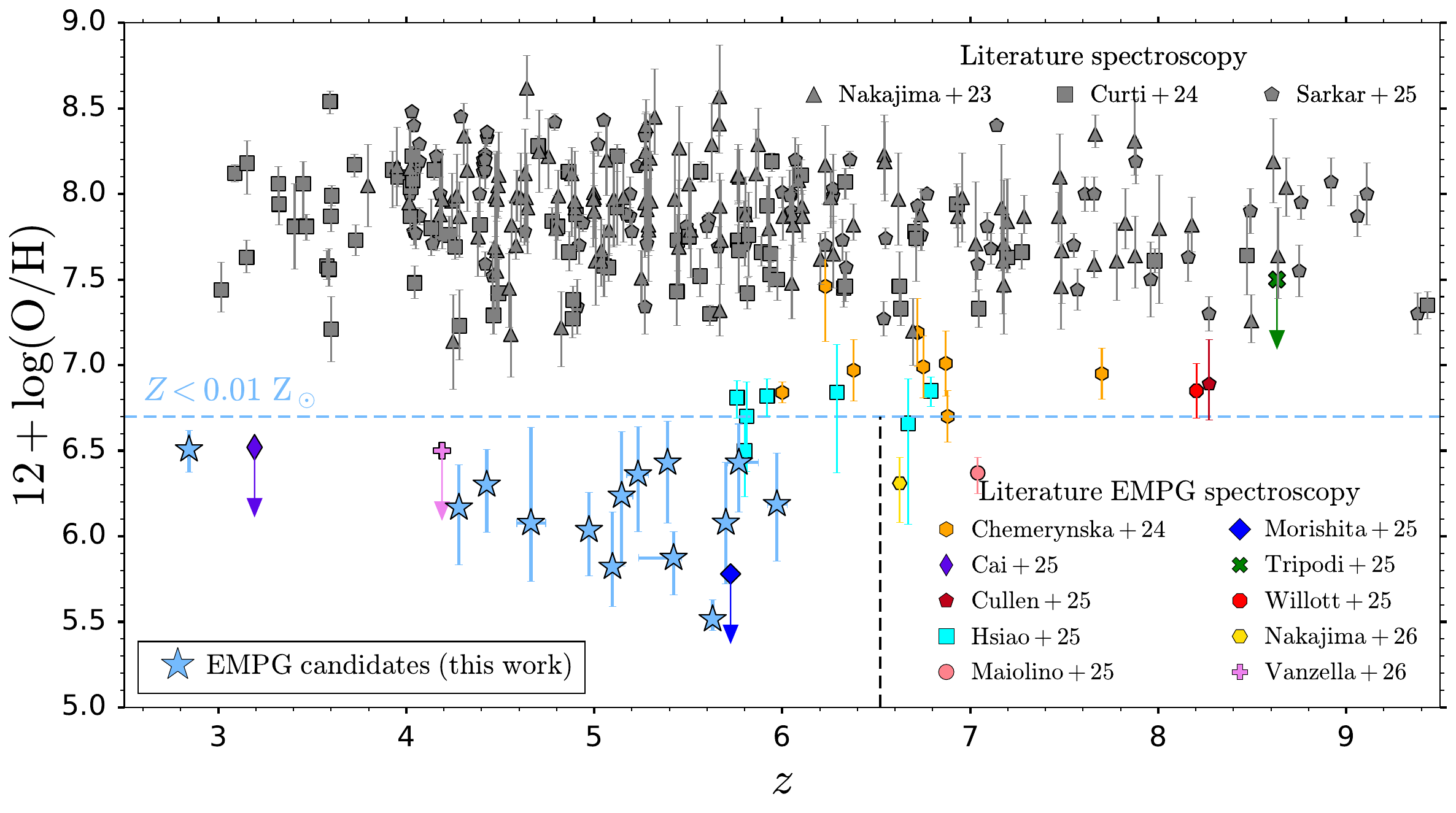}
\caption{Gas-phase metallicity traced by the oxygen abundance $12 + \log(\mathrm{O/H})$ against redshift. The photometrically-inferred metallicities of our final 22 EMPG candidates are displayed (light blue stars), spanning $2.84 < z < 5.97$. The light blue horizontal dashed line indicates our EMPG selection criterion: $Z < 0.01~\mathrm{Z}_\odot$. Spectroscopically-confirmed EMPGs are also shown, with gravitationally-lensed UNCOVER galaxies \citep[orange hexagons,][]{Chemerynska2024}, a serendipitously-discovered CAPERS galaxy \citep[purple thin diamond,][]{Cai2025}, a galaxy with extremely blue UV slope \citep[dark red pentagon,][]{Cullen2025}, SAPPHIRES-EDR galaxies \citep[cyan squares,][]{Hsiao2025}, the extremely metal-poor host galaxy of a lensed LRD \citep[pale red circle,][]{Maiolino2026b}, the most pristine galaxy currently known \citep[blue diamond,][]{Morishita2025}, the metal-poor host galaxy of a LRD \citep[green plus,][]{Tripodi2025}, a highly-magnified CANUCS galaxy \citep[red octagon,][]{Willott2025}, a highly-magnified strong \Lya\ emitter \citep[yellow hexagon,][]{Nakajima2026} and a highly-magnified UNCOVER galaxy \citep[pink plus,][]{Vanzella2026}. Such extremely metal-poor galaxies are typically missing (due to their rarity and faintness) from general high-redshift galaxy surveys (CEERS, ERO, GLASS and JADES, collectively, shown in grey), with metallicity measurements from \citet{Nakajima2023} (triangles), \citet{Curti2024} (squares) and \citet{Sarkar2025} (pentagons), which typically probe galaxies with much higher metallicities. Hence targeted NIRSpec follow-up of promising EMPG candidates is needed to accelerate our understanding of close-to-pristine star formation at high redshift.}
\label{fig:metallicity_redshift}
\end{figure*}

We report the general and metallicity-related properties of our 14 EMPG candidates in Table~\ref{tab:empg_candidates} and Table~\ref{tab:empg_candidates2}, respectively, ordered by increasing redshift. We show our 14 EMPG candidates (blue stars) in the metallicity--redshift plane in Fig.~\ref{fig:metallicity_redshift}. The bulk of our candidates are at $z > 5$, but we do find candidates as low as $z=2.84$. We also display spectroscopically-confirmed EMPGs from the literature as various coloured data points. These are: gravitationally-lensed UNCOVER galaxies targeted by NIRSpec \citep[orange hexagons]{Chemerynska2024}, a serendipitous EMPG candidate discovered by CAPERS/COSMOS NIRSpec spectroscopy \citep[purple thin diamond]{Cai2025}, a serendipitous EMPG targeted by EXCELS NIRSpec spectroscopy for its extremely blue UV slope \citep[dark red pentagon]{Cullen2025}, EMPGs identified by deep SAPPHIRES NIRCam slitless spectroscopy \citep[cyan squares]{Hsiao2025}, the extremely metal-poor host galaxy of the lensed LRD Abell2744-QSO1 \citep[pale red circle]{Maiolino2026b}, a highly-magnified and currently the most pristine galaxy known revealed by deep ALT NIRCam F356W slitless spectroscopy \citep[blue diamond]{Morishita2025}, the metal-poor host galaxy of a LRD \citep[green plus]{Tripodi2025}, a highly-magnified galaxy targeted for its large magnification \citep[red octagon]{Willott2025}, a highly-magnified galaxy first identified by \citet{Vanzella2023} through its strong \Lya\ emission from blind MUSE IFU spectroscopy \citep[yellow hexagon]{Nakajima2026} and a highly-magnified galaxy serendipitously discovered from UNCOVER NIRSpec data \citep[pink plus]{Vanzella2026}. Galaxies targeted by general high-redshift galaxy surveys are shown in grey, with metallicity measurements from \citet{Nakajima2023} (triangles), \citet{Curti2024} (squares) and \citet{Sarkar2025} (pentagons), collectively using data from the CEERS \citep{Finkelstein2025}, ERO \citep{Pontoppidan2022}, GLASS \citep{Treu2022} and JADES \citep{Bunker2024} programmes.

We see that thus far, very few EMPGs have been spectroscopically confirmed relative to the large number of galaxies targeted in general purpose spectroscopic surveys. This likely stems from the rarity and faintness of EMPGs, as we will discuss later. However, we do note that the metallicities shown (in grey) were all derived using local Universe strong-line calibrations. Owing to the generally higher ionisation parameters at high redshift, high-$z$ calibrations attribute less oxygen abundance to a given observed \OIII\ $\lambda 5007$/\Hb\ ratio \citep[see e.g.\@][]{Sanders2024}, lowering the inferred metallicity and thus the metallicity floor encountered by these general surveys. Nevertheless, targeted selection, such as highly-magnified sources which are intrinsically faint (and thus low mass and might be expected to be metal-poor), strong \Lya\ emitters which may be very young starbursts (with the strong \Lya\ enabling the detection of very faint galaxies that are otherwise challenging to detect with continuum photometry), peculiar sources with extremely blue UV slopes and so exotic stellar populations, or photometrically-inferred metal-poor galaxies, can favour the discovery of EMPGs. Furthermore, deep spectroscopy, either through long exposure times and/or significant gravitational lensing, is also reaping results. Our EMPG candidates, if spectroscopically confirmed, constitute a significant increase in the number of known EMPGs, possibly also pushing the metallicity frontier to even more pristine systems ($12 + \log(\mathrm{O/H}) < 5.78$).

\subsection{EMPG candidate ID 123650: A dusty and/or dense-gas starburst?} \label{subsec:123650}

In this section we discuss the NIRSpec/PRISM spectrum and photometry of EMPG candidate ID 123650. Motivated by this data, we outline how dusty starbursts and/or starbursts radiating into dense gas can potentially masquerade as EMPG candidates in photometry, possibly also exhibiting apparently strong \Ha\ emission, yet weak \OIII\ + \Hb\ emission. 

\begin{figure*}
\centering
\includegraphics[width=.8\linewidth]{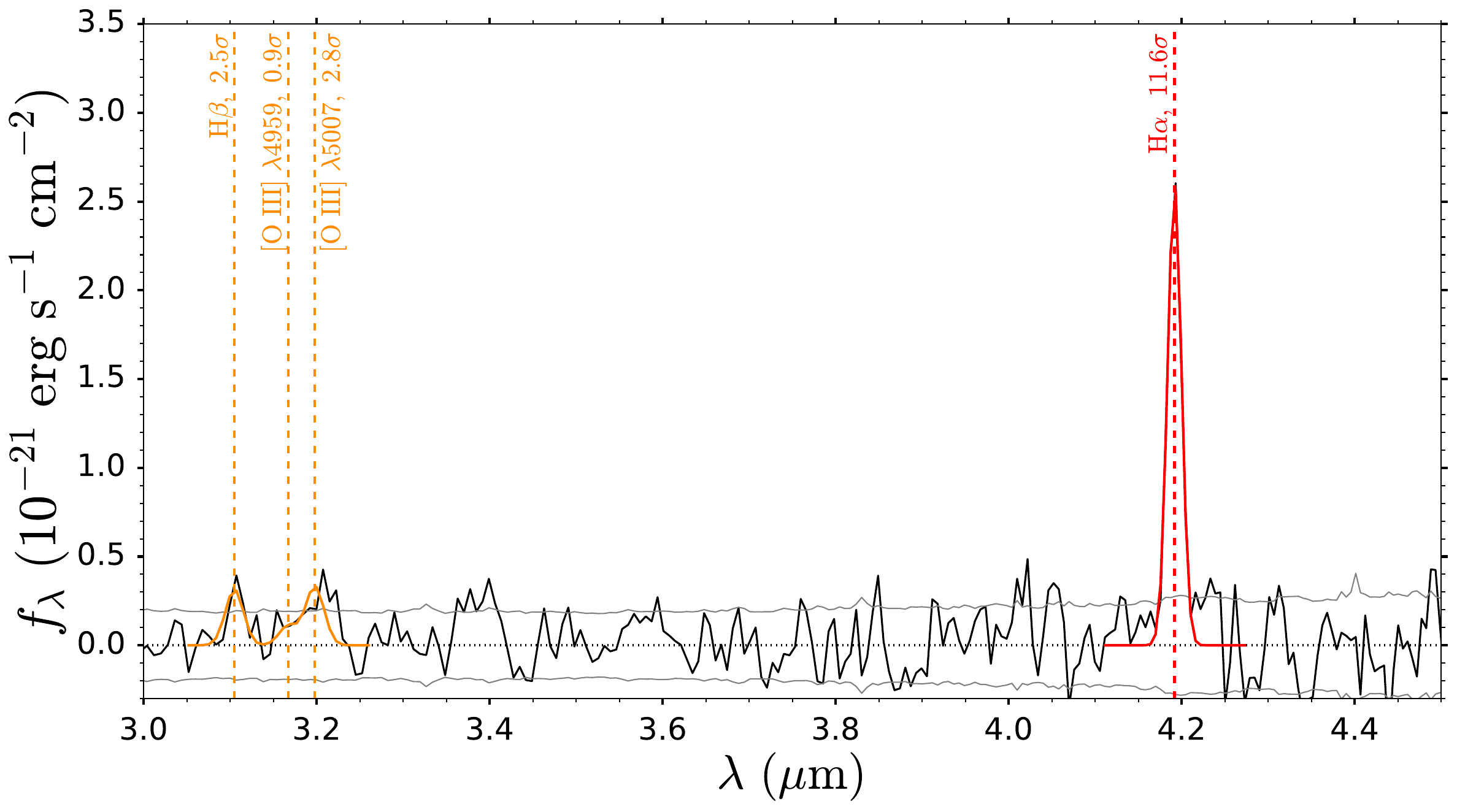} \\[4.5ex]
\includegraphics[width=.475\linewidth] {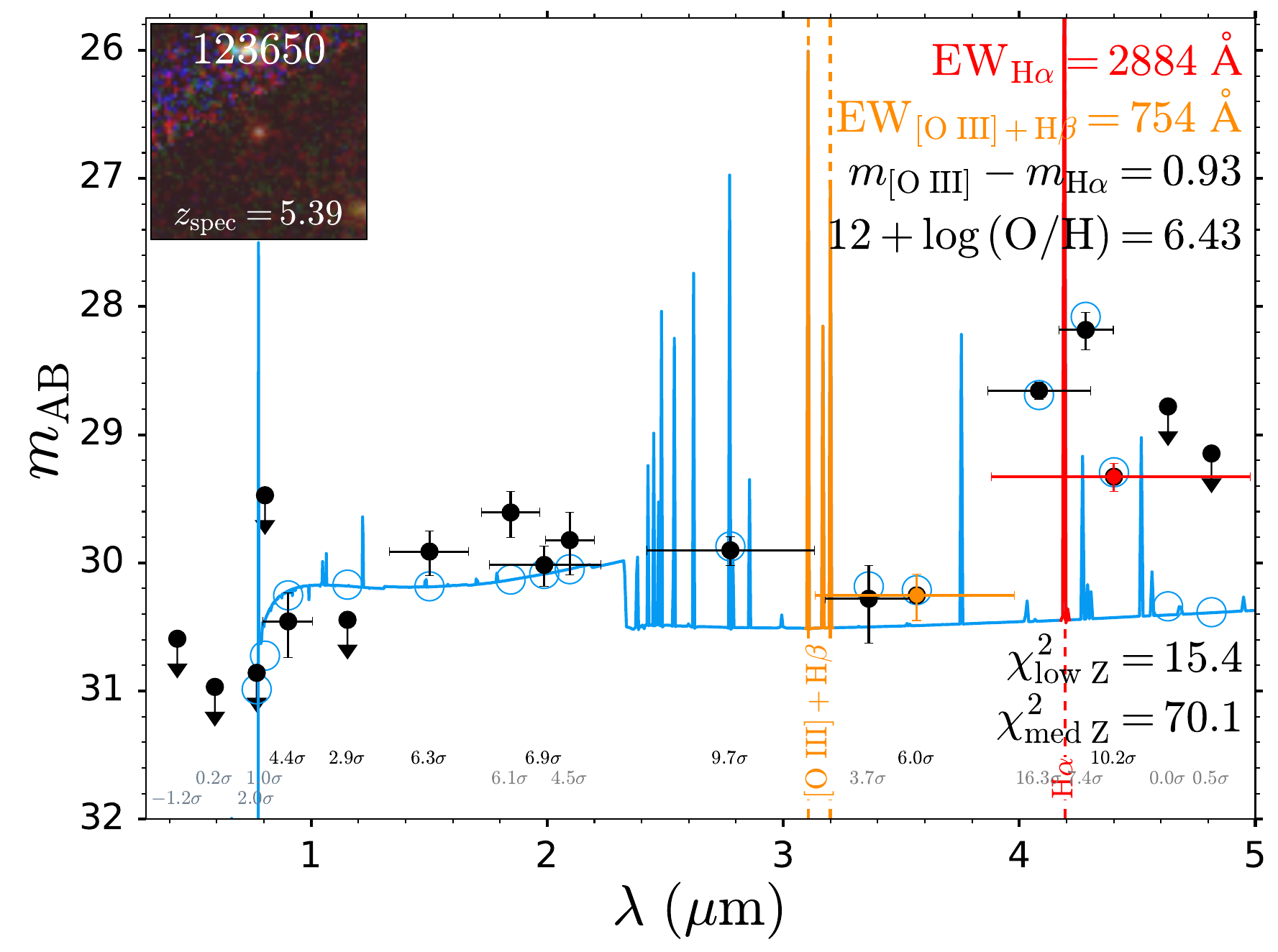} \hfill
\includegraphics[width=.475\linewidth]{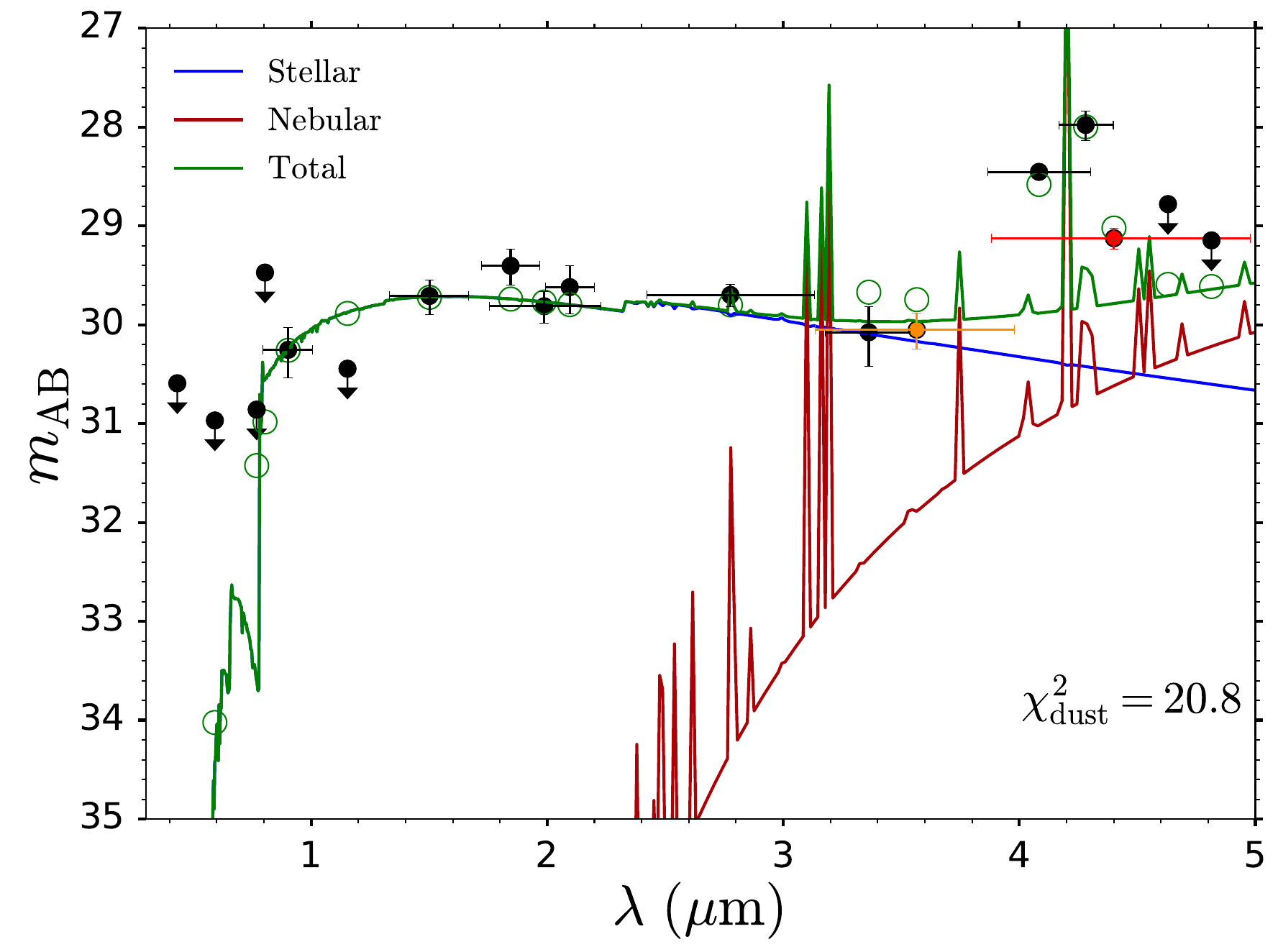}  
\caption{Dusty starbursts or starbursts radiating into dense gas can potentially masquerade as EMPG candidates in photometry, mimicking the strong \Ha, but weak \OIII\ + \Hb\ emission of extremely metal-poor galaxies. Top panel: NIRSpec PRISM spectrum (black, $\pm 1\sigma$ errors in grey) of EMPG candidate ID 123650 in the vicinity of \Ha\ (Gaussian fit highlighted in red) and \OIII\ + \Hb\ (highlighted in orange). Owing to the non-detection of the \OIII\ $\lambda5007$ and \Hb\ lines we cannot make conclusive statements about the gas-phase metallicity of ID 123650. However, the large Balmer decrement $\mathrm{H}\alpha/\mathrm{H}\beta > 5.33$ indicates dust reddening and/or dense gas effects are likely affecting its spectrum and photometry. Bottom-left panel: The photometry of EMPG candidate ID 123650 is well-described ($\chi^2 = 15.4$) by extremely metal-poor models. Bottom-right panel: Models incorporating separate dust attenuation for the stellar (blue, weakly attenuated) and nebular (dark red, strongly attenuated) components of a 1~Myr old instantaneous starburst (green, total) from \textsc{Bagpipes} can also provide a good fit to the photometry ($\chi^2 = 20.8)$. However, this requires an extremely steep dust attenuation law for the nebular component, with power-law index $\delta_\mathrm{neb}=-1.5$ modifying the \citet{Calzetti2000} law, as well as a still relatively low metallicity ($Z_\mathrm{gas} \approx 0.05~\mathrm{Z}_\odot$). This extreme nebular dust attenuation is at odds with the detection of \Lya\ in the 141~h MUSE eXtremely Deep Field spectrum \citep{Bacon2023} of ID 123650. Hence we favour the dense-gas scenario (with electron density $n_\mathrm{e} \gtrsim 7 \times 10^5$~cm$^{-3}$) for ID 123650, where the weak \OIII\ emission and weak \Hb\ emission are attributable to collisional de-excitation and resonant scattering, respectively.} 
\label{fig:123650}
\end{figure*}

We show the NIRSpec PRISM spectrum (black) of ID 123650 in the top panel of Fig.~\ref{fig:123650}, focusing on the spectral region in the vicinity of \Ha\ (highlighted in red) and \OIII\ + \Hb\ (highlighted in orange). We utilise a 3-pixel-wide extraction aperture (as opposed to the standard 5 pixels) to maximise the spectral SNR for this faint, compact source, following the JADES NIRSpec data reduction procedure outlined in \citet{Scholtz2026}. The $\pm 1\sigma$ errors are shown in grey. While \Ha\ is clearly detected, the \OIII\ + \Hb\ lines are not well-detected, as expected given the line fluxes estimated from SED-fitting the photometry relative to the PRISM line sensitivity limit in 7~h exposure time. We estimate line fluxes by fitting the \Ha, \Hb, and \OIII\ $\lambda\lambda4959,5007$ lines with Gaussian profiles using the \textsc{lmfit} package \citep{Newville2016}, comparing the average flux density in each spectral channel (rather than the Gaussian evaluated at the channel centre) to the data, optimising the fit for minimum $\chi^2$. To estimate line flux uncertainties, we fit 10000 realisations of the spectrum, each generated by randomly perturbing the flux density measurement in every spectral channel in the original spectrum by a normal distribution with zero mean and standard deviation given by the flux density error. Line flux errors are given by the standard deviation of the 10000 line flux measurements. 

We find that \Ha\ is confidently detected at $11.6\sigma$, yielding a spectroscopic redshift $z = 5.385 \pm 0.001$. On the other hand, \OIII\ $\lambda5007$ and \Hb\ are likely undetected, with line flux SNR = 2.8 and 2.5, respectively. Hence we cannot make concrete statements about the gas-phase metallicity of ID 123650, given the unreliable constraints on $R_3$ = \OIII\ $\lambda5007$ / \Hb. However, despite the \Hb\ non-detection, we can confidently infer that the $\mathrm{H}
\alpha/\mathrm{H}\beta$ line flux ratio is well above the dust-free case-B value (2.86), with $\mathrm{H}\alpha/\mathrm{H}\beta > 5.33$, adopting the $3\sigma$ upper limit on the \Hb\ line flux. This elevated Balmer decrement is either attributable to dust reddening or dense gas effects. We now explore whether these scenarios can also explain the photometry of ID 123650, without requiring extremely low metallicities to explain the strong \Ha, yet weak \OIII\ + \Hb\ emission (as shown in the bottom-left panel of Fig.~\ref{fig:123650}).

In the dust reddening scenario the lower limit on the Balmer decrement corresponds to $A_V > 2.15, 1.59$, assuming the \citet{Calzetti2000} and the relatively steeper SMC bar \citet{Gordon2003} dust attenuation laws, respectively. Clearly, dust reddening can reduce the line fluxes of \OIII\ and \Hb\ with respect to \Ha. However, equivalent widths, which our photometric EMPG selection is based on, are typically assumed not to be affected by dust reddening. If the continuum and line emission of a young starburst ($< 10$~Myr) are affected by the same dust attenuation, then the line equivalent widths are preserved because both the line flux and the neighbouring continuum level are decreased by the same factor, so metal-rich, dusty starbursts should not masquerade as EMPG candidates in photometry. However, if such a young starburst coincides with an unobscured population, then it is conceivable (though very extreme) that the continuum of the young starburst can dominate in the vicinity of \Ha\ (resulting in maximally strong equivalent widths, ${\sim}3000$~\AA), while the continuum of the unobscured population dominates in the vicinity of \OIII\ + \Hb\ (yielding comparatively very weak equivalent widths, ${\sim}750$~\AA), demanding an extremely steep dust attenuation law and/or severe dust attenuation for the continuum of the young starburst to change so substantially over such a small wavelength interval. Rather than invoking two stellar populations, it is possible to achieve a similar result by relaxing the assumption (commonly adopted in SED-fitting codes) that the continuum and line emission of a $< 10$~Myr old starburst are subject to the same amount of dust attenuation. Specifically, if the stellar continuum of the young starburst is less attenuated than the nebular line + continuum emission, then equivalent widths will not be preserved, actually decreasing with increasing attenuation. Such a scenario does indeed occur, such as in the Tarantula Nebula, where the OB stars are less dust-obscured along our sightline than the \ion{H}{II} region they are powering.

We fit the photometry of EMPG candidate ID 123650 with such a split dust model. Motivated by the extremely high \Ha\ equivalent widths inferred from the nominal EMPG model fit, which we consider robust given the combined F410M + F430M + F444W measurements of the line boost, we utilise maximally young 1~Myr old \textsc{Bagpipes} instantaneous starburst models. We separate the stellar and nebular (line+continuum) components, and apply different amounts of dust attenuation ($A_{V,*}$ and $A_{V, \mathrm{neb}}$) to these components. Furthermore, owing to the likely different star--dust and nebular--dust geometry, we also adopt different attenuation laws ($k_*(\lambda)$ and $k_\mathrm{neb}(\lambda)$) for the two components. Following \citet{Salim2018}, we utilise modified \citet{Calzetti2000} dust attenuation laws \citep[see also][]{Shivaei2025}, with $k(\lambda) \propto k_\mathrm{Calzetti}(\lambda)
\times \lambda^\delta$, where $\delta$ is the power-law index modifying the slope of the attenuation curve, where we adopt separate values ($\delta_*$ and $\delta_\mathrm{neb}$) for the two components. Note that negative $
\delta$ values steepen the dust attenuation law. We use \textsc{lmfit} \citep{Newville2016} to find the optimal fitting parameters yielding the minimum $\chi^2$ for a given input gas-phase metallicity. We find that this framework can yield an adequate fit to the photometry of ID 123650, with comparable $\chi^2 = 20.8$ to the EMPG scenario ($\chi^2 = 15.4)$, though still requires low metallicity ($Z_\mathrm{gas} \approx 0.05~\mathrm{Z}_\odot$). One can in principle photometrically distinguish between the EMPG and dusty starburst scenarios through deep medium-/wide-band photometry beyond \Ha\ establishing the absence (or presence) of a steeply rising red dusty continuum. 

We show the split dust model fit in the bottom-right panel of Fig.~\ref{fig:123650}, with the relatively unattenuated stellar component, the strongly dust-attenuated nebular component, and the total SED shown in blue, dark red and green, respectively. As depicted, to go from maximal \Ha\ equivalent widths (due to little dust attenuation) to minimal \OIII\ + \Hb\ equivalent widths (due to substantial dust attenuation), the nebular dust attenuation curve must be extremely steep for the attenuation effect to change so substantially over such a small wavelength range. Indeed, we require a very low nebular power-law index $\delta_\mathrm{neb} = -1.5$ (the lower limit of our fitting range) to achieve this \citep[for comparison, see e.g.\@][]{Shivaei2025, Sun2026}. Note that $\delta \approx -0.5$ for the SMC bar law \citep{Gordon2003}, already considered a steep attenuation law.  

A specific challenge for the dusty starburst scenario in the case of ID 123650 is the likely detection of \Lya\ in the ultra-deep 141~h MUSE spectrum taken through the MUSE eXtremely Deep Field survey \citep{Bacon2023}. Although assigned the lowest redshift confidence level in that survey (with ID 8425), the reported \Lya\ SNR is 7.3, with a redshift $z_{\mathrm{Ly\alpha}} = 5.3842$ that is fully consistent with our \Ha-based redshift $z_{\mathrm{H}\alpha} = 5.385 \pm 0.001$. Hence we consider the \Lya\ detection to be robust. With a reported \Lya\ line flux $F_{\mathrm{Ly}\alpha} = 1.0 \times 10^{-18}$~erg~s$^{-1}$~cm$^{-2}$, we find $\mathrm{Ly}\alpha/\mathrm{H}\alpha=1.92$. It seems very challenging for \Lya\ to be detected, and furthermore for it to exhibit such a large line ratio with \Ha, in the case of extreme dust attenuation discussed earlier. We note that the reported \Lya\ line flux places it well below ($\approx 3\times$) the PRISM $5\sigma$ detection limit, so is not seen in our spectrum.

Hence we consider an alternative scenario to explain both the high Balmer decrement and photometry of EMPG candidate ID 123650. At high gas densities, comparable to or exceeding the critical density of \OIII\ $\lambda5007$ ($7 \times 10^{5}$~cm$^{-3}$), collisional de-excitations substantially weaken the \OIII\ flux, decreasing as $(1 + n_\mathrm{e}/n_\mathrm{crit})^{-1}$. Furthermore, resonant scattering of \Hb\ into $\mathrm{Pa}\alpha$ and \Ha, which becomes more prominent at greater densities due to a larger fraction of neutral hydrogen in the $n=2$ state, weakens \Hb\ \citep{Yan2026}. In addition, the enhanced collisional excitation to $n=3$ at greater densities promotes \Ha\ emission. Hence the characteristic signature of extremely metal-poor galaxies, namely strong \Ha, yet weak \OIII\ + \Hb\ emission, can in principle be mimicked by young starbursts radiating into dense gas. Thus this scenario can qualitatively match both the photometry and the Balmer decrement signature in the spectrum of ID 123650. However, at least for dense gas surrounding accreting supermassive black holes \citep[believed to be applicable to Little Red Dots, e.g.\@][]{Inayoshi2025}, the \Ha\ equivalent widths are generally relatively low (several hundred \AA), unless the hydrogen volume density $n_\mathrm{H} = 10^{8\textrm{--}10}$~cm$^{-3}$ and the hydrogen column density $N_\mathrm{H} = 10^{23\textrm{--}24}$~cm$^{-2}$, achieving a maximum equivalent width of 2000~\AA\ rest-frame \citep{Yan2026}. The development of analogous models powered by young starbursts would help establish under which dense gas conditions the extreme \Ha\ equivalent width ($\approx 3000$~\AA) seen for ID 123650 can be achieved, also enabling quantitative comparisons to both its photometry and spectrum. Such models would also establish to which degree the \Lya\ flux is attenuated by neutral hydrogen gas, and whether this is compatible with the MUSE detection. 

We find it unlikely that ID 123650 is an AGN. We find no evidence for a broad-line component in the PRISM spectrum of \Ha, which is to be expected given the coarse PRISM spectral resolution and the relatively low line SNR. Both the extreme \Ha\ equivalent width and high Balmer decrement seem difficult to achieve for conventional Type-I AGN. These generally have very low \Ha\ equivalent widths \citep[${\sim}200$~\AA,][]{Maiolino2025}, as well as low Balmer decrements $\mathrm{H}\alpha/\mathrm{H}\beta \approx 3$ \citep{Dong2008} in the absence of dust reddening and/or dense gas effects. Furthermore, the bluer rest-frame optical counterparts to LRDs, referred to as Little Blue Dots \citep[LBDs,][]{Brazzini2026}, are thought to lack dense gas around their accreting supermassive black holes, so likely also have low \Ha\ equivalent widths and low Balmer decrements, so incompatible with ID 123650. Dense gas effects are probably applicable to LRDs, so these can have comparable Balmer decrements to ID 123650, though still are not expected to have extreme $\mathrm{EW_\mathrm{H\alpha}} \sim 3000$~\AA\ \citep{Yan2026}. Furthermore, ID 123650 does not exhibit a Balmer break in its photometry, so it seems unlikely to be an LRD.   

To summarise, dusty starbursts and dense-gas starbursts can possibly masquerade as EMPG candidates in photometry, potentially providing comparable quality photometric fits to extremely metal-poor models. Hence follow-up spectroscopy is needed to confirm the EMPG scenario. We deem the dense-gas scenario the most likely explanation for the combined photometry and spectrum of ID 123650, though it is possible that a combination of metal-poor, dense and dusty gas could be applicable, as is believed to be the case for the ``Pseudo Little Red Dot with no metal lines'' discussed by \citet{Caputi2026}. This source also exhibits signatures of strong \Ha, yet weak \OIII\ + \Hb\ in the photometry, together with a substantial Balmer decrement through a \Hb\ detection but lack of \OIII\ detection in PRISM spectroscopy, though with a slightly red $\Delta m_\mathrm{jump} > 0$ colour so would not be selected by our procedure. Deep grating spectroscopy of ID 123650 would enable \Hb\ to be detected at high significance. A high $R_3$ = \OIII\ $\lambda5007$ / \Hb\ ratio would establish high metallicity and thus perceived weak \OIII\ + \Hb\ emission in the photometry due to extreme dust attenuation. In contrast, a low $R_3$ would be compatible with both the extremely metal-poor and dense-gas scenarios. Deep rest-frame UV spectroscopy could distinguish between these cases by the lack (or detection) of metal lines (e.g.\@ \ion{O}{III}] and \ion{C}{IV}) with greater critical density than \OIII $\lambda5007$.

\section{Extremely metal-poor galaxy properties} \label{sec:properties}

Having outlined our extremely metal-poor galaxy selection procedure, as well as possible sources of contamination, we now discuss the properties of our final 14 EMPG candidates. We describe their metallicity-related properties in Section~\ref{subsec:metallicity}, their empirical properties in Section~\ref{subsec:empirical}, and their colours in Section~\ref{subsec:colours}. Furthermore, we discuss EMPG number statistics in Section~\ref{subsec:number}, and discuss the observability of EMPGs with \emph{JWST} imaging and spectroscopy in Section~\ref{subsec:observability}.

\subsection{Metallicity relations} \label{subsec:metallicity}

\begin{figure*}
\centering
\includegraphics[width=.9\linewidth]{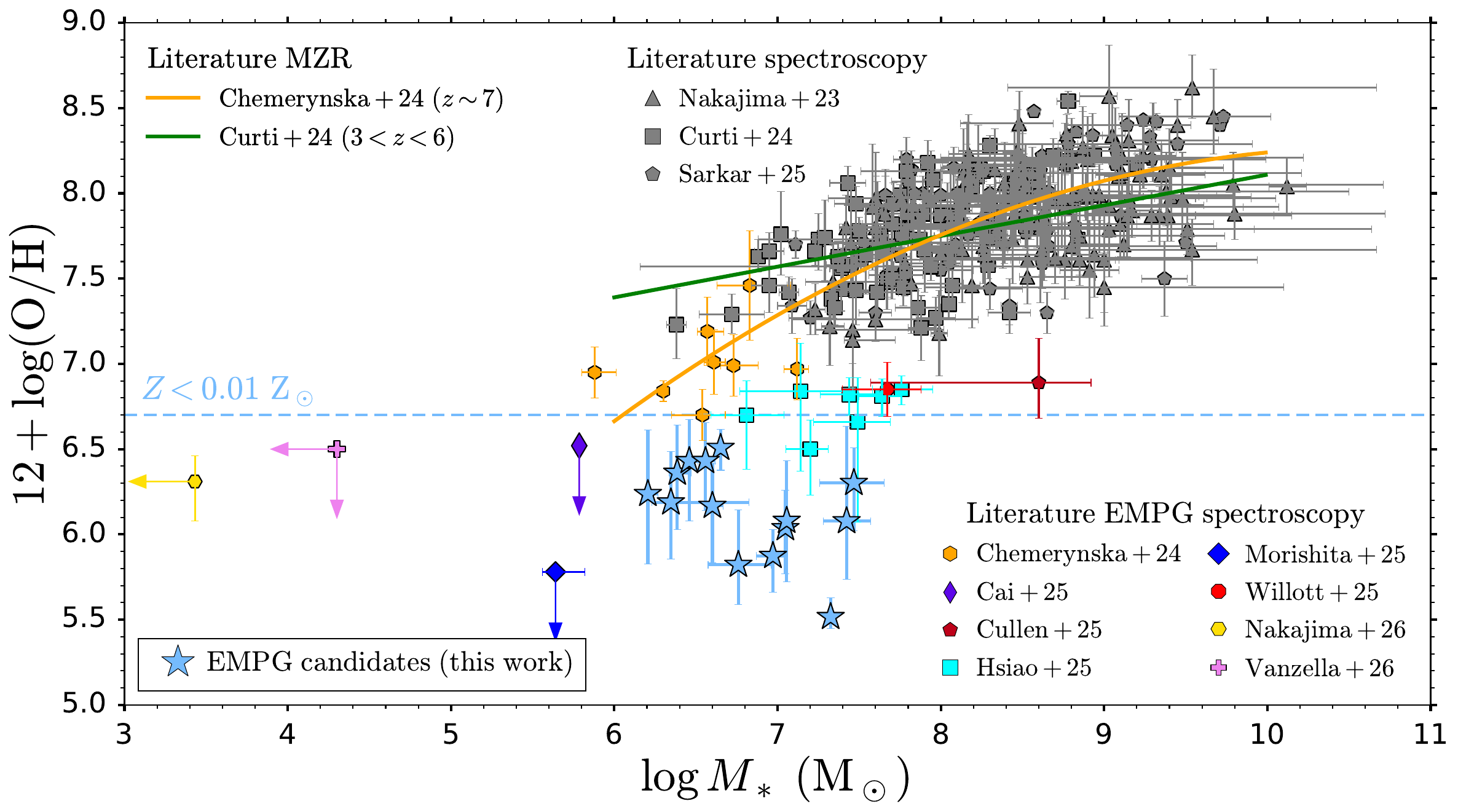}
\caption{The stellar mass--gas-phase metallicity plane, with symbols and colour coding as in Fig.~\ref{fig:metallicity_redshift}. Also shown are the observed mass--metallicity relations from \citet{Chemerynska2024} (orange) and \citet{Curti2024} (green). Our EMPG candidates are very low-mass dwarf galaxies (median $M_* = 10^{6.7}~\mathrm{M}_\odot$), likely constituting the low metallicity tail of the mass--metallicity relation extended to very faint systems, brought into view through the exceptionally deep JADES imaging.}
\label{fig:mass_metallicity}
\end{figure*}

We show our 14 EMPG candidates in the stellar mass--gas-phase metallicity plane in Fig.~\ref{fig:mass_metallicity} (blue stars). The symbols and colour coding are similar to Fig.~\ref{fig:metallicity_redshift}, now also showing the observed mass--metallicity relation from \citet{Chemerynska2024} (orange line) and \citet{Curti2024} (green). We infer our EMPG candidates to be very low-mass dwarfs, with stellar masses spanning $M_* = 10^{6.2\textrm{--}7.5}~\mathrm{M}_\odot$ (median $M_* = 10^{6.7}~\mathrm{M}_\odot$), brought into view out to $z=5.97$ thanks to the exceptionally deep JADES NIRCam imaging. Their exceptionally low metallicities (${\sim}0.1$--$1\%~\mathrm{Z}_\odot$) represent the extreme low metallicity tail of the mass--metallicity distribution pushed to very low stellar masses. Provided that typical gas-phase metallicities continue to fall with decreasing stellar mass (rather than saturating at some limiting value), we anticipate that even deeper imaging would reveal an even greater abundance of EMPGs ($Z < 0.01~\mathrm{Z}_\odot$), pushing down to still lower metallicities. Given conventional IMFs \citep[i.e.\@][]{Salpeter1955, Kroupa2001, Chabrier2003}, a $10^6~\mathrm{M}_\odot$ instantaneous starburst at $z\sim6$ has a peak UV brightness $m_\mathrm{UV} \sim 30$, thus roughly representing the lowest stellar mass threshold that deep NIRCam blank-field imaging can likely attain. The flux boost from strong gravitational lensing provides the only realistic pathway to study the even fainter, less massive, likely close-to-pristine dwarf galaxies. However, if such pristine star formation follows a more top-heavy IMF, which we suspect may be the case for our EMPG candidates given their exceptionally high $\xi_\mathrm{ion, obs}$ values, then the greater light-to-mass ratios mean that the actual masses (and star formation rates) of these galaxies are lower than estimated using conventional IMFs. 

\begin{figure}
\centering
\includegraphics[width=\linewidth]{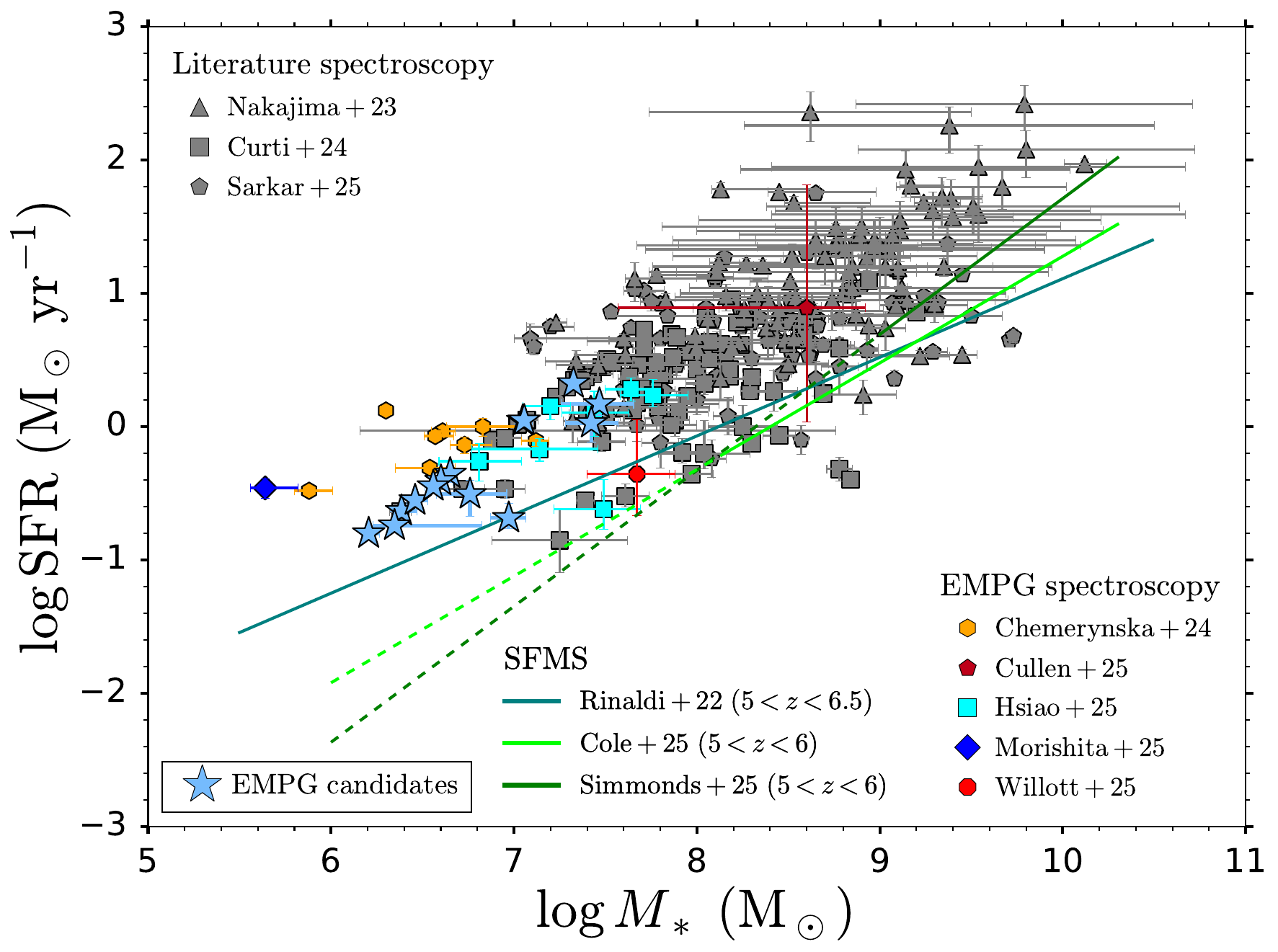}
\caption{The star formation rate (SFR) -- stellar mass plane, with symbols and colour coding as in Fig.~\ref{fig:metallicity_redshift}. Also shown are the star-forming main sequence SFR--$M_*$ relations from \citet{Rinaldi2022} (teal), \citet{Cole2025} (light green) and \citet{Simmonds2025} (dark green), with dashed lines indicating extrapolations beyond the stellar mass completeness limit. Our EMPG candidates are extreme starbursts, as expected from their strong \Ha\ equivalent widths and prominent Balmer jumps, being greatly elevated above the star-forming main sequence.} 
\label{fig:mass_sfr}
\end{figure}

To put the extremely low metallicities of our EMPGs in context, we also show their position in the SFR--$M_*$ plane in Fig.~\ref{fig:mass_sfr}, with the observed $5 < z < 6$ star-forming main sequence from \citet{Cole2025} and \citet{Simmonds2025} shown in light/dark green, respectively, as well as the \citet{Rinaldi2022} star-forming main sequence in teal. Unsurprisingly, our EMPGs (as well as others from the literature, various coloured data points) are well above the star-forming main sequence at high-redshift due to their starburst nature, in our case having been selected to have very strong \Ha\ EWs ($> 1500$~\AA) and prominent Balmer jumps ($\Delta m_\mathrm{jump} < -0.15$). We note that the sSFR$_{10}$ of our EMPGs tend to be around $10^{-7}$~yr$^{-1}$ ($= (M_*/(10^{7}~\mathrm{yr}))/M_*$), the limiting value determined from SED-fitting using an averaging timescale measurement of 10~Myr, corresponding to all the stellar mass being assembled in the past 10~Myr, reflecting their extreme starburst nature. The sSFRs from the literature spectroscopic samples can exceed this sSFR threshold, due to the SFR estimates being derived independently of the stellar masses using \Ha\ or \Hb\ line fluxes. Furthermore, we note that the SFRs of the spectroscopic samples tend to be biased high with respect to the photometrically-inferred star-forming main sequence, perhaps due to the requirement of sufficiently bright Balmer lines to make the spectroscopic SFR measurement. 

\begin{figure}
\centering
\includegraphics[width=\linewidth]{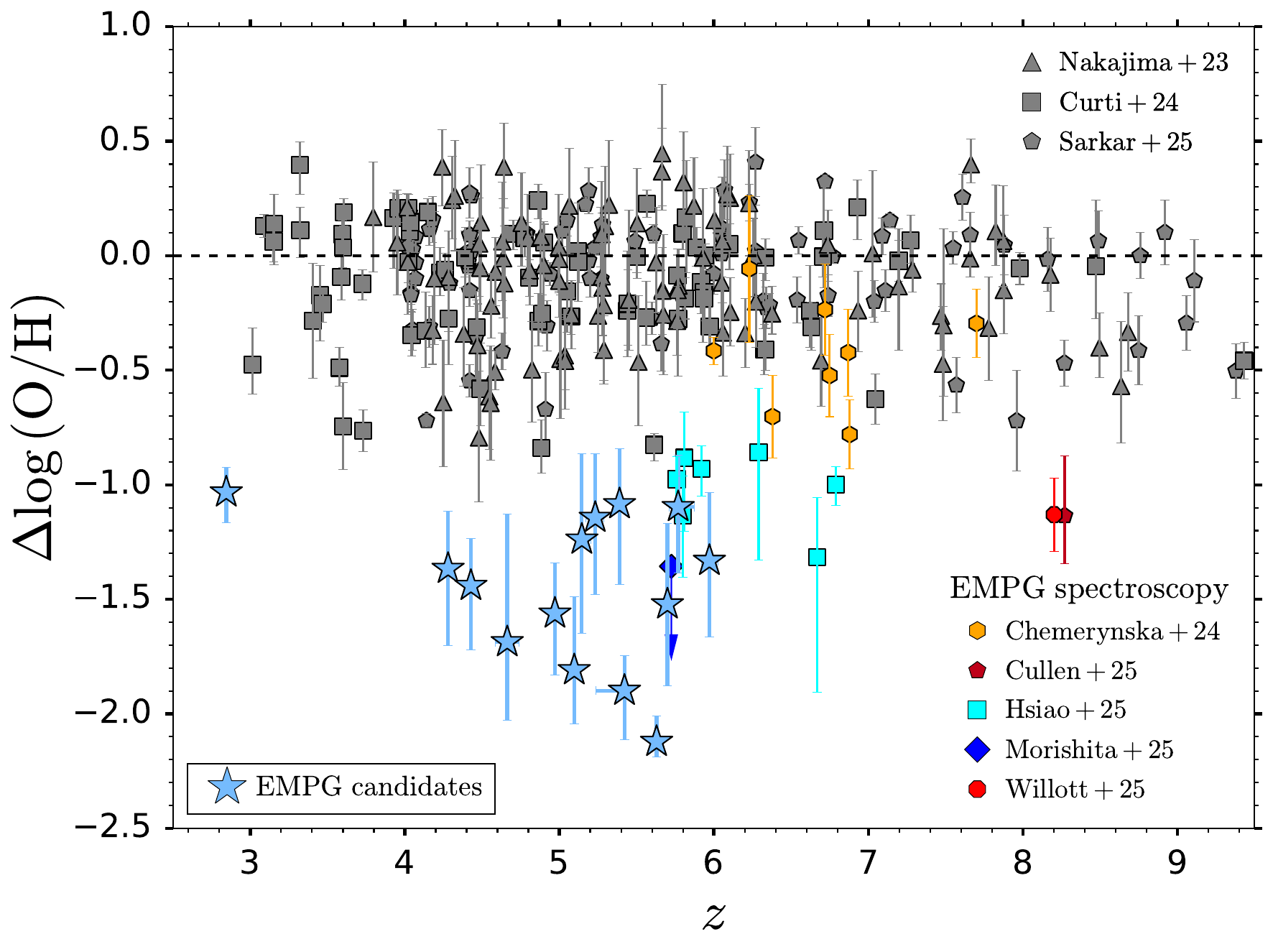}
\caption{Deviations $\Delta \log(\mathrm{O/H}) = (12 + \log(\mathrm{O/H}))_\mathrm{obs} - (12 + \log(\mathrm{O/H}))_\mathrm{pred}$ from the metallicity predicted \citep[using the formulation of][]{Andrews2013} by the fundamental metallicity relation, given the galaxy stellar mass $M_*$ and SFR. Our EMPG candidates are ${>}1$~dex below the FMR predictions, with these young, near-pristine starbursts thus representing an extreme departure from the more slowly-evolving equilibrium framework of the FMR, similar to literature EMPGs.}
\label{fig:fmr}
\end{figure}

Within the framework of the fundamental metallicity relation (FMR) relating stellar mass, star-formation rate and gas-phase metallicity \citep{Mannucci2010}, we would qualitatively expect our EMPGs to be metal-poor, due to both their low stellar masses and their relatively high star formation rates. However, as shown in Fig.~\ref{fig:fmr}, upon quantitatively comparing ($\Delta \log \mathrm{(O/H)}$) to the FMR-predicted gas-phase metallicity \citep[assuming][]{Andrews2013}, our EMPGs are estimated to have gas-phase metallicities that are $>1$ dex below the FMR predictions. This is again not unexpected, as these young starbursts (possibly undergoing one of their first star formation episodes) are extreme departures from the more slowly-evolving equilibrium framework of the FMR, with the literature EMPGs also tending to substantially deviate from the FMR predictions.  Indeed, even low-mass dwarf galaxies ($M_* < 10^9~\mathrm{M}_\odot$) in the local Universe do not follow the FMR \citep{Laseter2025}.

\subsection{Empirical properties} \label{subsec:empirical}

\begin{figure}
\centering
\includegraphics[width=\linewidth]{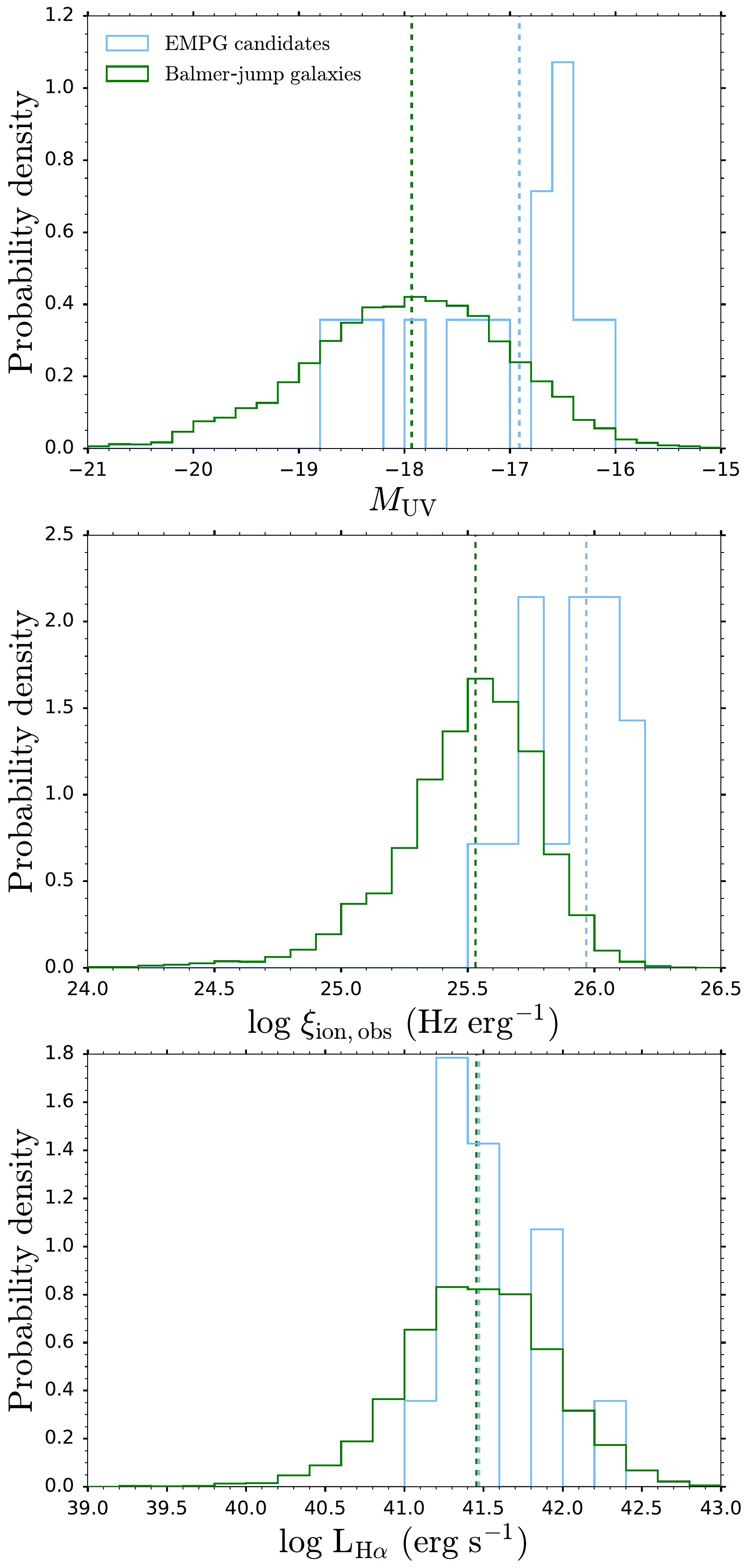}
\caption{Normalised histograms of empirical properties for EMPG candidates (light blue) and Balmer-jump galaxies (green). Top panel: EMPGs are typically very faint, with median (vertical dashed line) rest-frame UV absolute magnitude $M_\mathrm{UV} = -16.9$, demanding deep imaging to uncover. However, the rare bright tail ($M_\mathrm{UV} < -18 $) is accessible with shallower imaging. Middle panel: EMPGs are extremely efficient producers of ionising photons with median observed $\log\,( \xi_\mathrm{ion, obs} /\mathrm{(Hz\ erg^{-1})}) = 26.0$, thus exhibiting exceptionally strong nebular line and continuum emission, possibly powered by top-heavy star formation. Bottom panel: Despite their UV faintness, the EMPG candidates thus have intermediate \Ha\ luminosities with median $L_\mathrm{H\alpha} = 10^{41.5}$~erg~s$^{-1}$, facilitating future deep spectroscopic follow-up with NIRSpec.}
\label{fig:empirical_histograms}
\end{figure}

We now turn to the empirical properties of our EMPGs, as these are more directly accessible from the data, rather than depending on the IMF assumed. To put our EMPGs in context, we compare to the properties of general young starbursting Balmer-jump galaxies, defined to have $\Delta m_\mathrm{jump} < -0.15$ and rest-frame optical/NIR continuum (traced by the appropriate medium-band) SNR > 4 (i.e.\@ the green circles shown in Fig.~\ref{fig:ew_ew_selection}). It is for these galaxies that we believe our empirical colour-based assessment of line strengths is most reliable \citep[see also][]{Trussler2026}.

We show normalised histograms of the rest-frame UV absolute magnitude distribution $M_\mathrm{UV}$ for EMPGs (light blue) and Balmer-jump galaxies (green) in the top panel of Fig.~\ref{fig:empirical_histograms}. For Balmer-jump galaxies we empirically determine $M_\mathrm{UV}$ from the measurement of the apparent magnitude $m_\mathrm{UV}$ in the first wide-band filter fully redward of \Lya, also applying a correction to total flux density by scaling by the ratio between the flux densities in the Kron and 0.2~arcsec diameter apertures. EMPG properties are determined following a similar procedure, but derived from their median \textsc{Bagpipes} fits. We see that EMPGs are generally UV-faint, with a median $M_\mathrm{UV} = -16.9$, thus demanding deep NIRCam imaging to detect. They therefore mostly comprise the fainter end of the general starbursting Balmer-jump galaxy population, which has median $M_\mathrm{UV} = -17.9$. However, a rarer subset of EMPGs is brighter than $M_\mathrm{UV} = -18$, hence accessible in shallower imaging over larger areas, such as the $z=3.19$ EMPG candidate in the PRIMER/COSMOS field reported by \citet{Cai2025}. Thus extremely metal-poor star formation primarily occurs in faint galaxies, qualitatively in line with the very low stellar masses we were inferring for our EMPG candidates.

We show normalised histograms for the observed ionising photon production efficiencies in the middle panel of Fig.~\ref{fig:empirical_histograms}. As in \citet{Trussler2026}, \Ha\ line fluxes are empirically estimated using the photometric boost seen in the wide-band filter covering \Ha\ relative to the rest-frame optical/NIR continuum level traced by the medium band, taking into account the NIRCam wide-band filter throughput at the EAZY best-fit photometric redshift of the source. These line fluxes are converted into \Ha\ luminosities $L_{\mathrm{H}\alpha}$ according to the source redshift, and the ionising photon production rate $\dot{N}_\mathrm{ion} = 7.28\times 10^{11} L_\mathrm{H\alpha}$ is estimated assuming the standard case-B conversion rate at 10000~K \citep{Osterbrock2006}. The observed (i.e.\@ directly from the data, not correcting for potential dust attenuation) ionising photon production efficiency is then given by $\xi_\mathrm{ion,obs} = \dot{N}_\mathrm{ion}/L_{\nu,1500}$, where $L_{\nu,1500}$ is the monochromatic flux density in the first wide-band filter fully redward of \Lya. 

EMPGs are extremely efficient producers of ionising photons, with median $\log\, (\xi_\mathrm{ion, obs} /\mathrm{(Hz\ erg^{-1})}) = 26.0$, which is well above the median for Balmer-jump galaxies (25.5) and the canonical value of 25.2 \citep{Robertson2013}. Indeed, as discussed in \citet{Trussler2026} the maximum $\xi_\mathrm{ion, obs}$ achievable for normal starbursts following a regular IMF is $\approx 25.75$. Thus our EMPG candidates are likely undergoing top-heavy star formation with hot stars dominating the ionising spectrum, resulting in exceptionally strong nebular line and nebular continuum emission. Given the high $\log\, (\xi_\mathrm{ion, obs} /\mathrm{(Hz\ erg^{-1})}) > 25.6$ exceeding the threshold for nebular-dominated emission in \citet{Trussler2026}, our EMPG candidates may have a remarkably strong nebular continuum that outshines the stellar continuum in the rest-frame UV \citep{Cameron2024, Katz2025, Trussler2026}, possibly displaying a prominent UV downturn associated with two-photon continuum emission that could be revealed by future NIRSpec PRISM spectra, as seen in JADES GS-9422 \citep{Cameron2024}. We note that these remarkably high $\xi_\mathrm{ion, obs}$ values (beyond what is possible for dust-free regular-IMF starbursts) are achieved in our \textsc{Bagpipes} fitting with standard IMFs by invoking dust attenuation, which more greatly suppresses the UV luminosity relative to the \Ha\ emission in the optical, inflating the $\xi_\mathrm{ion, obs}$ value. The minimum, maximum and median V-band attenuation $A_\mathrm{V}$ arising outside \ion{H}{II} regions inferred for our set of 14 EMPG candidates are 0.01, 0.27, 0.16~mag, respectively (the values are $(1/0.44)\times$ larger inside \ion{H}{II} regions, namely 0.02, 0.61, 0.35~mag, respectively). We recognise that this dust attenuation is seemingly at odds with the extremely low metallicities ($Z < 1\%~\mathrm{Z}_\odot$) inferred for these galaxies, though do note that there is a case of high dust attenuation in a extremely metal-poor galaxy in the SPHINX$^{20}$ simulation \citep{Katz2023b}. We aim to reconcile this potential discrepancy by developing nebular-dominated spectral templates powered by top-heavy stellar populations in future work.

Despite their rest-frame UV faintness in imaging, EMPGs are intermediate in \Ha\ luminosity (bottom panel in Fig.~\ref{fig:empirical_histograms}), due to their exceptionally high ionising photon production efficiencies, with comparable median $L_\mathrm{H\alpha} = 10^{41.5}$~erg~s$^{-1}$ to Balmer-jump galaxies. Hence the Balmer line emission in our EMPG candidates will be accessible for deep spectroscopic follow-up, which we discuss further in Section~\ref{subsec:observability}. 

\subsection{Colours} \label{subsec:colours}

\begin{figure*}
\centering
\includegraphics[width=.55\linewidth]{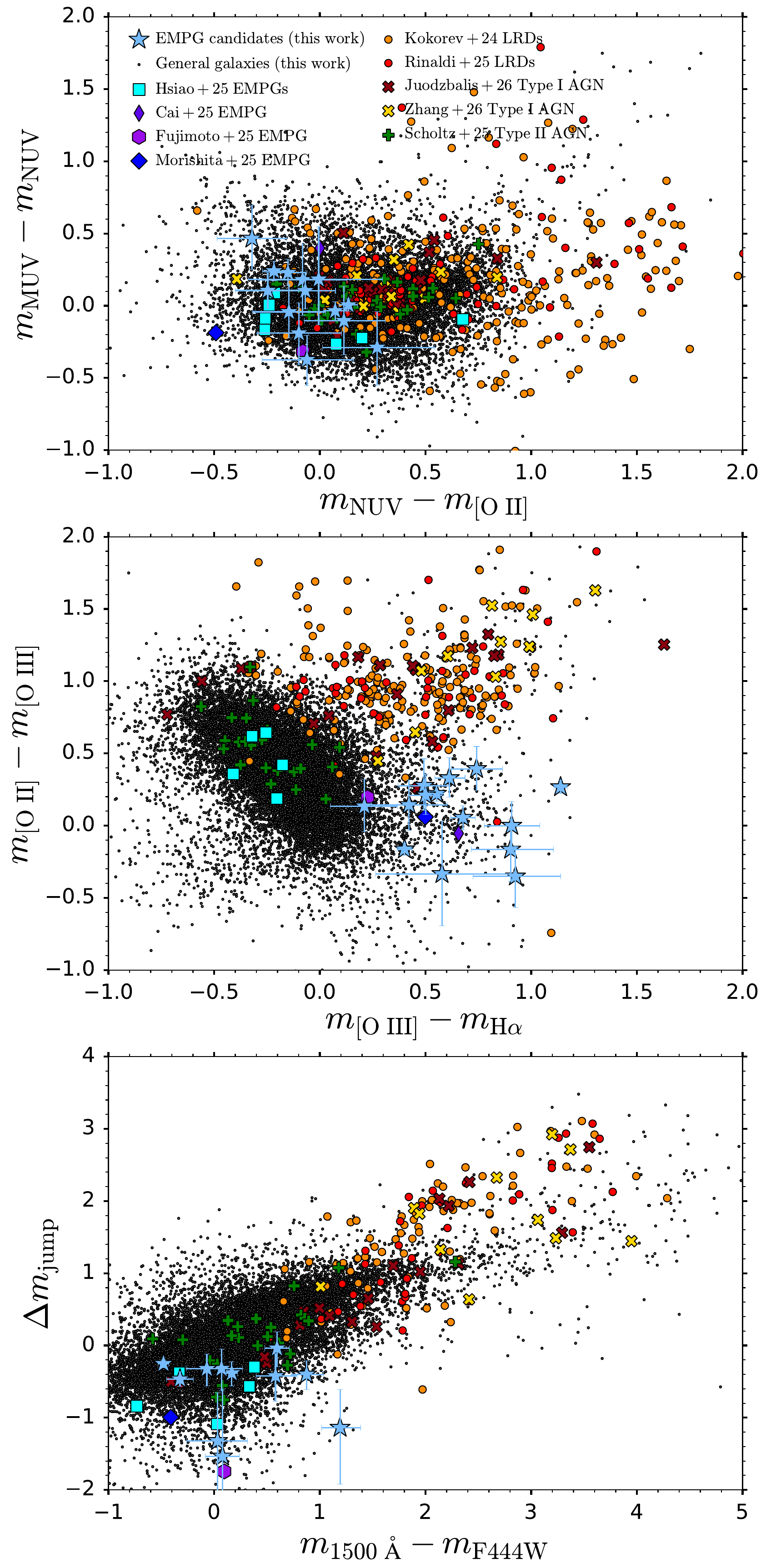}
\caption{Colour--colour diagrams of various rest-frame colours, with $m_\mathrm{H\alpha}$, $m_\mathrm{[O\ III]}$, $m_\mathrm{[O\ II]}$, $m_\mathrm{NUV}$, $m_\mathrm{NUV}$, corresponding to the apparent magnitude in the wide-band filters covering \Ha, \OIII, \OII\ (i.e.\@ the filter preceding \OIII), the `near-ultraviolet' (NUV, preceding \OII) and the `mid-ultraviolet' (MUV, preceding NUV). EMPGs are shown (similar colour coding and symbols to Fig.~\ref{fig:metallicity_redshift}), now also showing the \citet{Fujimoto2025} Pop III candidate (purple hexagon), as well as V-shape LRDs from \citet{Kokorev2024} and \citet{Rinaldi2025} (orange and red circles), broad-line selected Type I AGN from \citet{Juodzbalis2026} and \citet{Zhang2026} (dark red and yellow crosses), and Type II AGN from \citet{Scholtz2025} (green pluses). Top and middle panels: EMPGs have characteristically red $m_\mathrm{[O\ III]} - m_\mathrm{H\alpha}$ colours due to their strong photometric boosts by \Ha, yet correspondingly weak boosts by \OIII. Their otherwise relatively flat wide-band SED (with near-zero colours) sets them apart from LRDs and most broad-line AGN, which generally exhibit red colours due to their Balmer breaks and/or red continuum slopes. Type II AGN typically have colours resembling those of general galaxies (black dots). Thus EMPGs can in principle be identified using wide-band imaging alone, albeit at lower confidence and with looser metallicity constraints than with medium bands included. Bottom panel: Our EMPGs are selected to be young starburst galaxies with blue $\Delta m_\mathrm{jump} < 0$ Balmer jump colours, due to the deficit in rest-frame optical/NIR continuum level (probed by medium bands) relative to the UV. LRDs and Type I AGN typically have red $\Delta m_\mathrm{jump} > 0$ colours, indicative of Balmer breaks and/or red continuum slopes. They further typically exhibit very red $m_{1500\ \text{\AA}} - m_\mathrm{F444W}$ colours, the magnitude difference between the first wide-band filter fully redward of \Lya\ and F444W.}
\label{fig:colours}
\end{figure*}

Throughout this work, we demand the availability of medium-band photometry, as it aids in the reliable identification of EMPG candidates, providing tighter constraints on the photometrically-inferred gas-phase metallicities. The medium bands disentangle the line and continuum emission, allowing the \Ha\ and \OIII\ + \Hb\ equivalent widths to be unambiguously determined, narrowing the possible metallicity range. Furthermore, medium bands can provide independent measurements of line strengths, reinforcing the wide-band evidence for the metal-poor nature of our EMPG candidates. Additionally, Balmer jumps help establish that our EMPG candidates are young starbursts, rather than broad-line AGN/LRDs which are expected to have weaker jumps or even exhibit Balmer breaks. Finally, the emission line photometric boosts and Balmer jump continuum deficit probed by medium bands help to further narrow the photometric redshift range, ensuring our EMPG candidates are not edge cases where the perceived weak \OIII\ emission is due to \OIII\ emission at a wide-band filter edge (with low throughput) or within the NIRCam dichroic between F200W and F277W. Despite these great benefits, medium-band imaging has practical limitations, being generally more limited in footprint compared to wide-band imaging, providing key diagnostic information in narrower redshift intervals, and offering lower continuum sensitivity (though greater emission line sensitivity). Hence medium-band imaging is generally time-expensive, typically probing smaller cosmic volumes and limited to relatively brighter galaxies (if making continuum measurements, especially the Balmer jump) than wide bands. In this section we therefore assess the prospects for reliably identifying EMPGs using wide-band photometry alone, as was previously suggested for Pop III galaxies by \citet{Trussler2023}.

We show various colour--colour diagrams in Fig.~\ref{fig:colours}. We mostly consider rest-frame colours as these can be compared between galaxies across a range of redshifts. $m_\mathrm{H\alpha}$ and $m_\mathrm{[O\ III]}$ correspond to the apparent magnitudes in the wide-band filters covering \Ha\ and \OIII\ respectively, determined based off the source redshift. We further define apparent magnitudes in filters covering \OII, the `near-ultraviolet` NUV and the `mid-ultraviolet' MUV. These correspond to the wide-band filters successively blueward of the wide-band filter covering \OIII. For example, at $z=6$, the \Ha, \OIII, \OII, NUV and MUV filters are F444W, F356W, F277W, F200W, F150W, respectively. Our EMPGs are denoted by blue stars, with the general galaxies satisfying our wide-band SNR cuts (but not the EMPG criterion) shown as grey dots. EMPGs from the literature are shown as before: \citet{Hsiao2025} (cyan squares), \citet{Cai2025} (dark purple diamond), \citet{Morishita2025} (dark blue diamond), now also showing the \citet{Fujimoto2025} Pop III candidate (purple hexagon). Additionally, we also show colour-selected V-shape LRDs from \citet{Kokorev2024} (orange circles) and \citet{Rinaldi2025} (red circles), broad-line selected Type I AGN from \citet{Juodzbalis2026} (dark red crosses) and \citet{Zhang2026} (yellow crosses), and Type II AGN identified using emission line diagnostic diagrams from \citet{Scholtz2025} (green pluses). 

\subsubsection{$m_\mathrm{[O\ III]} - m_\mathrm{H\alpha}$ colour}

As shown in the middle panel of Fig.~\ref{fig:colours}, our EMPG candidates exhibit characteristically red $m_\mathrm{[O\ III]} - m_\mathrm{H\alpha}$ colours (range: 0.21--1.14, median: 0.60), as expected given our selection procedure (Fig.~\ref{fig:Ha_vs_OIII} and Fig.~\ref{fig:ew_ew_selection}). The literature EMPGs also tend to exhibit red $m_\mathrm{[O\ III]} - m_\mathrm{H\alpha}$ colours, with the exception of the slitless-selected EMPG candidates from \citet{Hsiao2025}, which have blue colours more in line with the bulk of the galaxy population. This suggests that either the \citet{Hsiao2025} EMPG candidates are not quite as metal-poor as the rest of the EMPG candidates, or that they possibly have very blue rest-frame optical slopes resulting in blue $m_\mathrm{[O\ III]} - m_\mathrm{H\alpha}$ colours despite the weak \OIII\ emission. We note that red $m_\mathrm{[O\ III]} - m_\mathrm{H\alpha}$ colours are not unique to EMPGs, as LRDs and Type I AGN can exhibit comparably red colours, either due to red rest-frame optical slopes \citep[e.g.\@][]{Trussler2023, Kokorev2024, Kocevski2025} and/or a lack of \OIII\ emission due to collisional deexcitation in BLR-dominated spectra. Notably Type II AGN tend to exhibit comparable colours to the bulk of the galaxy population, as expected given their only subtly different spectra to star-forming galaxies, being located in different regions of line diagnostic diagrams \citep[see e.g.\@][]{ArevaloGonzalez2025, Scholtz2025} which leave little resulting imprint on the wide-band photometry. 

\subsubsection{Further optical and UV colours}

Thus to discriminate between EMPGs and broad-line AGN/LRDs one needs to take into account further information contained within the SED. In particular, we see that EMPGs tend to have flat (i.e. close to zero) $m_\mathrm{[O\ II]} - m_\mathrm{[O\ III]}$ colours, owing to the lack of a prominent photometric boost by \OIII. This is in contrast to LRDs and most Type I AGN, which tend to have red $m_\mathrm{[O\ II]} - m_\mathrm{[O\ III]}$ colours, due to their red rest-frame optical slopes (with $\beta_\mathrm{opt} > -2)$ or \OIII\ contribution from their host galaxies. Thus EMPGs occupy a relatively unique location in the ($m_\mathrm{[O\ II]} - m_\mathrm{[O\ III]}$)--($m_\mathrm{[O\ III]} - m_\mathrm{H\alpha}$) plane, driven by their lack of \OIII\ emission. We note that there are rare exceptions, such as a \citet{Rinaldi2025} LRD (ID 1042541) which we believe has a misclassified redshift (confusing what is actually strong \OIII\ emission at $z \approx 7.4$ with strong \Ha\ emission but weak \OIII\ emission at $z=5.41$), and Type I AGN that are located on the periphery of the EMPG region. 

Taking into account further wide-band colours can increase the purity of the wide-band EMPG identification, removing possible AGN interlopers. As shown in the top panel of Fig.~\ref{fig:colours}, EMPGs have relatively flat $m_\mathrm{NUV} - m_\mathrm{[O\ II]}$ colours, due to the collective photometric boost by the weaker Balmer lines (\Hg, \Hd, \He,  etc.) in the vicinity of \OII\ hiding the continuum deficit associated with the Balmer jump in the \OII\ filter. In contrast, LRDs generally exhibit red $m_\mathrm{NUV} - m_\mathrm{[O\ II]}$ colours, which can range significantly, due to their variety in Balmer break strengths \citep[see e.g.\@][]{deGraaff2025, deGraaff2025b, Naidu2025b}. Type I AGN tend to have flat/red $m_\mathrm{NUV} - m_\mathrm{[O\ II]}$ colours, depending on the steepness of their continuum slopes. Furthermore, EMPGs tend to have flat $m_\mathrm{MUV} - m_\mathrm{NUV}$ colours, indicating $\beta \sim -2$, relatively similar to that of AGN/LRDs.

\subsubsection{Balmer-jump colour}

In the bottom panel of Fig.~\ref{fig:colours} we show the Balmer-jump colour $\Delta m_\mathrm{jump}$. Our EMPG candidates exhibit blue $\Delta m_\mathrm{jump} < 0$ colours indicative of Balmer jumps, as required by our selection procedure. In contrast, LRDs and Type I AGN generally have (very) red $\Delta m_\mathrm{jump} > 0$ colours, due to their strong Balmer breaks and/or red continuum slopes. This further motivates our medium-band-based Balmer-jump selection to remove possible AGN contaminants from our EMPG sample. We do note however that a few Type I AGN appear to exhibit a small Balmer jump. In some cases, from visual inspection, this may be due to source blending between the AGN and an adjacent star-forming galaxy in the small 0.2~arcsec diameter aperture, influencing the measurement. At least in the current Type I AGN sample with tentative Balmer jumps, none exhibit weak \OIII\ emission, so would not be mistaken for an EMPG. We note that such Type I AGN contamination can be a potential concern, as a BLR-dominated spectrum would lack \OIII\ emission (due to collisional deexcitation), just like for our EMPG candidates. Furthermore, some Type II AGN also exhibit Balmer jumps. As their BLR emission is hidden (or at least below the detection limit of the Balmer line spectroscopy), any potential lack of \OIII\ emission would be attributable to extremely low metallicities, in much the same way as EMPGs powered by star formation. Hence such Type II AGN would not be a contaminant concern. Finally, we also display $m_{1500\ \text{\AA}} - m_\mathrm{F444W}$, the colour between the bluest (first filter redward of \Lya) and reddest (F444W) wide-band filter available, highlighting the generally red SEDs associated with Type I AGN/LRDs, in contrast to EMPGs, hence motivating this colour cut in our initial selection procedure.

\subsubsection{Summary}

To summarise, EMPGs exhibit characteristically red $m_\mathrm{[O\ III]} - m_\mathrm{H\alpha}$ colours, due to their strong \Ha\ but relatively weak \OIII\ emission. Their otherwise relatively flat wide-band SED helps to discriminate them from LRDs and most AGN, with medium bands further revealing their Balmer jumps. Hence we assess that EMPGs likely can be identified using wide-band photometry alone \citep[as suggested by][]{Trussler2023}, albeit with looser gas-phase metallicity constraints and at lower confidence due to the lack of additional medium-band constraints on line strengths and photometric redshift.

\subsection{Number statistics} \label{subsec:number}

To put our EMPG candidates in a cosmological context, we now discuss their number statistics, in terms of the EMPG fraction and EMPG comoving number density.

\subsubsection{EMPG fraction}

 We define the EMPG fraction as the number of EMPG candidates within a given magnitude--redshift bin, divided by the total number of galaxies (i.e.\@ satisfying our wide-band SNR cuts, the general high-redshift galaxy sample) within that bin. We adopt absolute UV magnitude $M_\mathrm{UV}$ bins with $\Delta M = 1$ width. We further split galaxies in redshift according to the wide-band filter covering \Ha. These redshift intervals correspond to $2.73 < z \leq 3.45$ for F277W, $3.87 \leq z \leq 4.99$ for F356W, and $4.99 < z \leq 6.52$ for F444W. Note the prominent redshift gap ($3.45 < z \leq 3.87$) between the redshift intervals spanned by F277W and F356W, corresponding to redshifts where \OIII\ falls in the dichroic gap between the F200W and F277W NIRCam filters, and thus where we cannot make a proper photometric assessment of a lack of \OIII\ emission due to extremely low metallicity \citep[though in principle the lack of photometric boost by \SIII\ could still be used to make a weaker inference on a lack of metals in this redshift interval, see][]{Trussler2023}. Thus we avoid this redshift interval in our EMPG number statistics analysis.

\begin{figure}
\centering
\includegraphics[width=\linewidth]{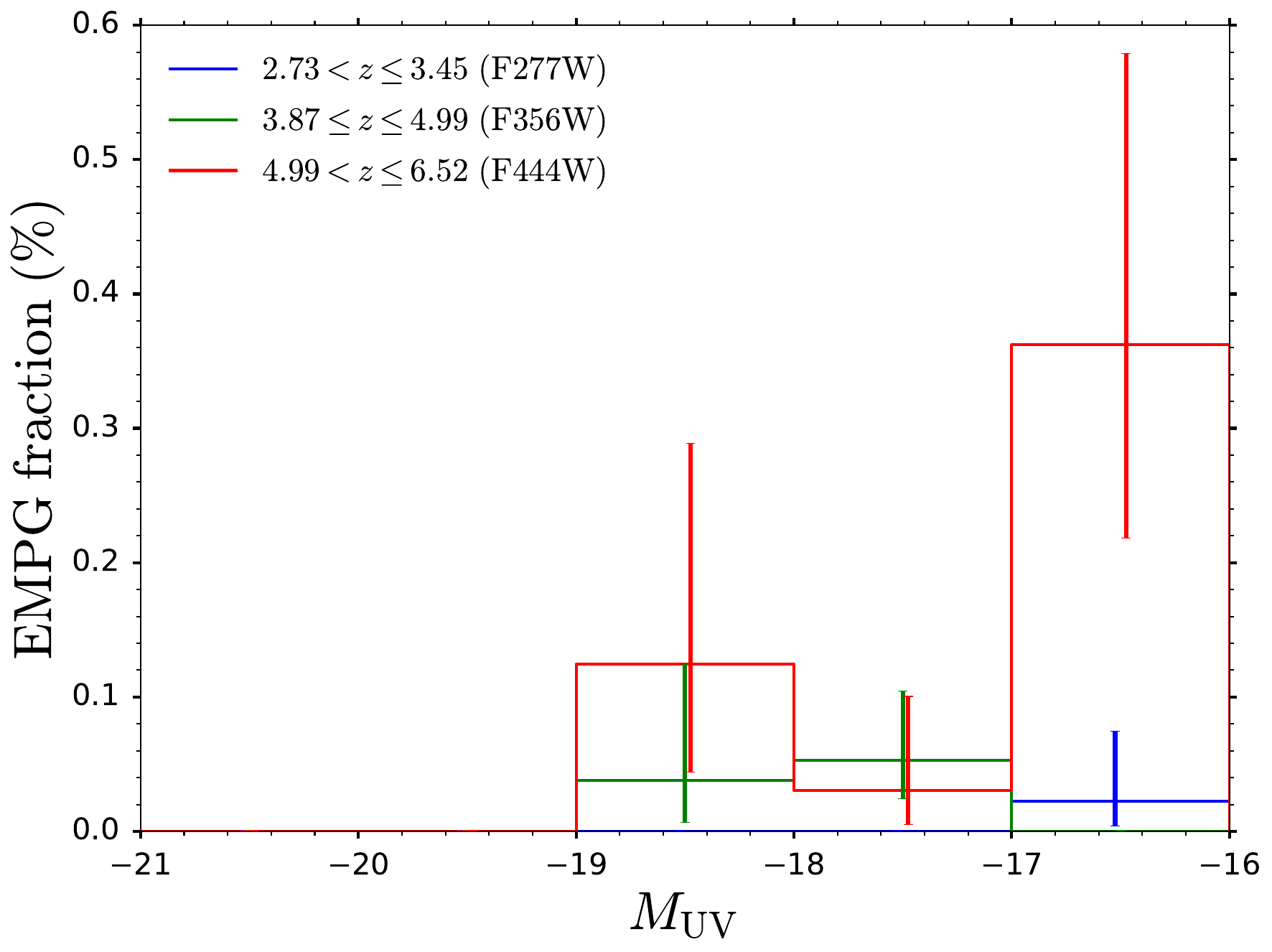}
\caption{The EMPG fraction (displayed as a percentage), defined as the number of EMPGs in a given magnitude--redshift bin, divided by the total number of galaxies within that bin (drawn from our general high-redshift galaxy sample), against absolute rest-frame UV magnitude $M_\mathrm{UV}$ (not completeness corrected). We adopt magnitude bins with $\Delta M = 1$ width, with the redshift bins (blue, green, red) corresponding to the intervals where \Ha\ resides in a given wide-band filter (F277W, F356W, F444W). The EMPG fraction tentatively increases with increasing $M_\mathrm{UV}$, as well as with increasing redshift. This trend may partly be affected by our $\mathrm{EW_{H\alpha}} > 1500$~\AA\ requirement. We anticipate that even deeper imaging (assisted by gravitational lensing) will reveal a high fraction of EMPGs in the ultra-faint regime ($M_\mathrm{UV} > -16$).}
\label{fig:empg_fraction}
\end{figure}

We show the EMPG fractions (as a percentage) in Fig.~\ref{fig:empg_fraction} for the redshift intervals tracing \Ha\ via F277W (blue), F356W (green) and F444W (red). The error bars on the EMPG fractions are given by propagating the Poisson errors on the number counts in quadrature. We see that even when pushing to very faint galaxies with $M_\mathrm{UV} = -16$ via the deep JADES blank-field imaging, EMPG galaxies are still very rare, constituting ${\sim}$0.02--0.4\% of galaxies, not corrected for completeness. From our limited EMPG statistics (14 galaxies in total), it appears that the EMPG fraction tends to rise with increasing $M_\mathrm{UV}$ (i.e.\@ as galaxies get fainter in the UV) and increasing redshift. This trend may be expected if the gas-phase metallicities of low-mass galaxies continue to decrease with decreasing stellar mass (thus becoming fainter), and if the mass--metallicity relation continues to decrease in normalisation and/or increase in metallicity scatter towards higher redshift. However, it should be noted that we further require our EMPG candidates to exhibit strong \Ha\ equivalent widths, so that the photometric boost by emission lines (or lack thereof) is clearly evident in the data. It is possible that the tentatively rising EMPG fractions we see may partly also be affected by burstiness trends \citep[if applicable, see e.g.\@][]{Boyett2024, Cole2025, Endsley2025, Simmonds2025}, rather than tracing metallicity trends alone. Nevertheless, our results suggest that deep imaging on cluster fields may reveal a high fraction of high-redshift EMPGs in the ultra-faint regime ($M_\mathrm{UV} > -16$), brought into view via strong gravitational lensing.

\begin{figure}
\centering
\includegraphics[width=\linewidth]{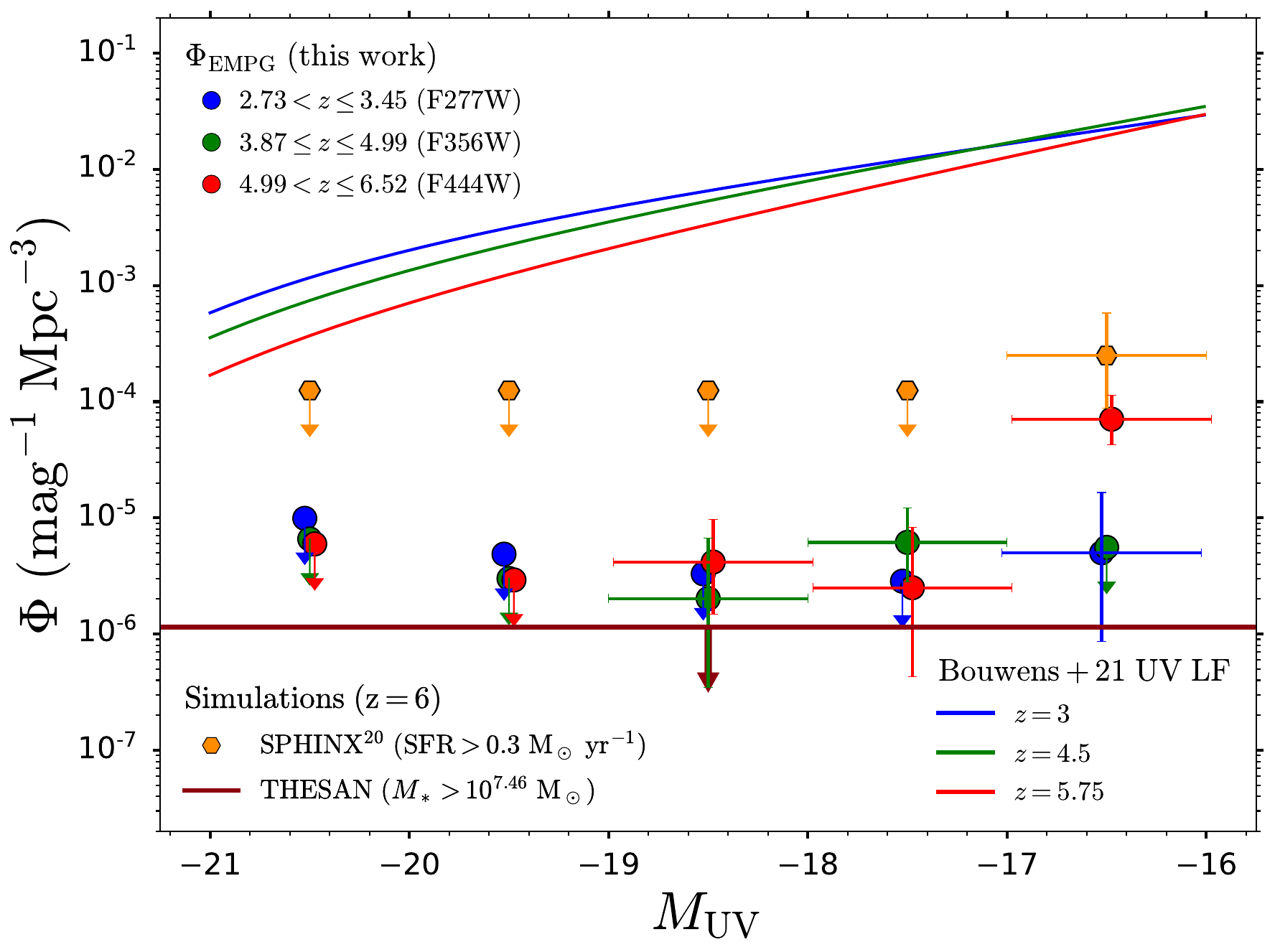}
\caption{Comoving number densities per unit magnitude interval against $M_\mathrm{UV}$. The observed EMPG comoving number densities $\Phi_\mathrm{EMPG}$ (blue, green, red datapoints) are crudely estimated by scaling the \citet{Bouwens2021} UV luminosity function (at $z=3,\ 4.5,\ 5.75$, respectively) by the EMPG fractions denoted in Fig.~\ref{fig:empg_fraction}. Downward arrows are where the EMPG fraction is zero, with comoving density upper limits estimated by assuming the fraction is at most $1/N_\mathrm{gal}$, where $N_\mathrm{gal}$ is the number of galaxies within that magnitude--redshift bin. Also shown are EMPG comoving number densities at $z=6$ from the SPHINX$^{20}$ ($20^3$~cMpc$^3$, orange) and THESAN ($95.5^3$~cMpc$^3$, dark red) simulations, with EMPGs defined by $Z_\mathrm{gas} < 0.01~\mathrm{Z}_\odot$. Note that these $\Phi_\mathrm{EMPG}$ estimates are both incomplete, owing to the minimum required SFR and stellar mass $M_*$, respectively. There are no EMPGs at all in THESAN, with extremely metal-poor star formation perhaps restricted to lower mass, fainter galaxies, as EMPGs start to emerge at the faint end of the smaller volume, finer mass resolution SPHINX$^{20}$ simulation.}
\label{fig:empg_number_density}
\end{figure}

\subsubsection{EMPG comoving number density}

We further consider the EMPG comoving number densities. We preface this by noting that our EMPG selection procedure is complex, designed so that we can be as confident as possible in the extremely metal-poor nature of our EMPG candidates, while also allowing for flexibility in the selection thresholds to account for uncertainty in the data. This selection therefore does not facilitate a clear one-to-one comparison with simulation predictions on EMPG statistics. Bearing this in mind, we perform a crude estimate of the EMPG comoving number density by multiplying the \citet{Bouwens2021} UV luminosity function by the EMPG fractions shown in Fig.~\ref{fig:empg_fraction}. We use the same magnitude--redshift bins as before, scaling the \citet{Bouwens2021} UV luminosity function evaluated at the magnitude bin centre, and at redshifts $z=3,\ 4.5,\ 5.75$, roughly corresponding to the midpoints of the redshift intervals spanned by the F277W, F356W, F444W filters. In the case of zero EMPG candidates within a given magnitude--redshift bin, i.e.\@ an EMPG fraction of zero, we estimate upper limits on the comoving number density by adopting an upper limit on the EMPG fraction given by $1/N_\mathrm{gal}$, where $N_\mathrm{gal}$ is the number of galaxies within that magnitude--redshift bin. We compare our observational results against the THESAN \citep{Garaldi2022, Garaldi2024, Kannan2022, Kannan2022b, Smith2022} and SPHINX$^{20}$ \citep{Rosdahl2018, Rosdahl2022, Katz2023b} simulations, corresponding to large ($95.5^3$~cMpc$^3$) and smaller cosmological boxes ($20^3$~cMpc$^3$, but with finer mass resolution) respectively. Owing to the finite mass resolution, we restrict our comparison to `resolved' (i.e.\@ with $\geq 50$ star particles) THESAN galaxies which have $M_* > 10^{7.46}~\mathrm{M}_\odot$. Furthermore, our comparison to SPHINX$^{20}$ is limited to simulated galaxies with SFR $> 0.3~\mathrm{M}_\odot~\mathrm{yr}^{-1}$ as these are provided in the public data release. Hence both our observed and the simulated EMPG comoving number densities (as a function of $M_\mathrm{UV}$) will be incomplete. Similar to the observations, we define EMPGs in the simulations as those having $Z_\mathrm{gas} < 0.01~\mathrm{Z}_\odot$.

We show our observed EMPG comoving number densities in Fig.~\ref{fig:empg_number_density} for the \Ha\ redshift intervals traced by F277W (blue), F356W (green) and F444W (red), together with the \citet{Bouwens2021} UV luminosity functions from which they are scaled (solid lines). Mirroring the rising EMPG fraction, the EMPG comoving number density is generally low (${\sim}10^{-5}$~cMpc$^{-3}$), but increases to ${\sim}10^{-4}$~cMpc$^{-3}$ at the faintest UV magnitude ($-17 < M_\mathrm{UV} < -16$) and highest redshift bin ($4.99 < z \leq 6.52$) accessible by the JADES NIRCam imaging. Comparing to the simulations at $z=6$, we see that there are no EMPGs at all in THESAN, with the dark red downward arrow indicating the upper limit on the EMPG comoving number density (${\sim}10^{-6}$~cMpc$^{-3}$), given the box volume. This complete lack of EMPGs likely stems from the coarser mass resolution in this large box simulation, restricted to $M_* > 10^{7.46}~\mathrm{M}_\odot$. Thus even when spanning large cosmic volumes, extremely metal-poor star formation may perhaps be restricted solely to the lower mass, fainter galaxies. Indeed, EMPGs start to emerge at the faint end of the much smaller volume (but finer mass resolution) SPHINX$^{20}$ simulation (orange), with comparable comoving number densities (${\sim}2\times10^{-4}$~cMpc$^{-3}$) in the $-17 < M_\mathrm{UV} < -16$ bin to our observational results (though both measurements are hampered by incompleteness). The lack of EMPG galaxies in brighter $M_\mathrm{UV}$ bins (shown as upper limits) likely indicates the need for larger simulation boxes with comparable mass resolution and physics modelling to SPHINX$^{20}$ to uncover the rarer EMPG candidates discovered in the large cosmic volume probed by JADES. 

Our observational results combined with the simulation predictions suggest that EMPGs may become increasingly common provided one pushes to sufficiently faint UV magnitudes and low stellar masses, only accessible observationally by strong gravitational lensing from galaxy cluster fields. Indeed, if the mass--metallicity relation continues to decline with decreasing stellar mass, then the EMPG fraction continues to rise with decreasing stellar mass. Combined with the increasing comoving number density of galaxies at lower stellar masses, the EMPG comoving number density would be expected to strongly rise as one progressively pushes further into the dwarf galaxy regime.

\subsection{Observability} \label{subsec:observability}

Finally, we close by discussing the observability of EMPGs with \emph{JWST}, as inferred from the imaging of our EMPG candidates, also providing forecasts for future EMPG spectroscopy. 

\subsubsection{Imaging}

\begin{figure}
\centering
\includegraphics[width=\linewidth]{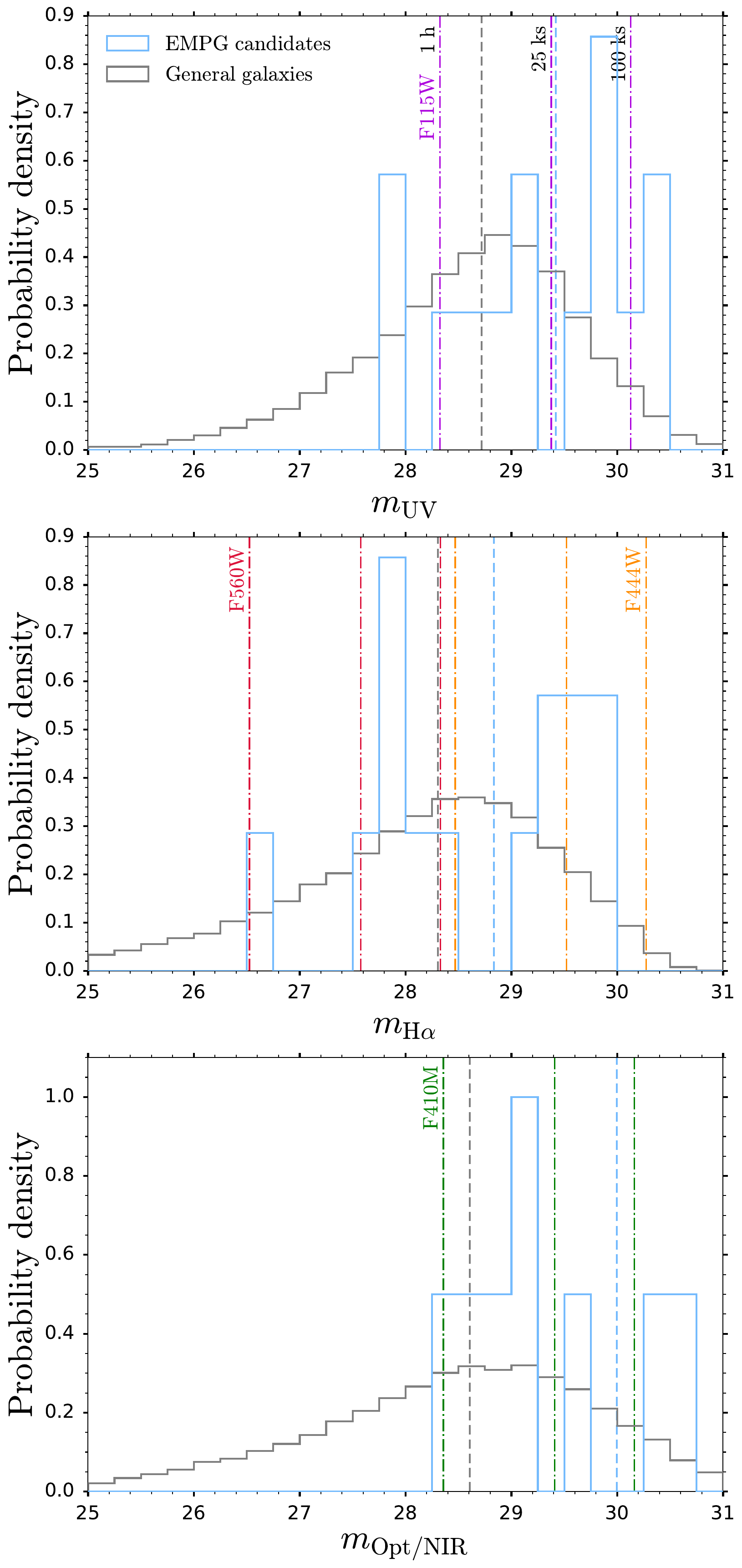}
\caption{Apparent magnitude distributions for EMPGs (light blue) and our general high-redshift galaxy sample (grey), showing median values (dashed lines) and $5\sigma$ imaging sensitivity limits \citep[0.2~arcsec diameter circular apertures for NIRCam,][]{Eisenstein2026} in 1~h (medium depth), 25~ks (= 7~h, deep) and 100~ks (= 28~h, ultra-deep) integrations (dot-dashed lines). Top panel: Apparent magnitude distribution $m_\mathrm{UV}$ in the first wide-band filter fully redward of \Lya. With median $m_\mathrm{UV} = 29.4$, EMPGs are typically very faint, accessible in deep 8~h integrations. Middle panel: Wide-band filter covering \Ha. While EMPGs (with median $m_\mathrm{H\alpha} = 28.8$) are readily detectable with NIRCam (${\sim}2$~h), MIRI detections at $z > 6.6$ will be challenging, owing to the $\approx 2$~mag lower sensitivity \citep{Alberts2024} and $\approx 4\times$ smaller field of view, thus requiring targeted observations. Bottom panel: Medium-band measurements of the rest-frame optical/NIR continuum, with median $m_\mathrm{Opt/NIR} = 30.0$, typically require ultra-deep $20$~h integrations.}
\label{fig:apparent_magnitude}
\end{figure}

We show the apparent magnitude (in 0.2~arcsec diameter apertures, with point source aperture corrections applied) distribution $m_\mathrm{UV}$ in the first wide-band filter fully redward of \Lya\ in the top panel of Fig.~\ref{fig:apparent_magnitude} for our EMPG candidates (light blue), as well as for the general high-redshift galaxy sample (i.e.\@ galaxies satisfying our wide-band SNR cuts, grey). Purple vertical dot-dashed lines indicate the $5\sigma$ depth achieved in 0.2~arcsec diameter apertures with F115W in 1 h (representing a medium-depth programme), 25 ks (= 7 h, deep) and 100 ks (= 28 h, ultra-deep), with flux density $f_\nu$ depths scaled by $1/\sqrt{t}$ from the JADES depths reported in \citet{Eisenstein2026}. As discussed earlier, EMPGs are generally very faint, with median $m_\mathrm{UV} = 29.4$ (indicated by the vertical dashed light blue line), accessible in deep 8~h integrations. However, there is a substantial spread in UV brightness, with the bright tail detectable in medium-depth imaging, while the currently observed faint tail requires ultra-deep observations. We anticipate that progressively deeper imaging will reveal progressively more EMPGs in the ultra-faint regime.

\begin{figure*}
\centering
\includegraphics[width=\linewidth]{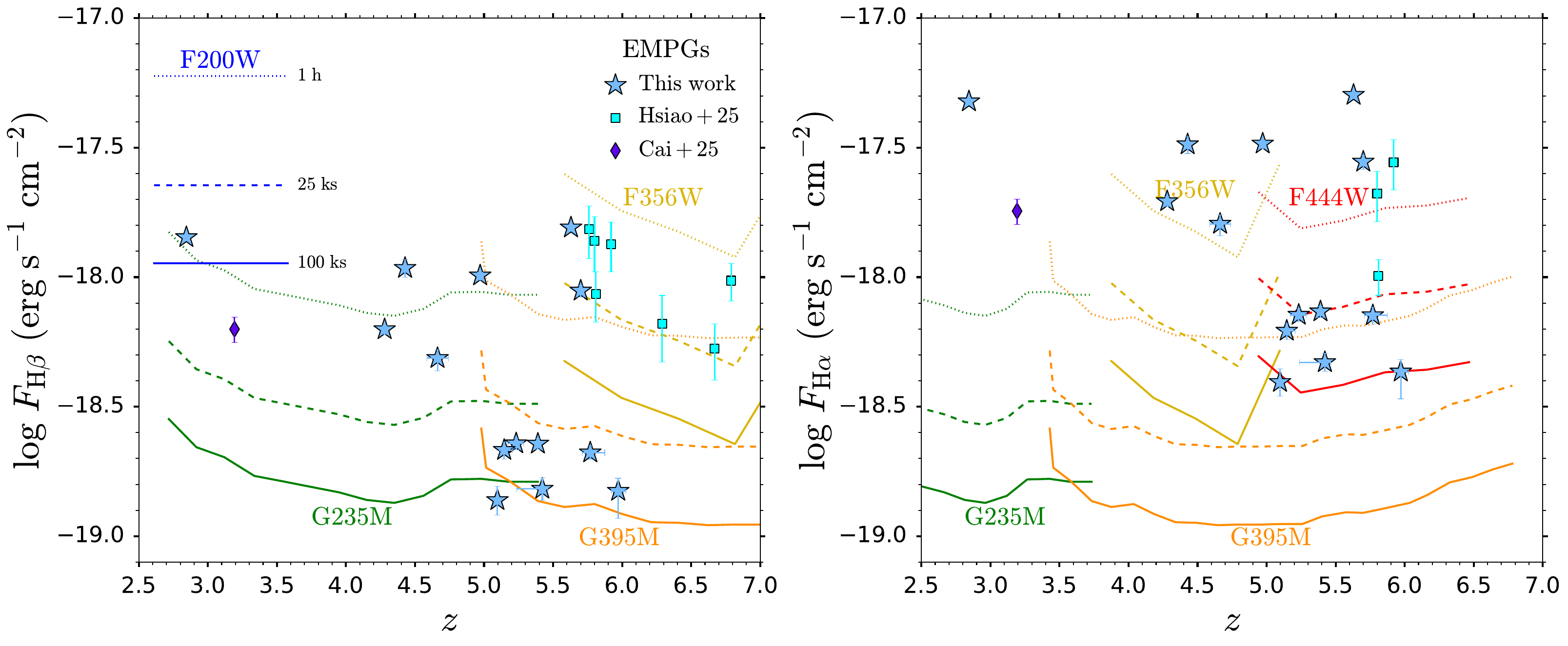}
\caption{Spectroscopic observability of EMPGs with \emph{JWST}. Line fluxes of EMPGs are displayed (symbols), together with $5\sigma$ line sensitivities for NIRISS slitless F200W \citep[blue, from NGDEEP,][]{Pirzkal2024}, NIRCam slitless F356W and F444W \citep[gold and red, from SAPPHIRES-EDR,][]{Sun2025}, and NIRSpec G235M and G395M \citep[green and orange, from JADES,][]{Eisenstein2026}, with 1~h (dotted), 25~ks (= 7~h, dashed) and 100~ks (= 28~h, solid) integrations. Left panel: Ultra-deep NIRCam F356W slitless spectroscopy can $5\sigma$ detect \Hb\ in the brightest EMPGs. The sensitivity of the NIRSpec gratings is needed to detect the faintest EMPGs (without gravitational lensing), also placing the strongest upper limits on the gas-phase metallicity in the case of \OIII\ $\lambda 5007$ non-detections. Right panel: Ultra-deep NIRCam slitless spectroscopy can deliver spectroscopic redshifts via $5\sigma$ \Ha\ detections for even the faintest EMPGs, providing valuable metallicity constraints (analogous to medium-band imaging) by combining \Ha\ line flux measurements with wide-band photometry.}
\label{fig:flux_redshift}
\end{figure*}

We show the apparent magnitude distribution $m_\mathrm{H\alpha}$ in the wide-band filter covering \Ha\ in the middle panel of Fig.~\ref{fig:apparent_magnitude}. Owing to the strong photometric boosts by \Ha\ (by selection), our EMPGs are typically substantially brighter in this band with median $m_\mathrm{H\alpha} = 28.8$, generally detectable in only ${\sim}2$~h integration time with NIRCam F444W (orange). However, extending the photometric search for EMPG candidates based off their strong \Ha, but weak \OIII\ emission will be challenging above $z > 6.6$, where \Ha\ gets redshifted into the MIRI regime. The much lower MIRI F560W sensitivity \citep[$\approx$ 2 mag, red lines, scaled from][]{Alberts2024} demands $\approx 40\times$ longer integration times compared to NIRCam F444W. Coupled with the fainter source magnitudes due to the greater redshift (decreasing in brightness by $\approx 0.5$~mag from $z=5.5$ to $z=7.5$), as well as the $\approx 4\times$ smaller MIRI imaging field of view compared to NIRCam, \Ha\ detections of EMPGs will be limited, unless MIRI pointings are optimised. This can be achieved by targeting $z > 6.6$ galaxies that are either highly magnified (greatly reducing integration time) and/or promising EMPG candidates with seemingly weak \OIII\ emission, indicated by flat $m_\mathrm{[O\ II]} - m_\mathrm{[O\ III]}$ colours (see Fig.~\ref{fig:colours}) together with a Balmer jump/photometric excess from medium bands \citep[ruling out the mini-quenched/napping/smouldering scenario,][]{Looser2024, Looser2025, Cole2025, Trussler2025b}.

We show the apparent magnitude distribution $m_\mathrm{Opt/NIR}$ in the medium-band measurement of the rest-frame optical or NIR continuum level in the bottom panel of Fig.~\ref{fig:apparent_magnitude}. We prioritise showing the rest-frame optical continuum measurement if available, otherwise the rest-frame NIR measurement is shown. Owing to their Balmer jumps (by selection), EMPGs are faintest in this filter, with median $m_\mathrm{Opt/NIR} = 30.0$. Coupled with the lower continuum sensitivity of medium bands (due to spanning a narrower spectral range), it is particularly challenging to detect EMPGs in this way, typically requiring $20$~h integrations, assuming the F410M sensitivity (green). Hence the rest-frame optical/NIR medium-band continuum measurement represents the limiting factor in our EMPG selection. This demanding exposure time can be alleviated by removing the $5\sigma$ detection requirement (as we have done). Due to the deep JADES medium-band imaging, we are able to achieve reliable measurements ($> 3\sigma$) down to very faint magnitudes. Alternatively, one can rely on the increased emission line sensitivity of medium bands (due also to their narrower spectral range) to push to fainter systems, using the joint photometric excesses by \Ha\ in medium--wide filter pairs (e.g. F410M and F444W at $z = 5$) to place coarser constraints on the rest-frame optical continuum level \citep[as we have done, see also][]{Trussler2026}.

\subsubsection{Spectroscopy}

We further show forecasts for the spectroscopic observability of EMPGs in Fig.~\ref{fig:flux_redshift}. The \Hb\ line fluxes for our EMPG candidates estimated using \textsc{Bagpipes} are shown in the left panel. We also show $5\sigma$ line sensitivities in 1~h (dotted), 25~ks (= 7~h, dashed) and 100~ks (= 28~h, solid) integrations for various spectroscopic observing modes with \emph{JWST}, all scaled by $1/\sqrt{t}$ from the reported sensitivities for NIRISS F200W wide field slitless spectroscopy (WFSS) \citep[blue, from NGDEEP,][]{Pirzkal2024}, NIRCam F356W and F444W WFSS \citep[yellow and red, from SAPPHIRES,][]{Sun2025} and NIRSpec/G235M and G395M \citep[green and orange, from JADES,][]{Eisenstein2026}.

We first consider the prospects of detecting EMPGs through blind slitless spectroscopy. Deep (7~h) to ultra-deep (28~h) NIRCam/F356W spectroscopy is capable of detecting \Hb\ for the brightest EMPGs, as was done by \citet{Hsiao2025} (blue cyan squares) with 9.4~h SAPPHIRES-EDR data. The ratio $R_3 = \mathrm{[O\ III]}\ \lambda 5007 / \mathrm{H}\beta$ of the \OIII\ $\lambda 5007$ and \Hb\ line fluxes can then be used together with strong-line calibrations to estimate the gas-phase metallicity of the EMPG. In the case of \OIII\ non-detections, upper limits on the gas-phase metallicity can be placed. However, if one wishes to detect \Hb\ in the fainter EMPGs, or placer tighter upper limits on the gas-phase metallicity requiring ${\gg}5\sigma$ \Hb\ detections, then this is not achievable via slitless spectroscopy without strong gravitational lensing \citep[see the highly magnified pristine source from][]{Morishita2025}.

Now, ultra-deep slitless spectroscopy will be capable of $5\sigma$ detections of \Ha\ in even the faintest EMPG candidates (right panel). If paired with NIRCam imaging, the spectroscopic redshifts will rule out edge cases where the apparent lack of \OIII\ emission in the photometry could have been attributable to \OIII\ falling in the NIRCam dichroic gap or on the edge of a wide-band filter, thus aiding in the reliable identification of EMPGs. Additionally, the \Ha\ line flux measurement will, similar to medium-band imaging, help disentangle emission line and continuum contributions to wide-band photometry, thus establishing the rest-frame optical continuum level and \Ha\ equivalent width, tightening the constraints on the metallicity of the EMPG candidate.

If viable EMPG candidates have already been identified from medium-band selections, wide-band-only selections or even shallower slitless spectroscopy, then the most efficient way to confirm them would be via the enhanced emission line sensitivity of NIRSpec grating spectroscopy\footnote{The NIRSpec PRISM offers substantially (${\sim}2.5\times)$ to slightly (${\sim}1.1\times$) less line sensitivity \citep[depending on wavelength, see][]{Eisenstein2026} with possible blending of the \OIII\ and \Hb\ lines \citep{Chevallard2019}, though would provide valuable additional insights on the nebular conditions and ionising spectrum via continuum measurements of the Balmer jump and possibly two-photon continuum, as well as from rest-frame ultraviolet emission lines (at higher redshifts).}. Ultra-deep (28~h) integrations would be sufficient to $\approx5\sigma$ detect \Hb\ in all of our EMPG candidates. Assuming a \OIII\ $\lambda 5007$ non-detection, so assigning a $2\sigma$ line flux upper limit, the resulting upper limit on the gas-phase metallicity for $R_3 = 2/5$ is $Z \lesssim 0.4\%~\mathrm{Z}_\odot$, i.e.\@ $12 + \log (\mathrm{O/H}) \lesssim 6.3$ \citep{Sanders2024}.

One final spectroscopic option is to combine the blind statistics of slitless spectroscopy with the sensitivity of the NIRSpec gratings. This is dense-shutter spectroscopy \citep{D'Eugenio2026}, an ambitious and highly efficient use of NIRSpec that maximises the number of open shutters on the NIRSpec MSA used to target galaxies, resulting in massively-multiplexed, highly-complete spectroscopy. This allows EMPGs to be blindly targeted without a photometric pre-selection, perhaps most useful in the case of wide-band-only imaging where it can be more difficult to identify secure EMPG candidates, also offering a simple selection function with which to compare the observed EMPG statistics against simulations. Moreover, this mode bypasses the need for expensive MIRI imaging to identify EMPG candidates at $z > 6.6$, offering an alternate route to discovering the faintest pristine galaxies all the way out to $z=10$ (before \OIII\ is redshifted out of the G395M grating at 5.5~\textmu m).

\section{Conclusions} \label{sec:conclusions}

We photometrically identify 14 extremely metal-poor galaxy candidates with $Z < 1\%~\mathrm{Z}_\odot$ at $2.5 < z < 6.5$, made possible by the extensive deep JADES NIRCam medium-band imaging. EMPGs are identified based off strong photometric boosts by \Ha, yet correspondingly weak boosts by \OIII\ + \Hb, likely requiring extremely low metallicity to explain the lack of \OIII\ emission, resulting in characteristically red $m_\mathrm{[O\ III]} - m_\mathrm{H\alpha}$ colours in the wide-band filters covering \OIII\ and \Ha. Medium bands greatly aid in the confident identification of EMPG candidates, also yielding tighter constraints on the inferred gas-phase metallicities. The medium bands disentangle the line and continuum emission, provide constraints on the rest-frame optical/NIR continuum level, establish the presence of Balmer jumps, and tighten the photometric redshift constraints. 

We utilise the medium-band optical/NIR continuum measurements to empirically determine emission line equivalent widths from photometric boosts in wide-band filters. Preliminary EMPG candidates are identified based off low \OIII\ + \Hb\ equivalent widths relative to \Ha. This represents an extension from the wide-band Pop III selection of \citet{Trussler2023}, which was restricted to the youngest pristine starbursts, to older non-pristine EMPG starbursts. In cases where the empirical estimates of line equivalent widths are deemed unreliable, we select EMPG candidates based off their red $m_\mathrm{[O\ III]} - m_\mathrm{H\alpha}$ wide-band colours. We further require our EMPG candidates to exhibit strong Balmer jumps, likely indicating that they are young starbursts, rather than LRDs which have Balmer breaks or broad-line AGN which tend to have redder UV--optical colours due to their red continuum slopes. We then establish the most secure EMPG candidates by fitting their full NIRCam + HST/ACS photometry with \textsc{Bagpipes}, utilising low metallicity ($Z_\mathrm{star} < 0.03~\mathrm{Z}_\odot)$, intermediate metallicity ($0.03 < Z_\mathrm{star}/\mathrm{Z}_\odot < 0.20$), and in some cases, edge-case models (to account for \OIII\ falling in the NIRCam dichroic gap between F200W and F277W). Our final 14 EMPG candidates are inferred to have close-to-pristine metallicities (${\sim}0.001\textrm{--}0.01~\mathrm{Z}_\odot$), strongly favouring the low metallicity fit ($\Delta \chi^2 > 9$) over the intermediate metallicity and edge-case models. However, motivated by the NIRSpec PRISM spectrum of EMPG candidate ID 123650, which exhibits a substantial Balmer decrement $\mathrm{H}\alpha/\mathrm{H}\beta > 5.33$, we discuss how dusty starbursts and starbursts radiating into dense gas remain a contaminant concern, as well as extremely dust-obscured AGN for candidate ID 1008821.  

From SED-fitting with close-to-pristine models, we infer our EMPG candidates to be very low-mass dwarfs (median $M_* = 10^{6.7}~\mathrm{M}_\odot$), constituting the extreme low metallicity tail of the mass--metallicity relation. The EMPGs are starburst galaxies, characterised by strong Balmer jumps and high \Ha\ equivalent widths, placed well above the star-forming main sequence. They are $> 1$~dex more metal-poor than would be expected given their stellar mass and star formation rate, being near-pristine young starburst galaxies, representing an extreme departure from the slowly-evolving equilibrium framework of the fundamental metallicity relation. The EMPG candidates are generally very UV-faint, with median $M_\mathrm{UV} = -16.9$, brought into view thanks to the exceptionally deep JADES imaging, though the rare, brighter tail would be detectable with shallower imaging from wider programmes. The EMPG candidates are extremely efficient producers of ionising photons, with median $\log\, (\xi_\mathrm{ion, obs} /\mathrm{(Hz\ erg^{-1})}) = 26.0$. To power this extreme emission, they possibly harbour top-heavy stellar populations, their exceptionally high $\xi_\mathrm{ion, obs}$ compatible with being nebular-dominated, indicating a possibly dominant nebular continuum in the rest-frame UV. Despite their UV faintness, the EMPG candidates have intermediate \Ha\ luminosities (median $L_\mathrm{H\alpha} = 10^{41.5}$~erg~s$^{-1}$), favouring their detection in deep \emph{JWST} spectroscopy. EMPGs are currently rare even in the deep JADES imaging, constituting ${\sim}0.02\textrm{--}0.4\%$ of $2.5 < z < 6.5$ galaxies at $-19 < M_\mathrm{UV} < -16$. We anticipate that even deeper imaging (likely boosted by strong gravitational lensing) will reveal progressively more EMPG candidates.

Owing to their faintness, it is challenging to detect the optical/NIR continua of EMPGs with medium bands. Hence we explore the prospects of identifying EMPG candidates using wide-band imaging only. In principle, EMPGs can be selected based off their red $m_\mathrm{[O\ III]} - m_\mathrm{H\alpha}$ colours and otherwise relatively flat SEDs. This sets them apart from contaminant LRDs and most BLR AGN which generally exhibit redder SEDs. However, such wide-band-only EMPG selections will likely come at the cost of lower confidence (smaller $\Delta \chi^2$ with standard models), as well as looser constraints on the inferred gas-phase metallicity.

The photometric identification of EMPGs based off their strong \Ha\ but weak \OIII\ emission will be difficult beyond $z > 6.6$. Due to the lower MIRI imaging sensitivity (due to the higher background), exposure times will be extraordinarily long (${\sim}40\times$ longer than NIRCam), not to mention the fainter source brightness due to the greater redshift and the $4\times$ smaller MIRI field of view. Hence EMPG detections will be limited, unless MIRI pointings are optimised. This can be achieved by targeting highly-magnified galaxies and/or promising EMPG candidates which appear to exhibit weak \OIII\ emission through flat $m_\mathrm{[O\ II]} - m_\mathrm{[O\ III]}$ colours together with a Balmer jump / line boost in medium bands. 

From the line luminosities inferred for our EMPG candidates, we forecast future spectroscopic prospects with \emph{JWST}. Deep (${\sim}28$~h) NIRCam slitless spectroscopy can blindly identify the brightest EMPGs through strong \Hb\ but weak \OIII\ emission. It can also yield spectroscopic redshifts for the faintest EMPGs through \Ha\ detections. Efficient follow-up of EMPG candidates demands the line sensitivity of NIRSpec grating spectroscopy, being able to deliver \Hb\ detections for the faintest EMPGs, also providing the most stringent upper limits on gas-phase metallicities. Massively-multiplexed, highly-complete dense-shutter spectroscopy with the NIRSpec MSA provides an alternative route to discovering the faintest pristine galaxies out to $z=10$, without requiring deep existing medium-band / MIRI imaging to identify secure candidates.

Our 14 EMPG candidates constitute a significant increase in the number of known EMPGs, possibly also pushing the metallicity frontier to even more pristine systems. We deem it unlikely that their weak \OIII\ emission is attributable to extremely high metallicities and/or extremely low ionisation parameters. However, despite our best selection efforts, it is plausible that the \OIII\ emission is suppressed due to collisional deexcitation in a BLR-dominated spectrum and/or emission originating from very dense gas powered by star formation. In the former scenario (unlikely as our candidates 
exhibit extremely high $\mathrm{EW_{H\alpha}} > 1500$~\AA), high SNR spectroscopy could confirm/refute the presence of broad \Ha\ emission. In the latter scenario \citep[plausible as evidence is building for dense gas in \ion{H}{II} regions at high redshift, e.g.\@][]{Topping2024, Topping2025, Caputi2026}, deep rest-frame UV spectroscopy could establish the presence/absence of metal lines (e.g.\@ \ion{O}{III}] and \ion{C}{IV}) with greater critical densities. If genuinely near-pristine, we still do not anticipate our EMPG candidates to be metal-free Pop III galaxies, rather that their metallicity is sufficiently low that their \OIII\ emission is below our current detection threshold. Future deep NIRSpec grating spectroscopy could confirm the lack of \OIII\ emission, tightening the upper limits on the gas-phase metallicity. Furthermore, deep rest-frame UV spectroscopy could reveal a two-photon continuum downturn, non-negligible \ion{He}{II} $\lambda 1640$ emission and distinct metal--metal line ratios (if detectable), providing unique insights on the near-pristine IMF and primordial nucleosynthesis pattern, bringing us closer to the nature of the first stars than ever before. 

\section*{Acknowledgements}
We thank the referee for their useful comments which helped to improve this article. JAAT thanks Lisa Kewley for helpful discussions that aided in the interpretation of the results. JAAT acknowledges support from the Simons Foundation and \emph{JWST} programs 2959 and 3215. Support for programs 2959 and 3215 was provided by NASA through a grant from the Space Telescope Science Institute, which is operated by the Association of Universities for Research in Astronomy, Inc., under NASA contract NAS 5-03127. DJE is supported as a Simons Investigator and by JWST/NIRCam contract to the University of Arizona, NAS5-02105. AJB and JC acknowledge funding from the "FirstGalaxies" Advanced Grant from the European Research Council (ERC) under the European Union’s Horizon 2020 research and innovation programme (Grant agreement No.\@ 789056). AJC and JW gratefully acknowledge support from the Cosmic Dawn Center through the DAWN Fellowship. The Cosmic Dawn Center (DAWN) is funded by the Danish National Research Foundation under grant No. 140. SC acknowledges support by European Union’s HE ERC Starting Grant No. 101040227 - WINGS.
CJC acknowledges support from the ERC Advanced Investigator Grant EPOCHS (788113). ECL acknowledges support of an STFC Webb Fellowship (ST/W001438/1). FDE and RM acknowledge support by the Science and Technology Facilities Council (STFC), by the ERC through Advanced Grant 695671 ``QUENCH'', and by the UKRI Frontier Research grant RISEandFALL. EE, JMH, ZJ, BDJ, BER, FS and CNAW acknowledge support from \emph{JWST}/NIRCam contract to the University of Arizona, NAS5-02105. JMH also acknowledges support from JWST Programs 3215 and 8544. TJL gratefully acknowledges support from the Swiss National Science Foundation through a Postdoc Mobility Fellowship and from the \emph{JWST} Program 5997. RM also acknowledges funding from a research professorship from the Royal Society. BER also acknowledges support JWST Program 3215. FS also acknowledges support for \emph{JWST} program \#2883, 4924, 5105, 6434 provided by NASA through grants from the Space Telescope Science Institute, which is operated by the Association of Universities for Research in Astronomy, Inc., under NASA contract NAS 5-03127. ST acknowledges support by the Royal Society Research Grant G125142. H\"U acknowledges support by the Max Planck Society through the Lise Meitner Excellence Program. H\"U acknowledges funding by the European Union (ERC APEX, 101164796). Views and opinions expressed are however those of the authors only and do not necessarily reflect those of the European Union or the European Research Council Executive Agency. Neither the European Union nor the granting authority can be held responsible for them. The research of CCW is supported by NOIRLab, which is managed by the Association of Universities for Research in Astronomy (AURA) under a cooperative agreement with the National Science Foundation. Funding for this research was provided by the Johns Hopkins University, Institute for Data Intensive Engineering and Science (IDIES).

This work is based on observations made with the NASA/ESA \emph{Hubble Space Telescope} (\emph{HST} and NASA/ESA/CSA \emph{James Webb Space Telescope} (\emph{JWST}), obtained from the Mikulski Archive for Space Telescopes (MAST) at the Space Telescope Science Institute (STScI), which is operated by the Association of Universities for Research in Astronomy, Inc., under NASA contract NAS 5-03127 for \emph{JWST}, and NAS 5–26555 for \emph{HST}. JADES DR5 includes NIRCam data from JWST programs 1176, 1180, 1181, 1210, 1264, 1283, 1286, 1287, 1895, 1963, 2079, 2198, 2514, 2516, 2674, 3215, 3577, 3990, 4540, 4762, 5398, 5997, 6434, 6511, and 6541. The authors acknowledge the teams of programs 1895, 1963, 2079, 2514, 3215, 3577, 3990, 6434, 6541 and 8060 for developing their observing program with a zero-exclusive-access period. The authors acknowledge use of the {\it lux} supercomputer at UC Santa Cruz, funded by NSF MRI grant AST 1828315. This work is based on observations taken by the MUSE-Wide Survey as part of the MUSE Consortium.

This research made use of Astropy,\footnote{http://www.astropy.org} a community-developed core Python package for Astronomy \citep{astropy2013, astropy2018}.

\section*{Data Availability}

The \emph{JWST}/NIRCam imaging data and photometry used in this work are publicly available through the JADES Data Release 5 (DR5), see \citet{Johnson2026} and \citet{Robertson2026}. Any remaining data underlying the analysis in this article will be shared on reasonable request to the first author.



\bibliographystyle{mnras}
\bibliography{main.bib} 






\bsp	
\label{lastpage}
\end{document}